\documentclass[conference]{IEEEtran}
\usepackage{ifpdf}
\usepackage{amsthm}
\usepackage{cite}
\usepackage{comment}
\usepackage{xcolor}
\newtheorem{definition}{Definition}

\ifCLASSINFOpdf
   \usepackage[pdftex]{graphicx}
 
\else
 
   \usepackage[dvips]{graphicx}
 
\fi
\usepackage[cmex10]{amsmath}
\usepackage{algorithmic}
\usepackage{array}
\usepackage{mdwmath}
\usepackage{mdwtab}
\usepackage{eqparbox}
\usepackage[font=footnotesize]{subfig}
\usepackage{fixltx2e}
\usepackage{stfloats}
\usepackage{url}

\newcommand{\Cc}{{\cal C}}

\newcommand{\Mc}{{\cal M}}

\newcommand{\Pc}{{\cal P}}

\newcommand{\Vc}{{\cal V}}

\begin{document}
\title{PoF: Proof-of-Following for Vehicle Platoons}

\vspace{-0.8in}

\author{\IEEEauthorblockN{Ziqi Xu\IEEEauthorrefmark{1}, Jingcheng Li\IEEEauthorrefmark{1}, Yanjun Pan, Loukas Lazos, Ming Li}
\IEEEauthorblockA{Department of Electrical and Computer Engineering\\
University of Arizona\\
\{zxu1969, jli2972, yanjunpan, llazos, lim\}@email.arizona.edu}
\and
\IEEEauthorblockN{Nirnimesh Ghose}
\IEEEauthorblockA{School of Computing\\
University of Nebraska–Lincoln\\
nghose@unl.edu}
}

\IEEEoverridecommandlockouts
\makeatletter\def\@IEEEpubidpullup{6.5\baselineskip}\makeatother
\IEEEpubid{\parbox{\columnwidth}{
    Network and Distributed Systems Security (NDSS) Symposium 2022\\
    24-28 April 2022, San Diego, CA, USA\\
    ISBN 1-891562-74-6\\
    https://dx.doi.org/10.14722/ndss.2022.23077\\
    www.ndss-symposium.org
}
\hspace{\columnsep}\makebox[\columnwidth]{}
}

\maketitle

\renewcommand{\thefootnote}{\fnsymbol{footnote}}
\footnotetext[1]{These authors contributed equally to this work.}

\begin{abstract}
Cooperative vehicle platooning significantly improves highway safety, fuel efficiency, and traffic flow. In this model, a set of vehicles move in line formation and coordinate acceleration, braking, and steering using a combination of physical sensing and vehicle-to-vehicle (V2V) messaging. The authenticity and integrity of the V2V messages are paramount to safety. For this reason, recent V2V and V2X standards support the integration of a PKI. However, a PKI cannot bind a vehicle's digital identity to the vehicle's physical state (location, velocity, etc.). As a result, a vehicle with valid cryptographic credentials can impact platoons from a remote location. 

In this paper, we seek to provide the missing link between the physical and the digital world in the context of vehicle platooning. We propose a new access control protocol we call Proof-of-Following (PoF) that verifies the following distance between a candidate and a verifier. The main idea  is to draw security from the common, but constantly changing environment experienced by the closely traveling vehicles. We use the large-scale fading effect of ambient RF signals as a common source of randomness to construct a {\em PoF} primitive. The correlation of large-scale fading is an ideal candidate for the mobile outdoor environment because it exponentially decays with distance and time. We evaluate our PoF protocol on an experimental platoon of two vehicles in freeway, highway, and urban driving conditions. We demonstrate that the PoF withstands both the pre-recording and following attacks with overwhelming probability. 
\end{abstract}

\section{introduction}

Cyber-physical systems (CPSs) enable a plethora of technological innovations that will dramatically improve everyday life. One prime CPS example is  autonomous driving systems (ADSs) for  coordinating a set of autonomous vehicles (AVs) safely, securely, and efficiently  \cite{ADSs,platoon2020}. In ADS, multiple connected vehicles use on-board sensors and vehicle-to-vehicle (V2V) communications to coordinate their actions and improve on safety, fuel-efficiency, traffic flow, and driving convenience \cite{bian2019reducing}. When applied on a single lane, this cooperation model is referred to as cooperative adaptive cruise control (CACC) and can be used to form semi-autonomous, or autonomous vehicle platoons \cite{turri2016cooperative,lyamin2016study}. Several works have shown that the V2V messages exchanged between platoon members can significantly reduce the platoon following distance (from 2 seconds to as much as 0.5 seconds), without compromising the platoon safety \cite{ song2021organization, luu2020spacing}.

However, the complex integration of multi-modal physical sensing, computation, and communication creates a particularly challenging environment to safeguard. The safety of the platoon relies on the veracity of the V2V messages exchanged between platoon members, as falsified messages about acceleration, location, and  velocity can lead to life-threatening accidents, damage to high-value cargo, and monetary loss \cite{zhang2020distributed, ko2021approach}.  The key security questions for a platooning application are: (a) who is authorized to participate in the platoon and  how is the identity of the platoon members verified? (b) how is the integrity of the V2V messages guaranteed? (c) how is the veracity of V2V messages validated?

\begin{figure}[t]
\centering
\includegraphics[width=0.8\linewidth]{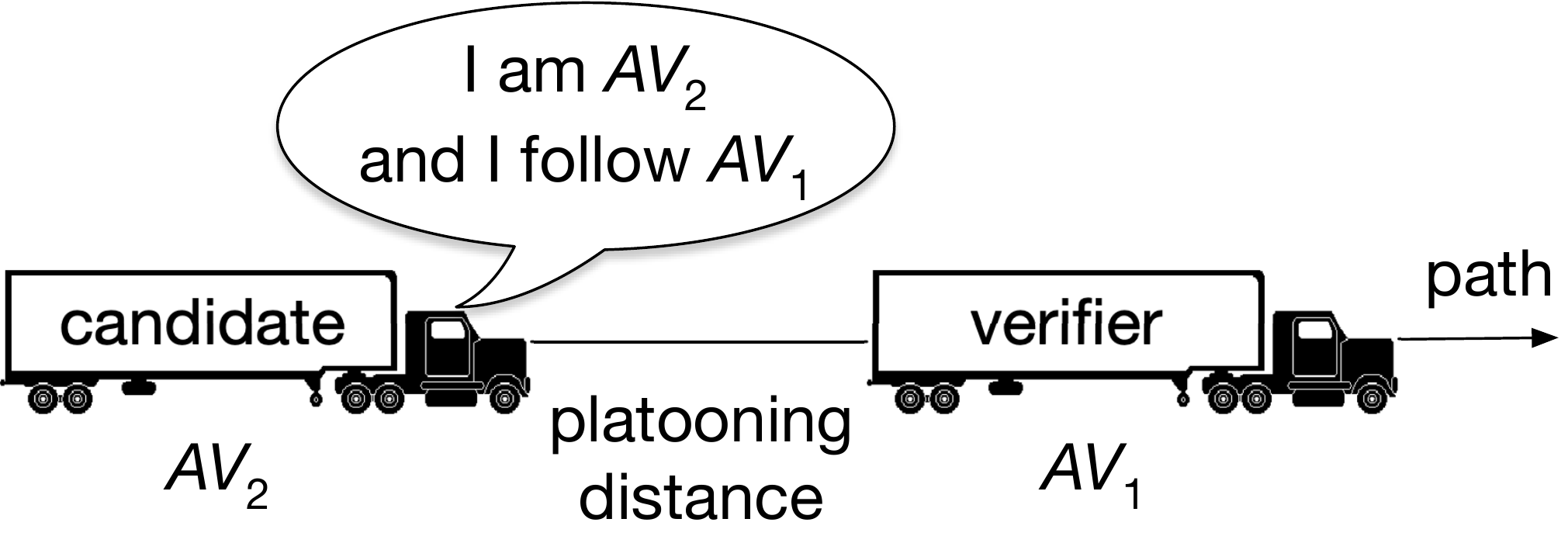}
\vspace{-0.1in}
\caption{Platooning of $AV_1$ and $AV_2.$ The $AV_1$ acts as a verifier to validate $AV_2$'s claim that it follows the platoon.
}
\vspace{-0.2in}
\label{fig:scenario}
\end{figure}

Whereas some of these problems can be addressed with  traditional information security methods (e.g., source authentication and message integrity), others such as access control and verification of V2V messages cannot be achieved cryptographically. To demonstrate this shortcoming, consider the scenario of Fig.~\ref{fig:scenario} where $AV_1$ is followed by $AV_2.$ 
Existing  wireless  standards, including the  IEEE 1609.2 for V2V communication \cite{IEEE:WAVE} and  the more recent 3GPP TS 33.185 for Cellular Vehicle-to-Everything \cite{secureV2X} recommend the use of a public key infrastructure (PKI). Using the PKI, the two vehicles can mutually authenticate and exchange messages whose integrity and confidentiality are guaranteed.

However, {\em the PKI cannot bind a vehicle's digital identity to the vehicle's physical location and state,} allowing for the impersonation of ``ghost'' vehicles \cite{bissmeyer2012central, han2017convoy}, injection of false data from remote locations, and ultimately jeopardizing the safety and efficiency of the platoon. Note that even if $AV_1$ uses its physical sensors to cross-validate the information contained in $m$, this verification cannot serve as a valid proof.  For instance, even if $AV_1$ detects a following vehicle, $AV_1$ has no means to bind the detected vehicle with $m$. 

{\bf Proof-of-following.} In this paper, we seek to provide a new form of access control, which we call proof-of-following (PoF). A PoF aims at binding the digital identity of a candidate vehicle with the property of following a mobile verifier within typical platooning distance, referred to as {\em the following distance.} PoF primitives prevent malicious vehicles that do not follow the platoon from remotely injecting messages either via long-range V2V communication or the V2I infrastructure. We emphasize that admitted platoon members that follow the platoon closely can still potentially inject false messages. The significance of the PoF lies in restricting access to physically platooning  members  only, thus substantially increasing the adversarial effort for scaling a false injection attack. Without a PoF, a remote adversary could potentially join and impact many platoons simultaneously from a single remote location.

A PoF protocol provides similar access control to distance bounding protocols \cite{brands1993distance, tippenhauer2015uwb, Avoine2018survey} and proximity verification methods \cite{mathur2011proximate,miettinen2014context,hayashi2013casa, conti2020context} with notable differences. A distance bounding protocol verifies that a prover is located within bounded distance from the verifier at one time instance without taking into account mobility and time. A PoF protocol continuously verifies a physical distance bound over time while the involved entities are moving. Although a PoF can be implemented as a repeated application of distance bounding, we are exploring a looser form of verification where the distance bound does not need to be strictly met at every time instance. This model readily corresponds to a vehicle platooning application where the distance between the platooning vehicles could naturally fluctuate.  Moreover, distance bounding protocols require UWB communications and custom hardware that has been optimized to minimize the modulation symbol size and any processing delays to remain secure \cite{tippenhauer2015uwb}.

\begin{figure}[t]
\centering
\includegraphics[width=0.95\linewidth]{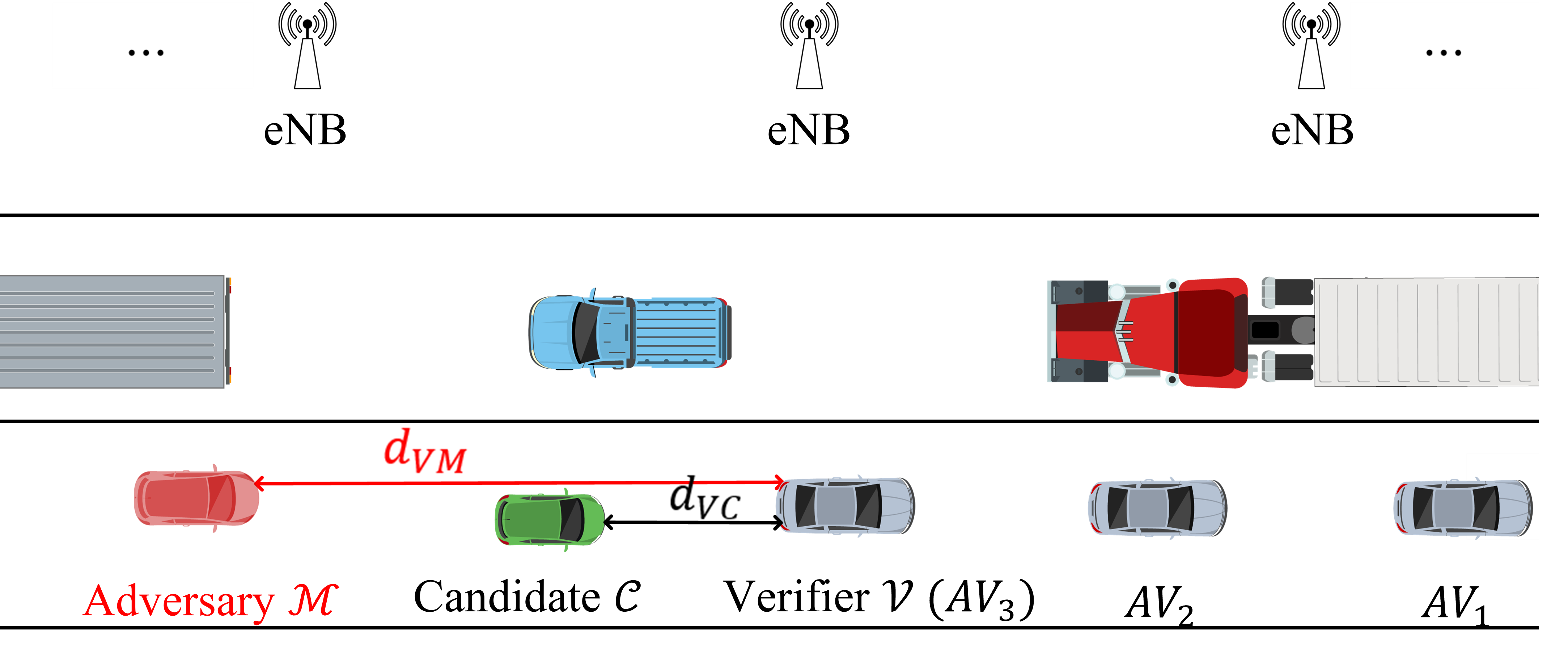}
\caption{A platoon of three vehicles with $AV_3$ acting as a verifier. The candidate and the verifier execute a PoF by sampling the ambient RF signals transmitted by the LTE eNBs. 
}
\label{model}
\vspace{-0.2in}
\end{figure}

{\bf Main idea of PoF.} The main idea of our PoF is inspired by a common car game called ``I spy''. In I spy, one player (the spy) chooses a visible object and announces it to other players with some attribute description (first letter, color, size). The other players have to guess the spied object. The game is ideal for car journeys because the visible objects are continuously renewed. Similar to the common vision of co-travelers in the I spy game,  {\em if the candidate and verifier vehicles are platooning, they should see (sense) the same  environment.} Security is drawn from the rapidly changing environment due to motion. Although several different modalities can be used to sense the environment, we opt to measure ambient RF signals. Specifically, our PoF protocol exploits the large-scale fading characteristics of RF propagation to correlate the moving paths of the platoon members.  By continuously sampling ambient RF signals from cellular towers, platoon members verify that they sense the same RF environment. The main idea is demonstrated in Fig.~\ref{model}.

 The use of ambient RF signals from the cellular infrastructure has several notable advantages. From a practical perspective, our method operates in-band using only a cellular receiver. It does not require any additional sensors such as cameras, LiDAR, etc. A cellular transceiver is likely to be standard equipment given the global momentum for the adoption of the Cellular-V2X (C-V2X) 3GPP standard \cite{3GPP, secureV2X, qualcommV2X}. From a security perspective, RF signals decorrelate rapidly with distance and time, especially when mobility is involved \cite{gudmundson1991correlation, szyszkowicz2010feasibility}. Thus, predicting the instantaneous RF environment other than pre-recording signals along a route or following at a large distance becomes impossible.

\textbf{Contributions.} Our main contributions are  as follows. 
\begin{itemize}
    \item We define the Proof-of-Following (PoF) security primitive for performing physical access control in the context of vehicle platooning. We develop a PoF protocol which enables a candidate vehicle prove to a verifier that it follows the verifier within the following distance. The PoF protocol  binds the ``following'' physical property to the candidate's digital identity. The protocol enables new vehicles to join a platoon and also the continuous verification of platooning for existing members. 
    
    \item Our PoF protocol exploits the large-scale RF propagation characteristics to correlate the motions of the candidate and the verifier.  We are the first to exploit the large-scale fading property (mainly due to shadowing) as a new modality. It can accommodate typical platooning distances (tens of meters \cite{luu2020spacing}), and we show it is suitable for outdoor mobile settings due to the unpredictability of the surrounding environment. Besides the RF spatial correlation, we also utilize the temporal correlation.

    \item We demonstrate the security of our PoF protocol against  an attacker that pre-records the RF environment along the route of the platoon,   one that follows the platoon but at a longer distance, and one that partially follows the platoon. In addition, we show that our protocol is secure against Man-in-the-Middle attacks when the verifier's identity is known to the candidate, and adapt our protocol to deal with unknown verifiers using commitments with a delayed opening phase.
    
    \item We experimentally evaluate the performance and security of our protocol against our adversary model using a USRP radio-equipped candidate-verifier vehicle testbed in urban, freeway, and highway driving settings. In such realistic conditions, we demonstrate that the PoF withstands both the pre-recording and following attacks with overwhelming probability. 
    
\end{itemize}

{\bf Limitations:} Our PoF protocol has two limitations. First, it can only verify following for distances up to some bound. This is because RF measurements decorrelate exponentially with distance and approach zero after such bound. We emphasize that this is not a limitation for platooning applications because the benefits of platooning can only be reaped if the formation is tight (typically less than 25m)  \cite{platoon2020,turri2016cooperative}. 

Second, our PoF construct does not verify the relative vehicle positioning. Though positioning is important, we have left the construct  more general to verify that the candidate is around the verifier rather than exactly behind it. This allows for more flexibility in the application domain. Note that an adversary would have to practically follow the platoon in close distance to be able to pass a PoF test which is equivalent to actually being part of the platoon. That is, if the adversary can pass a PoF test, the adversarial effort of being anywhere around the platoon is similar so it has no reason to not actually follow. From our general construct, relative positioning methods can be further developed. For instance, if multiple verifiers are incorporated, the candidate must be located in the intersection of the respective distance bounds.

\textbf{Paper organization:} Section \ref{sec:related work} presents related work. In Sec. \ref{sec:model}, we state the system and threat models. We present the PoF protocol in Sec. \ref{sec:System design}, and  experimentally evaluate the PoF in Sec. \ref{sec:evaluation}. Future directions are discussed  in Sec.~\ref{sec:discussion} and in Sec.~\ref{sec:conclusion}, we conclude.

\section{Related Work}
\label{sec:related work}

\textbf{Physical context verification for connected vehicles/platoons.}
Specific to vehicular  applications, various methods have been proposed to verify claimed physical properties of vehicles \cite{so2019physical,Nguyen2020Enhancing,Kamel2020Simulation,sun2017data, sun2020svm}. For example, secure localization/tracking  \cite{Nguyen2020Enhancing} or motion verification approaches \cite{sun2017data, sun2020svm}, which  check the consistency between a vehicle's claim with  other measurable features of wireless signals (e.g.,  angle-of-arrival or Doppler shift). However, the problem of misbehavior detection is different from  platoon membership verification, and verifying  the exact location of a vehicle may be too taxing.

The works closest to ours are those directly addressing platoon membership authentication \cite{han2017convoy, juuti2017stash, vaas2018get}.  
Han $et\ al.$ \cite{han2017convoy}  leveraged the physical context to authenticate a new candidate. They observed that platooning vehicles will record similar vertical accelerations due to uneven road conditions. However, this approach does not prevent record and replay attacks since the road surface condition rarely changes. Vaas $et\ al.$ \cite{vaas2018get} and Juuti $et\ al.$ \cite{juuti2017stash} used driving trajectory as a proof for platoon membership, which compares a  candidate vehicle's future route with a trusted vehicle in the platoon. After being promoted as a co-presence vehicle, the platoon then authenticates its V2V messages. However, the trajectory can be predictable, especially by a following afar adversary.  Compared with above works, our scheme can defend against both replay and following-afar attacks. 

\textbf{Distance bounding.} In  distance bounding (DB) \cite{brands1993distance, tippenhauer2015uwb, Avoine2018survey}, a verifier $\mathcal{V}$ interacts with a prover $\mathcal{P}$ to ensure that $\Pc$ is no further than a distance $d$ from $\Vc$.  The general idea of DB constructions is to engage the two parties in a challenge-response protocol such that the round-trip time measured over a fast bit exchange can be bound \cite{Avoine2018survey}. However, realization of DB protocols is challenging because secure ranging systems with nanosecond accuracy are required. Tippenhauer {\em et al.}  designed a secure DB system that can achieve cm level accuracy \cite{tippenhauer2015uwb}. However, their design required a custom UWB transceiver with minimum processing delay of 100ns and a shortened modulation symbol size to eliminate early detect/late commit attacks. Our proposed PoF method operates in-band with commercial-off-the-shelf V2X transceivers.

In theory, a PoF can be implemented by the repeated application of a DB  protocol. Like the PoF, a DB protocol provides a form of location-based access control between a prover (candidate) and a verifier. However, a DB protocol verifies {\em an instance}  of the relative location relationship between the candidate and the verifier. In the PoF protocol presented in our work, the candidate collects RSS values over a period of time (in the order of minutes), as opposed to one bit exchange. Repeated executions of the PoF allow for a continuous verification of the PoF property.

\textbf{Physical context-based proximity verification.} The underlying idea of context-based proximity verification  is to leverage common context that is observable by two of more devices in close proximity to establish a shared secret and authenticate the devices, while an adversary that is located far away cannot obtain a similar context. Works in this domain can be divided into two broad categories; in-band RF methods \cite{mathur2011proximate,shi2013bana,wu2018survey} or out-of-band methods using other modalities such as sound, light, temperature, etc. \cite{schurmann2011secure, miettinen2014context, han2018you, li2020t2pair}. In-band RF methods leverage the small-scale fading of wireless signals to verify the co-presence of devices within a very short distance. This is because small-scale fading is mainly caused by multi-path distortion which quickly decorrelates with distance. Typical distances are a few wavelengths (e.g., the wavelength is 12.5cm at 2.4GHz). Therefore, in-band methods mainly find application in indoor/static environments and are not suitable for vehicle platoons.

Out-of-band methods use a variety of modalities, such as ambient luminosity \cite{miettinen2014context}, audio \cite{schurmann2011secure}, etc., to establish proximity. While they do not have the restriction of limited proximity range of the small-scale  RF fading,  they require the devices to be equipped with  the same sensing hardware. Recently, the problem of context-based device pairing for heterogeneous Internet of Things (IoT) devices was studied by Han $et\ al.$ \cite{han2018you} and Li $et\ al.$ \cite{li2020t2pair}, where devices may not share the same sensing interface. However, one challenge of all the out-of-band approaches is that the sensing modality may lack enough entropy in the outdoor setting (e.g. change of luminosity during the day). We emphasize that out-of-band methods have only been tested in confined indoor settings with clear physical separations (e.g., walls) between the adversary and the legitimate parties. Such separations may not hold true in the mobile outdoor setting. For instance, a far-away vehicle could still sample the same luminosity with a verifier travelling hundreds of meters ahead. At the same time, a valid candidate and verifier could sample drastically different ambient sound environments even when they platoon.

In contrast, we are the first to exploit the large-scale fading (mainly due to shadowing), which is a new modality and only requiring a common RF interface. It can accommodate typical following distances (several tens of meters) and we show it is suitable for outdoor mobile settings due to the unpredictability of the surrounding environment. Besides the spatial correlation of the large-scale fading, we also utilize the temporal correlation, which is another novel aspect.

\section{Models and Assumptions}
\label{sec:model}

\subsection{Platooning Model}
Although a PoF primitive is general and can be applied to various mobile scenarios where verification of following is necessary, we explore it in the context of a vehicle platooning application. Platoons are led by a manually-operated or autonomous vehicle, which is followed by autonomous or semi-autonomous vehicles \cite{guanetti2018control}. Platoon members coordinate driving by sensing the physical environment and also exchanging control messages that contain motion state information such as acceleration, velocity, steering, etc.  \cite{jia2015survey}. Vehicles may be equipped with sensors (e.g., cameras, radar or LiDAR), and run control algorithms such as cooperative adaptive cruise control (CACC) \cite{turri2016cooperative,lyamin2016study} to maintain a fixed distance.

To secure the platoon operation, the V2V messages are protected using cryptographic primitives. According to the  C-V2X  communication standard (3GPP TS 33.185 \cite{secureV2X}), V2X communication is supported by a PKI that provides each vehicle  a private/public key pair and a digital certificate as a proof of identity. These credentials can be used to establish trust among the platoon vehicles. We assume that digital signatures are used to prove the source authenticity of messages. Key management of digital identities and platoon secrets is beyond the scope of this work. A PoF involves the following entities.
 
\textbf{Candidate ($\mathcal{C}$)}:
The candidate vehicle wishes to join a moving platoon by sending a join request to the platoon verifier. The candidate is in possession of a public/private key pair $(pk_\mathcal{C}, sk_\mathcal{C})$ and a certificate $cert_\mathcal{C}$ that is issued by a trusted certificate authority. The candidate vehicle is not allowed to receive or transmit any platoon coordination messages before it completes a PoF with the verifier.

\textbf{Verifier ($\mathcal{V}$)}: 
 The verifier is an existing platoon member that is responsible to verify the digital identity of the candidate and that he indeed physically follows the platoon. The verification process may involve the verifier alone or require interaction with other platoon members. Typically, the role of the verifier is assumed by the last platoon vehicle. Once a candidate is admitted, its public key is added to the list of platoon members by all the other vehicles in the platoon. The verifier is given a public/private key pair $(pk_\mathcal{V}, sk_\mathcal{V})$ and a certificate $cert_\mathcal{V}.$

\subsection{Threat Model}
\label{sec:threat model}

\textbf{Attacker goals and capabilities.} We consider an external attacker $\mathcal{M}$ who attempts to pass a {\em PoF} verification without following the platoon. The ultimate goal of the adversary is to be admitted into the platoon and inject falsified coordination messages. The attacker is assumed to be in possession of a valid public/private key pair $(pk_M, sk_M)$ and a certificate $cert_M$ issued by a trusted certificate authority. Further, the adversary can control the communication channel between $\Cc$ an $\Vc$ and inject, replay, modify, or delete messages of his own choosing. We consider three adversary models.

\begin{enumerate}

    \item {\em Remote adversary.} A remote adversary is stationed at some location away from the moving platoon and uses the existing infrastructure (cellular tower or road side units) to communicate with the platoon. The adversary is aware of the platoon's route in advance and in real time. The adversary can use this knowledge to traverse and observe (e.g., measure the RF environment) the platoon's route ahead of time. He requests to join the platoon, pretending to be a vehicle that follows the platoon.

    \item {\em Following-afar adversary.} A following-afar adversary tails the platoon from a long distance that does not meet the following distance requirement, but still allows him to communicate with the platoon. As an example, the adversary could be within a few hundred meters from the platoon. The adversary is also aware of the platoon's route and can traverse it ahead of time. Moreover, since the adversary follows the platoon from afar, he can obtain more up-to-date RSS measurements in real time.

    \item {\em Partially-following adversary.} A partially-following adversary follows the platoon within the following distance only for a fraction of time and then trails the platoon from a far distance or becomes a remote adversary.

\end{enumerate}

\begin{figure}[t]
\centering
    \includegraphics[width=0.95\linewidth]{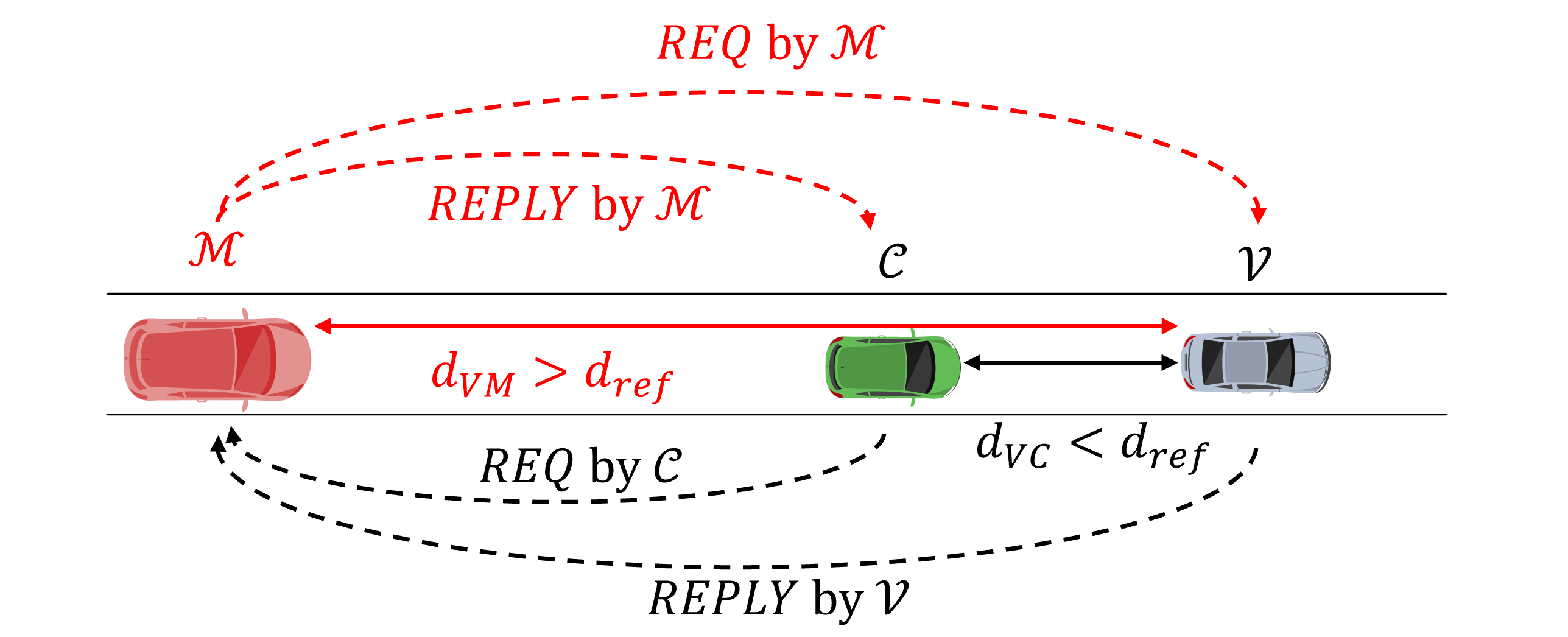}
\vspace{-0.1in}
\caption{The topology of a MiTM attack. $\mathcal{M}$ is beyond the following distance $d_{ref}.$}
\label{fig:MitM_topology}
\vspace{-0.3in}
\end{figure}

\textit{Man-in-the-middle attacks.} For all three attack models, the adversary can launch man-in-the-middle (MiTM) attacks to gain admittance to the platoon. During a MiTM attack, $\mathcal{M}$ maintains parallel sessions with the candidate and the verifier in an attempt to pass a PoF verification while not following the platoon. A MiTM attack is shown in Fig.~\ref{fig:MitM_topology}. A candidate $\mathcal{C}$ requests to join the platoon represented by verifier $\mathcal{V}.$ The adversary intercepts the request and  opens parallel sessions with $\mathcal{C}$ and $\mathcal{V}$ in an attempt to be admitted to the platoon.

\textit{DoS attacks.} We do not consider DoS attacks in which the adversary attempts to deny $\mathcal{C}$ from joining the platoon. Such attacks do not violate the PoF property.

\subsection{Proof-of-Following}
\label{sec:definitions}
To formally define the following property, we first give a definition of a route for a moving vehicle.

\begin{definition}
{\bf Route:} A route $\mathcal{L}_X$ of a moving object $X$ is represented as a set of $n$ time-ordered positions $\mathcal{L}_X=(\ell_X(1), t(1))\rightarrow (\ell_X(2), (t(2))\rightarrow\ldots\rightarrow(\ell_X(n), t(n)),$ where each position $\ell_X(i)$ is the object's geospatial coordinate $(x_X(i), y_X(i))$ at time $t(i),$ with $1\leq i \leq n$, and $t(i)<t(j)$ for $i<j$. 
\end{definition}

Based on the route definition, we  now provide the   definition of following.

\begin{definition}
{\bf Following:} Let a verifier $\mathcal{V}$ move on a route $\mathcal{L}_\mathcal{V}$ and a candidate move on a route $\mathcal{L}_\mathcal{C}$. The candidate $\mathcal{C}$ is said to follow verifier $\mathcal{V}$ if the Euclidean distance between $\mathcal{V}$ and $\mathcal{C}$ for each time $t(i)$ is bounded by
\[
||\ell_{\mathcal{V}}(i) - \ell_{\mathcal{C}}(i)|| \leq d_{ref},~~~\forall~i \leq n,
\] 
where $d_{ref}$ is a desired following distance bound.
\label{Following}
\end{definition}

\begin{definition}
{\bf Proof-of-Following:} A PoF is a protocol executed between a verifier $\mathcal{V}$ and a candidate $\mathcal{C}$. If the candidate $\mathcal{C}$ always follows  $\mathcal{V}$ (i.e.,  $||\ell_V(i) - \ell_C(i)|| \leq d_{ref}$, $\forall~i \leq n$),  $\mathcal{V}$ outputs accept. If $\mathcal{C}$ always does not follow  $\mathcal{V}$ (i.e.,  $||\ell_V(i) - \ell_C(i)|| > d_{ref}$, $\forall~i \leq n$), $\mathcal{V}$ outputs reject. 
\label{PoF}
\end{definition}

Note that our {\em PoF} definition is a relaxed one as it only differentiates between the two extreme cases (always follows and always does not follow). The most strict PoF definition would output reject if the candidate $\mathcal{C}$ does not follow  $\mathcal{V}$ for any of the $n$ time points (rather than all).  However, even for legitimate following vehicles, some error or brief violation of following should be tolerated (e.g., when vehicles are human-driven, or not following using a strict CACC algorithm). 

Because of this, our PoF definition is not simply a generalization of repeated distance bounding tests over a discretized route mobile setting. Note also that the definition does not place any restriction on the location sampling rate, which can be adjusted based on the application scenario. Moreover, the definition does not specify the relative positioning between the two moving objects. That is, the candidate can be around the verifier (either leading or following). This is to allow for a more general definition, which can be further restricted based on the application requirements. 

Another aspect of the definition is that it only requires one-way PoF, from $\Cc$ to $\Vc.$ This provides protection of the platoon from non-following candidates. However, it does not protect candidates from joining fictitious verifiers. The same PoF definition can be applied with the roles of $\Cc$ and $\Vc$ reversed to allow for mutual verification of the PoF property.

\section{A PoF Construct based on the RF Environment}
\label{sec:System design}

\subsection{Main Idea}
\label{sec:motivation}

The chief idea of constructing our PoF primitive is to exploit the randomness of the continuously changing environment due to mobility to prove continuous vehicle proximity. The selected modality for perceiving the environment should satisfy two important criteria.

\begin{enumerate}
    \item The environment should exhibit spatial and temporal decorrelation. 
    \item The environment should exhibit high entropy and should not be repeatable. 
    \end{enumerate}
Several modalities such as sound, vision, and RF may meet the two criteria. For instance, ambient sound while travelling on a freeway decorrelates rapidly with distance. Moreover, it varies significantly with time at the same location. 

We have opted to exploit both the spatial and temporal correlation of ambient wireless signals. Specifically, a legitimate candidate who is closely following the platoon will observe a similar RF environment as the verifier. Besides, the RF environment is dynamic with time and location due to the constant change of the physical environment and the motion of other vehicles. The temporal variation (short channel coherence time) can prevent an adversary from pre-recording ambient RF signals and replaying them to a traveling platoon. Moreover, the RF modality is widely available for outdoor scenarios. Vehicles will already be retrofitted with cellular receivers to support the C-V2X standard \cite{secureV2X}. In our PoF, vehicles exploit the ambient RF signals transmitted by cellular base stations (eNBs) along the traversed route.

\subsection{Rationale and Feasibility Study}
\label{sec:Feasibility study}

In this section, we conduct a feasibility study on exploiting large-scale fading as a PoF modality. 

{\bf Why large-scale fading?}  
Large-scale fading is the result of signal attenuation due to signal propagation through  large distances and diffraction around large objects in the propagation path. 
The wireless signal propagation loss can be represented using the well-known log-distance path loss model \cite{recommendation1997guidelines}:
\begin{align}
    L(d_{tr}) =\overline{L}(d_0 ) + 10\beta log \left(\frac{d_{tr}}{d_0}\right)+ X_{\sigma}, 
    \label{pathlossmodel}
\end{align}
where $L(d_{tr})$ is the propagation loss (or large-scale fading), $d_{tr}$ is the distance between the transmitter (TX) and receiver (RX), $d_0$ is the reference distance, $\overline{L}(d_0)$ is
the path loss at $d_0$,  $\beta$ is the path loss exponent, and $X_{\sigma}$ is
the  shadow fading. 

Since the large-scale  fading  is impacted by terrain configuration between the TX and RX, it  brings randomness and unpredictability as the vehicles move.  It is more stable when two closely-located vehicles sense ambient RF signals from far-away cellular base stations because the distance and diffraction from a base station to the two vehicles  is approximately the same.  Moreover, large-scale fading in mobile outdoor scenarios  decorrelates more gracefully with distance and time than small-scale fading \cite{gudmundson1991correlation, szyszkowicz2010feasibility}. 

\begin{figure}[t]
\centering
\includegraphics[width=0.9\linewidth]{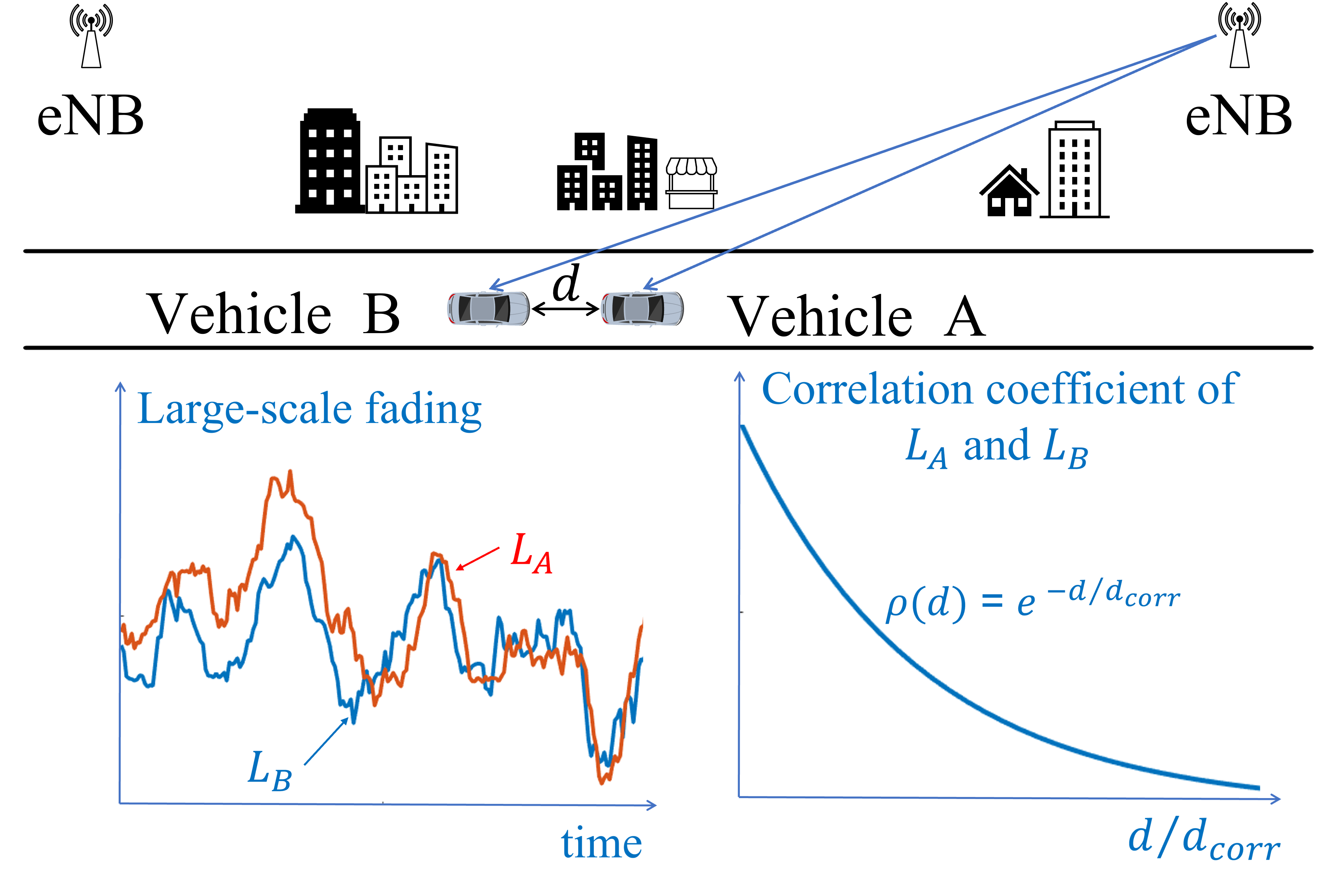}
\vspace{-0.1in}
\caption{Large-scale fading correlation model.}
\label{fig:ccmodel}
\vspace{-0.2in}
\end{figure}

Several models have been proposed  to capture the spatial decorrelation of large-scale fading \cite{gudmundson1991correlation, senarath2007multi, szyszkowicz2010feasibility}. The exponential model is the one that has been most widely adopted \cite{gudmundson1991correlation,recommendation1997guidelines,guan2015measurements}. In Fig. \ref{fig:ccmodel}, let two vehicles $A$ and $B$ simultaneously measure the large-scale fading from the same base station, denoted by $L_{A}$ and $L_{B}$, respectively. The correlation  $\rho_d$ between $L_{A}$ and $L_{B}$ is expressed as
\begin{align}
    \rho_d=e^{- d/d_{corr}},
    \label{exp model}
\end{align} 
where $d_{corr}$ is  the decorrelation distance, which depends on the physical environment \cite{algans2002experimental, he2014shadow}. From this model, we expect the correlation to be high for vehicles with distances smaller than $d_{corr}$, but to drop significantly for larger distances. Several $d_{corr}$ values have been empirically determined for different mobile environments  \cite{algans2002experimental, he2014shadow}.

{\bf Experimental validation.} To validate the spatial and temporal  correlation properties of large-scale fading, we collected  measurements of LTE signals in a freeway environment, which is the most relevant for platooning applications. The data was collected by driving a platoon of two vehicles $A$ and $B$ and simultaneously measuring the RSS (Note that, since RSS (in dB) is the difference between the transmit  power  and path loss, measuring RSS is equivalent  to measuring the fading). from eNBs. The location and timestamp of each sample was also recorded to allow for time sample alignment and the computation of the separation $d.$   Figure~\ref{fig:ccmodel} shows the topology used in the experiments, along with a sample realization of the measured large scale fading samples and the correlation model in eq. \eqref{exp model}. The experimental setup is described in detail in Sec.~\ref{sec:setup}. We tested the following two main hypotheses.

\begin{enumerate}

    \item \textit{Spatial correlation decreases with distance.} Here, we seek to verify the correlation model in \cite{gudmundson1991correlation} and determine the decorrelation distance $d_{corr}$.

    \item \textit{Temporal correlation decreases with time.} Here, we seek to verify that the correlation of RF signals collected at the same location but different times decreases with the time difference. 

\end{enumerate}

\begin{figure}[t]
\centering
    \includegraphics[width=0.7\linewidth]{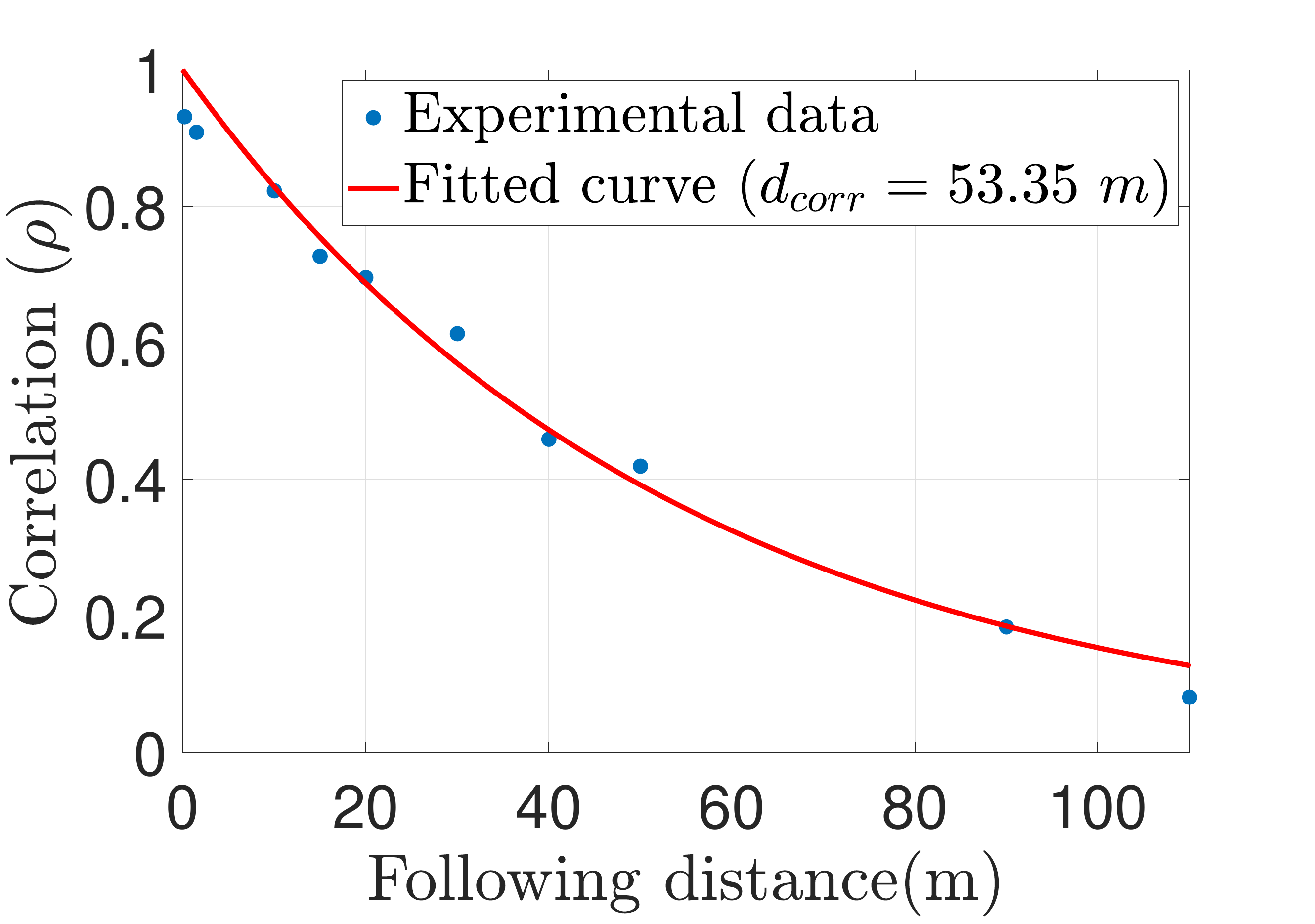}
\caption{Correlation as a function of the following distance $d$ for two vehicles driving in a freeway environment.}
\label{fig:spatial_corr}
\vspace{-0.2in}
\end{figure}

To extract the large-scale fading and filter  out the small-scale fading, we apply an $M$-point moving RSS average $\overline{\gamma}_A(i)$. We then compute the Pearson's correlation coefficient $\rho$ as the correlation metric, defined as:
\begin{align}
    \rho=\frac{\sum\limits_{i=1}^{n}(\overline{\gamma}_A(i)-\overline{\gamma_A})(\overline{\gamma}_B(i)-\overline{\gamma_B})}{\sqrt{\sum\limits_{i=1}^{n}(\overline{\gamma}_A(i)-\overline{\gamma_A})^2}\sqrt{\sum\limits_{i=1}^{n}(\overline{\gamma}_B(i)-\overline{\gamma_B})^2}},
\label{eq:CC}
\end{align}
where $\overline{\gamma_A}$ and $\overline{\gamma_B}$ are the mean values of the RSS moving average over $n$ RSS entries for $A$ and $B$, respectively.

\subsubsection{Spatial Correlation Decreases with Distance}
\label{sec:spatial correlation}

To validate this hypothesis, two vehicles were driven with following distance $d$ on a freeway at 30mph. A total of 6,000  samples were collected at a sampling rate of 20Hz for each $d$ (5 min duration). The samples were organized to subsets of 1,200 samples (1 min duration) and the correlation $\rho$ was computed over each subset, using a 40-point moving average.

Figure~\ref{fig:spatial_corr} shows the correlation $\rho$ averaged over all subsets, as a function of the vehicle separation $d.$ The fitting of the theoretical curve obtained from the exponential model in \eqref{exp model} yielded a decorrelation distance $d_{corr}=53.35m$. This is in the same order of a typical platooning distance bound, indicating that correlation would be an ideal metric to implement the PoF primitive. Note that although the average correlation  fits to a deterministic model, the RSS samples used to compute the correlation are constantly changing with mobility and have high entropy (which we will show in Appendix A).

\subsubsection{Temporal correlation decreases over time}

\begin{figure}[t]
\centering
\setlength{\tabcolsep}{-5pt}
\begin{tabular}{cc}
    \includegraphics[width=0.55\linewidth]{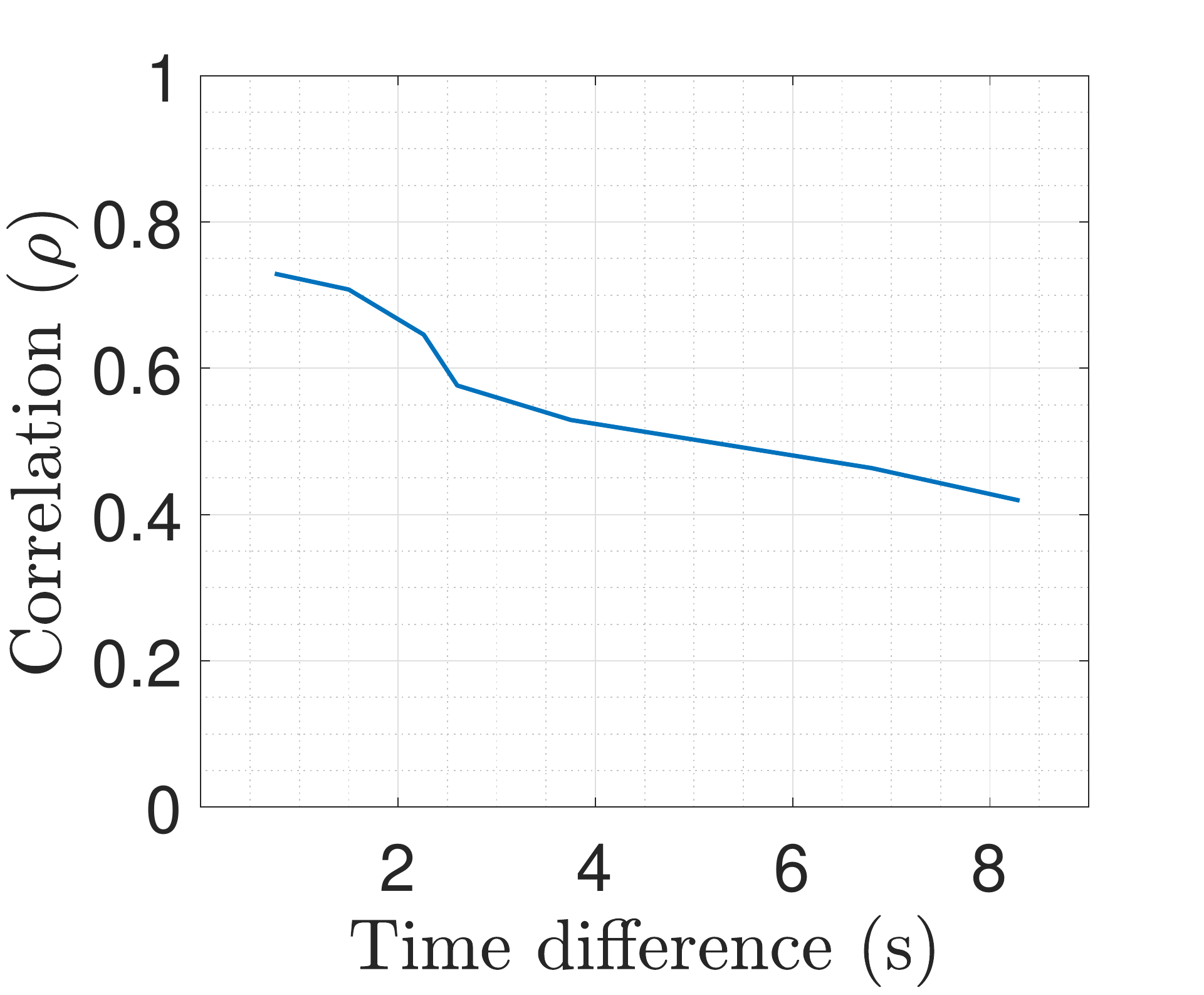} &
    \includegraphics[width=0.55\linewidth]{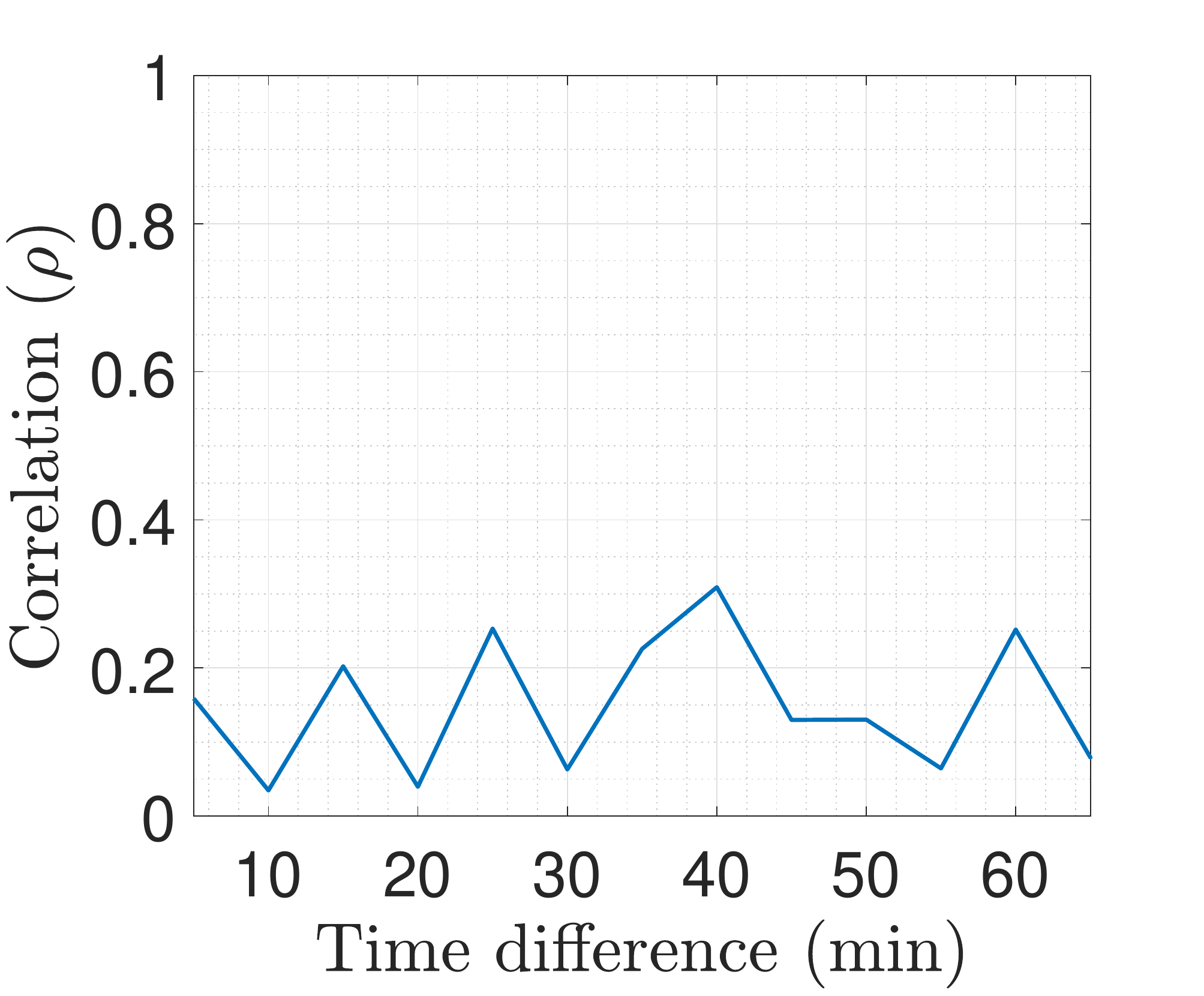} \\
    (a) short time scale & (b) long time scale
    \end{tabular}
\caption{Correlation at the same location.}
\label{fig:temporal_corr}
\vspace{-0.2in}
\end{figure}

To validate the temporal decorrelation hypothesis, we collected LTE transmission samples over the same route but at different times. We drove vehicle $A$ and $B$ platooning over the same freeway segment and computed the correlation $\rho$ between samples collected by the two vehicles, but aligned to the same locations. That RSS samples aligned to the same location but different time, where the difference equals the time separation of the two vehicles. The two vehicles were moving at 30Mph and had fixed time difference from 1s to 9s (13m to 112m). Due to the absence of CACC, fixed separation was achieved by engaging the cruise control on both vehicles and performing many trial runs.
Figure~\ref{fig:temporal_corr}(a) shows the  correlation $\rho$ as a function of the time difference between sampling of the same location. We observe that the temporal correlation drops to fairly low values after a few seconds. This is an important property to prevent pre-recording attacks, where the adversary traverses the platoon route ahead of time to collect historic RSS data and use these data to defeat a PoF verification. Figure~\ref{fig:temporal_corr}(b) shows the temporal correlation for a longer timescale.

From the experimental evaluation of the the spatial and temporal correlation of the large-scale fading effect, we can conclude that it is a good candidate to differentiate a following vehicle as $\rho$ drops to low values for separations larger than the platooning distance (beyond 50m) and also remains low between samples collected at even just a few seconds apart (this is inline with the typical channel coherence time of  outdoor channels for large-scale fading). 

\subsection{Proof-of-Following  Protocol}
\label{sec:authentication steps}

\subsubsection{Overview}
The high level idea of our PoF   protocol is to compute the correlation between the smoothed RSS samples gathered by the candidate and verifier  (corresponding to large-scale fading), and compare it with a threshold $\tau$. Suppose the correlation function $\rho(d)$ follows the deterministic model in Eq. \eqref{exp model},  given a required distance threshold $d_{ref}$, setting $\tau =  e^{-\frac{d_{ref}}{d_{corr}}}$ will make all vehicles following within $d\leq d_{ref}$ pass the PoF test ($\rho(d)\geq \tau$) and all others will fail. However,  the correlation  model  expressed  in Eq.  (2)  is for  the  average  correlation, and  in practice certain  correlation  instances will fluctuate around the average, due to changes of the environment/terrain as vehicles move.   A single test designed according to the   correlation model will only provide a weak probabilistic guarantee for passing or failing the test at different distances (the CDF of the correlation would be required). Thus, we opted  to  organize  the  RSS  data  in $K$ shorter  correlation  tests  instead  of  a  single  long  test. Repeated tests is a standard mechanism for driving the adversary's probability of passing verification towards zero. 

\subsubsection{Protocol Details}

\begin{figure*}[t]
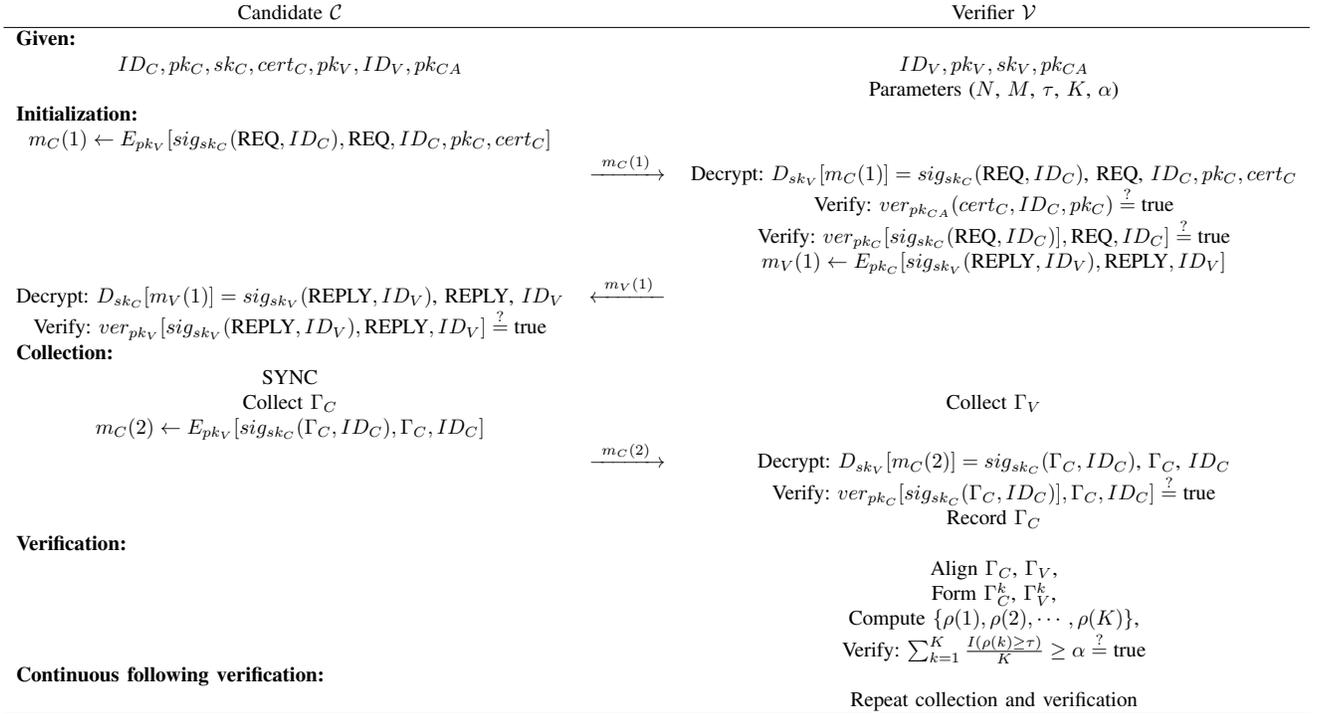

        \centering
        \scalebox{0.79}{
        \begin{tabular}{ccc}
            Candidate $\mathcal{C}$& &  Verifier $\mathcal{V}$  \\
            \hline
            \multicolumn{1}{l}{\bf Given:} & &\\
             $ID_{C},pk_{C},sk_{C},cert_C, pk_{V}, ID_V, pk_{CA}$ & &$ID_{V},pk_{V}, sk_{V}, pk_{CA}$ \\
              & & Parameters ($N$, $M$, $\tau$, $K$, $\alpha$)\\

            \multicolumn{1}{l}{{\bf Initialization:}}& &\\
             $m_C(1) \leftarrow E_{pk_V}[sig_{sk_C}(\text{REQ}, ID_C), \text{REQ}, ID_C, pk_C, cert_C]$ &  &  \\
            
            &$\xrightarrow{~m_{C}(1)~}$& Decrypt: $D_{sk_V}[m_C(1)] = sig_{sk_C}(\text{REQ}, ID_C)$, $\text{REQ}$, $ID_C, pk_C, cert_C $\\
            
            & & Verify:  $ver_{pk_{CA}}(cert_C,ID_C,pk_C)\overset{?}{=} \text{true}$\\
            
            & & Verify: $   ver_{pk_C}[sig_{sk_C}(\text{REQ}, ID_C)],\text{REQ}, ID_C] \overset{?}{=} \text{true}$  \\
            
            & &  $m_V(1) \leftarrow E_{pk_C}[sig_{sk_V}(\text{REPLY}, ID_V), \text{REPLY}, ID_V]$\\
            
            Decrypt: $D_{sk_C}[m_V(1)] = sig_{sk_V}(\text{REPLY}, ID_V)$, $\text{REPLY}$, $ID_V$   &  $\xleftarrow{~m_{V}(1)~}$&\\
            
            Verify: $   ver_{pk_V}[sig_{sk_V}(\text{REPLY}, ID_V),\text{REPLY}, ID_V] \overset{?}{=} \text{true}$ & &\\
            
            \multicolumn{1}{l}{{\bf Collection:}}& &\\
            SYNC & & \\
            Collect $\Gamma_C$ & & Collect $\Gamma_V$\\
             
             $m_C(2) \leftarrow E_{pk_V}[sig_{sk_C}(\Gamma_C, ID_C), \Gamma_C, ID_C]$ &  &\\
            
             &$\xrightarrow{~m_{C}(2)~}$& Decrypt: $D_{sk_V}[m_C(2)] = sig_{sk_C}(\Gamma_C, ID_C)$, $\Gamma_C$, $ID_C$\\
            
            & & Verify: $   ver_{pk_C}[sig_{sk_C}(\Gamma_C, ID_C)],\Gamma_C, ID_C] \overset{?}{=} \text{true}$ \\

          & & Record $\Gamma_C$
          \\ 
          
          \multicolumn{1}{l}{{\bf Verification:}}& &\\
          & & Align $\Gamma_C$, $\Gamma_V$,\\ 
           
          & & Form $\Gamma_C^k$, $\Gamma_V^k$,\\
           
          & & Compute \{$\rho(1), \rho(2), \cdots, \rho(K)$\},\\
           
          & & Verify: $\sum_{k=1}^{K} \frac{I(\rho(k) \geq \tau)}{K} \geq \alpha \overset{?}{=} \text{true}$ \\
          \multicolumn{1}{l}{{\bf Continuous following verification:}}& &\\ 
          & & Repeat collection and verification\\ 
            \hline
        \end{tabular}}
         \caption{The PoF verification protocol. The identity of the verifier is known to $\mathcal{C}$, as $ID_V$ and $pk_V$ are   given parameters.}
     \label{algorithm}
    \vspace{-0.2in}
\end{figure*}

 A {\em PoF} consists of four phases, namely \textit{initialization}, \textit{collection}, \textit{PoF verification}, and \textit{continuous following verification}. The steps of the {\em PoF} are shown in Fig.~\ref{algorithm}.  Note  that $\Cc$ knows a priori which platoon he is going to join. We treat the case of an unknown verifier with an amended PoF protocol to account for MiTM attacks. To facilitate platoon discovery, the verifier $\mathcal{V}$ periodically broadcasts its ID along with his credentials (public key and certificate). 
\medskip

\noindent
\textbf{\textit{Initialization phase.}}  
\begin{enumerate}
    \item Candidate $\mathcal{C}$ sends a platoon join request message REQ to the verifier $\mathcal{V}$. The message is signed with $sk_C$ and then encrypted with $pk_V.$ The candidate prepends the verifier's ID to REQ.
    
    \item The verifier $\mathcal{V}$ decrypts the message with $sk_V$ and verifies the signature with $pk_C$. If verification passes, $\mathcal{V}$ triggers a proof-of-following verification by sending a reply message REPLY. The message is signed with $sk_V$ and then encrypted with $pk_C.$ The message also contains a) the start and end times of RSS sampling, and b) the sampled frequency and sampling rate.
    
    \item $\mathcal{C}$ decrypts and verifies the signature of $\mathcal{V}.$ It then records the start and end times of the collection phase. Loose clock synchronization is achieved via the GPS clocks. 
\end{enumerate}
\textbf{\textit{Collection phase.}} In this phase, $\mathcal{V}$ and $\mathcal{C}$ sample a common frequency between the start and end times. 
\begin{enumerate}
  \setcounter{enumi}{3}
  \item The verifier and the candidate simultaneously collect RSS samples $\Gamma_{V}$ and $\Gamma_{C}$, respectively.
  \[
  \Gamma_{V}=\{(\gamma_V(1),\ t_V(1)), (\gamma_V(2),\ t_V(2)), \cdots,\},
  \] 
  \[
  \Gamma_{C}=\{(\gamma_C(1),\ t_C(1)), (\gamma_C(2),\ t_C(2)), \cdots, \},
  \]
  where $\gamma_X(i)$ is the RSS sample collected by vehicle $X$ at time $t_X(i)$. 
  \item The candidate reports his recording $\Gamma_{C}$ to  the verifier signed and encrypted. The verifier decrypts $\Gamma_{C}$ and verifies the signature. If verification passes, it moves to the PoF verification phase.
\end{enumerate}
\textbf{\textit{PoF verification phase.}}
In this phase, $\mathcal{V}$ verifies the ``following'' claim of the candidate by computing the RF correlation between the reported RSS measurements $\Gamma_{C}$ and its own recorded measurements $\Gamma_{V}$. 

\begin{enumerate}
  \setcounter{enumi}{5}

    \item The verifier aligns $\Gamma_{V}$ and $\Gamma_{C}$ using the respective timestamps. This is done by aligning the first sample (the two vehicles use the same sampling rate). $\mathcal{V}$ then updates the RSS sets $\Gamma_{V}$ and $\Gamma_{C}$ to 
    \begin{align*}
    & \Gamma_{V}=\{\gamma_V(1), \gamma_V(2), \cdots \}, & \Gamma_{C}=\{\gamma_C(1), \gamma_C(2),\cdots\} ,   
    \end{align*}
     where each $\gamma_V(i)$ is time-aligned with sample $\gamma_C(i).$  
  
  \item The verifier separates $\Gamma_{V}$ and $\Gamma_{C}$ into $K$ subsets of size $N$ samples. Let $\Gamma^k_{X}$ denote the $k$-th subset of set  $\Gamma_{X}$. 
  
  \item The verifier computes the correlation $\rho(k)$ for the subsets $\Gamma^k_{C}$ and $\Gamma^k_{V}$ for $k=1, 2, \cdots, K$ using eq. \eqref{eq:CC}. The verifier obtains $K$ correlation values $\rho(1), \rho(2),\ldots, \rho(K).$
  
  \item The verifier compares each correlation value $\rho(k)$ with a passing threshold $\tau$. if a fraction $\alpha$ ($0 \leq \alpha \leq 1$) of correlation values exceed  the passing threshold $\tau$, the verifier ACCEPTS. Otherwise, the verifier REJECTS. That is, the verification test is passed if 
  \[
  \sum_{k=1}^K \frac{I(\rho(k)\geq \tau)}{K} \geq \alpha,
  \]
  where $I(\cdot)$ is the indicator function.
\end{enumerate}

\textbf{\textit{Continuous following verification phase.}} 
\begin{enumerate}
  \setcounter{enumi}{9} 
  \item If the candidate passes the {\em PoF} verification, it is accepted in the platoon. Continuous following verification can be achieved by repeating the collection and verification phases continuously.
\end{enumerate}

\subsection{Security Analysis}

In this section, we analyze the security of our PoF protocol against different types of adversaries.

\subsubsection{Remote and Following-afar Adversaries} A remote or following-afar adversary can request to join a platoon represented by verifier $\Vc$ by sending a join request
\[
\resizebox{.95\hsize}{!}{$m_M(1)\leftarrow E_{pk_V}[sig_{sk_M}(REQ,ID_M),REQ,ID_M, pk_M, cert_M].$}
\]
The verifier $\Vc$ verifies the identity of $\Mc$ (by verifying the public key using the certificate and sends REPLY $m_{V}(1)$ to $\mathcal{M}$ which also indicates the start and end times of the RSS collection, as well as the probed frequency. At this point, the adversary has two choices: (a) to use RSS data that is pre-recorded on the path traveled by the platoon, or (b) to collect real time data at his current location (for the following-afar adversary). Then $\mathcal{M}$ sends 
\[
 m_M(2)\leftarrow E_{pk_V}[sig_{sk_M}(\Gamma_M, ID_M), \Gamma_M, ID_M]
\]
to $\mathcal{V}$ for authentication. To pass the PoF verification, the set $\Gamma_M$ provided by $\Mc$ must satisfy the correlation test applied by $\Vc.$ However, since $\Mc$ is not within the following distance or $\Gamma_M$ was collected a long time before $\Gamma_V$, the two RSS data sets decorrelate according to Eq. \eqref{exp model} and Fig. \ref{fig:temporal_corr}. So both remote adversary and following-afar adversary will be rejected. We experimentally evaluate the correlation achieved at different distances and for different environments in Sec.\ref{sec:evaluation}. 

\subsubsection{Partially-following Adversary} 
A partially-following adversary can similarly initiate the PoF process by sending an $m_M(1)$ message to $\mathcal{V}$. It can pass a fraction of $K$ tests when it is within following distance  but fail the rest of tests when he moves far from $\Vc$. The fraction of correlation tests passed by the adversary depend on the fraction of time that $\Mc$ follows $\Vc.$ Theoretically, if the adversary is within a following distance for a fraction $\alpha$ of the RSS collection time, then he should pass a fraction $\alpha$ of the $K$ tests, thus being admitted in the platoon. In practice, a larger fraction of time may be needed because some tests fail even for valid candidates due to the RSS randomness. In Sec. \ref{sec:partially-following}, we evaluate the passing rate of a partially following adversary in our experimental setting. Note that when continuous authentication is employed, the partially-following adversary has to periodically approach $\Vc$ to be retained as a platoon member. 


\subsubsection{MiTM Attacks} 
In a MiTM attack, the adversary attempts to be admitted to the platoon when a valid candidate initiates a join request with the verifier. The adversary is not within following distance of $\Vc$. We analyze two instances of the attack. In the first instance, $\Cc$ is aware of the verifier that he attempts to join. This is implied in our PoF protocol, as $pk_V$ and $cert_V$ are given to $\Cc$ according to the assumptions we have made for the PoF protocol (see Fig.~\ref{algorithm}). We further treat the case where the candidate does not know the verifier a priori, but responds to a broadcast of a nearby verifier.

\begin{figure}[t]
  \centering
  \includegraphics[width =0.95\columnwidth]{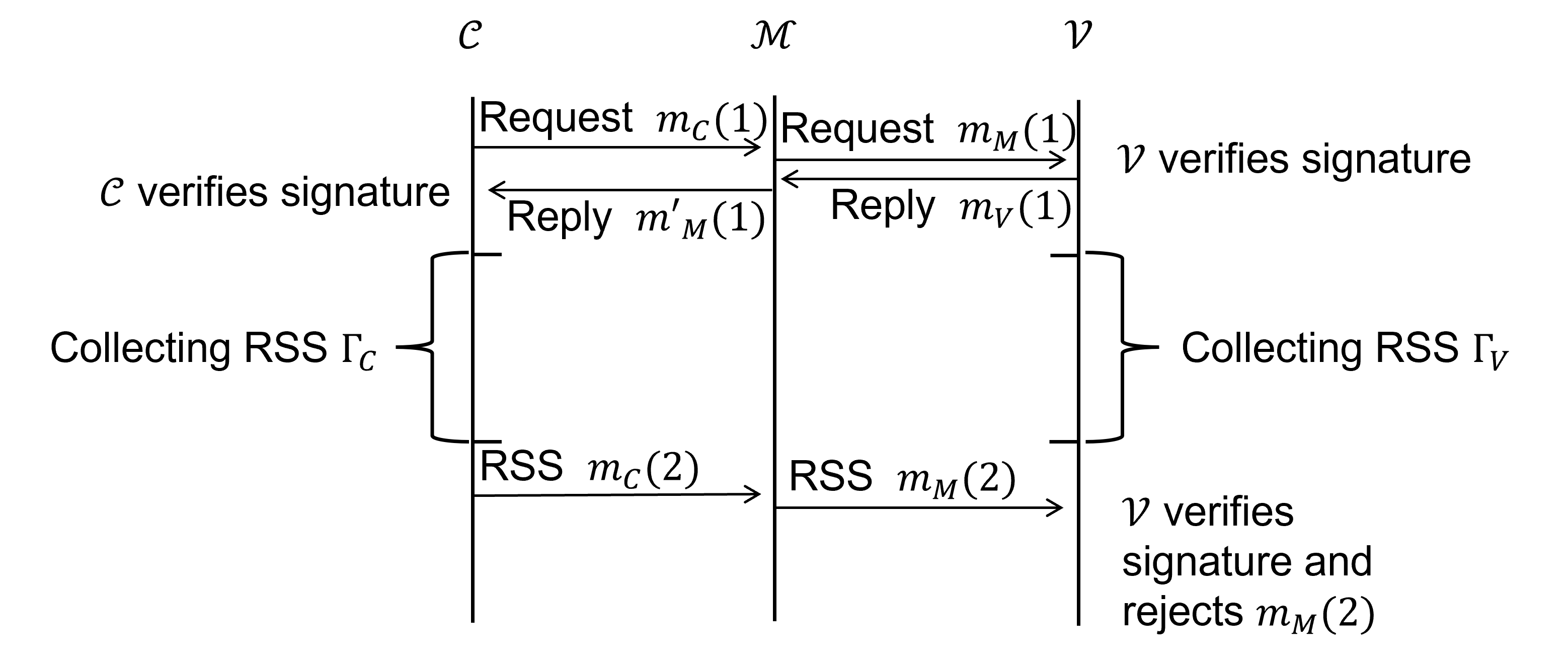}
  \caption{A MiTM attack on the PoF protocol when $ID_V$ and $pk_V$ are known to $\mathcal{C}$. Because $m_C(2)$ is encrypted with $\mathcal{V}$'s public key, the adversary $\mathcal{M}$ cannot obtain $\Gamma_C$ and construct a valid $m_M(2)$.}
  \label{fig:MitM}
  \vspace{-0.2in}
\end{figure}
\medskip

{\bf Known verifier.}  Let $(ID_V, pk_V, cert_V)$ be given to $\mathcal{C}$ via some out-of-band channel before the PoF protocol execution (e.g., $\Cc$ is instructed to join a specific $\Vc$). In this case, spurious forwarding from a MiTM adversary is prevented by the authenticated encryption used to transmit the RSS value set to the verifier. For ease of illustration,  Figure~\ref{fig:MitM} demonstrates a possible MiTM attack.  Let the candidate initialize the protocol by sending the join request message $m_C(1).$ Note that $m_C(1)$ is encrypted with $pk_V$ so only $\mathcal{V}$ can decrypt it.  The adversary can attempt to initiate parallel sessions by eliminating $m_C(1)$ from the channel and injecting his own message
\[
m_M(1)\leftarrow E_{pk_V}[sig_{sk_M}(REQ,ID_M),REQ,ID_M]
\]
to the verifier. After receiving the reply $m_V(1)$, if the adversary replaces $m_V(1)$ with his own message 
\[
m'_M(1)\leftarrow E_{pk_C}[sig_{sk_M}(REPLY,ID_M),REPLY,ID_M],
\]
$\Cc$ will abort because the signature verification will fail ($\Cc$ expects a message signed by $\mathcal{V}$ and verified with $pk_V$). Alternatively, $\Mc$ can relay $m'_M(1)\leftarrow m_V(1)$ and move forward to the stage of transmitting the collected RSS sets. Upon collection of $\Gamma_C$, the candidate sends message 
 \[
 m_C(2)\leftarrow E_{pk_V}[sig_{sk_C}(\Gamma_C, ID_C), \Gamma_C, ID_C]
 \]
which is intercepted by $\mathcal{M}$. Because $m_C(2)$ is encrypted with $\mathcal{V}$'s public key, $\mathcal{M}$ cannot recover  $\Gamma_C.$ As a result, $\mathcal{M}$ fails to construct message $m_M(2)$ that would contain $\Cc$'s RSS samples. The only recourse is to include his own samples which are not correlated with $\Vc$'s samples, as $\Mc$ is beyond the following distance, leading to a REJECT by the verifier. \medskip
 
{\bf Unknown verifier.} If the verifier's identity is not known to $\mathcal{C}$ a priori, the candidate must undergo a discovery phase to probe for nearby verifiers. This scenario is relevant when vehicles are forming platoons in an ad hoc manner to improve fuel efficiency and safety, without necessarily belonging to the same organization. If the verifier is unknown, $\Mc$ can launch a spoofing attack on $\Cc$ and convince him that $\Mc$ is a valid verifier. At the same time $\Mc$ can open a parallel session with the legitimate $\Vc$, requesting to join the platoon. The candidate will provide $\Mc$ with his RSS set $\Gamma_C$, which $\Mc$ can immediately use to pass the RSS correlation test at $\Vc.$

To prevent a MiTM attack, we amend our PoF protocol to include a commitment  scheme \cite{halevi1996practical} with a delayed opening phase that renders the RSS samples obtained by the adversary stale. The key idea is rather than revealing the RSS samples, $\Cc$ presents a commitment on those samples to the verifier. The commitment is opened with a delay to complete the PoF verification. A MiTM adversary spoofing a verifier cannot open the commitment of $\Cc$ due to the hiding property. The delayed opening phase forces $\Mc$ to commit without knowing $\Gamma_C,$ or commit late with RSS samples that are not aligned with those of $\Vc.$ Thus, $\Vc$ will reject $\Mc$'s join request. Due to space limitations, we present the amended PoF protocol and a detailed security analysis in Appendix \ref{Appendix-pof}.
 
\subsubsection{Mutual PoF Verification} 
We emphasize that our PoF protocol does not provide any proof of platooning to the candidate, as it implements access control to existing platoons. This opens the possibilities for verifier spoofing where the candidate may join a verifier that he does not follow. This type of attack can be thwarted by extending the delayed commitment scheme in both directions. The verifier would also be required to commit to its own RSS samples and reveal them in a delayed opening phase.

\section{Evaluation}
\label{sec:evaluation}

In this section, we evaluate the correctness and soundness of the PoF. We describe our experimental testbed, we demonstrate how to select the test parameters, and present experimental results in freeway, urban, and highway environments.

\subsection{Testbed}
\label{sec:setup}

 We developed two setups based on the NI USRP platform \cite{NI}. The first setup was employed in the freeway and urban driving experiments whereas the second setup was employed in the highway driving experiments. 

\begin{figure}[t]
\centering
\includegraphics[width=0.9\columnwidth]{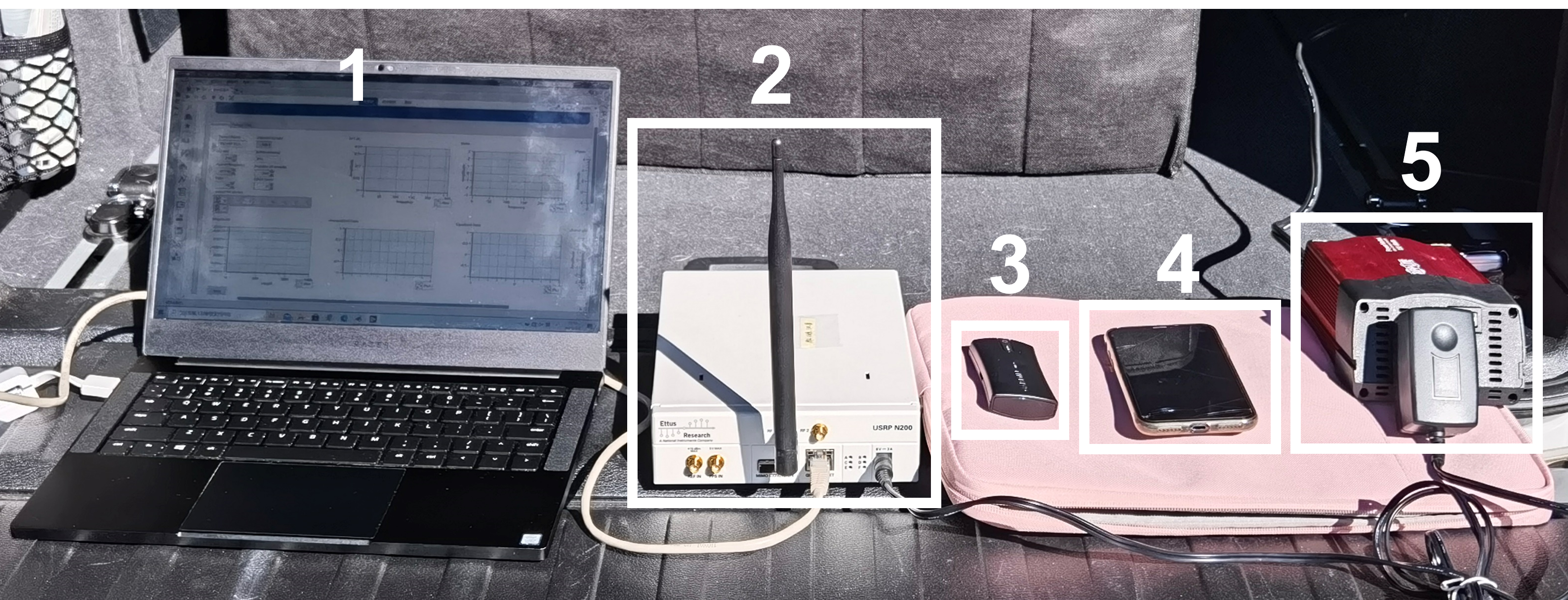}
\caption{The RSS data collection equipment consists of (1) laptop recording GPS and RSS data, (2) USRP N200 with a VERT900 antenna  working as the LTE receiver, (3) Garmin Bluetooth GPS receiver, (4) iPhone X that acts as a hotspot to the laptop and connects via Bluetooth to the GPS receiver, (5) a battery power station that supplies power to the USRP.}
\label{fig:setup}
\vspace{-0.1in}
\end{figure}

\textbf{Setup 1.} We used a Nissan Sentra and a BMW X5 acting as $\mathcal{V}$ and $\mathcal{C}$, respectively. The two vehicles had cruise control capabilities (not adaptive), but were otherwise manually operated.  We placed at the trunk of each vehicle the equipment shown in Fig. \ref{fig:setup}.  A USRP N200 radio device was connected to a VERT900 antenna. The USRP was programmed to implement an OFDM receiver for LTE signals. It  operated at 1.972GHz with a 4MHz bandwidth, which is the frequency used for personal communications service (PCS) in LTE.  We set the gain of the antenna to 10dB and the sample rate to 20Hz. A Razer blade stealth laptop was connected to the USRP for recording the RSS data. The laptop was also connected to a GPS receiver to record positioning information at 5Hz sampling rate. The synchronization between the RSS and GPS data was achieved via the laptop clock.

\textbf{Setup 2.} In setup 2, we formed a two-vehicle platoon for driving in  a highway environment. Here, $\Vc$ (Honda Pilot) led the platoon with cruise control engaged, whereas $\Cc$ (Toyota RAV-4) followed $\Vc$ with {\em adaptive cruise control} engaged. The candidate was equipped with  a LiDAR to measure the distance to $\Vc$. This allowed for an easier and more accurate control of the separation distance between the two vehicles at highway speeds. Although setup 2 is superior to setup 1 from a platooning perspective, it was not always available to us to conduct the experiments that spanned many hours and days, so we limited it to highway experiments were maintaining constant distance presents more challenges. The data collection setup was identical to that of Setup 1, with the central frequency set to 875MHz with 4 MHz bandwidth and the antenna gain was 20 dB. The new frequency was selected based on the signal availability at the specific part of the highway were experiments were conducted.

\begin{figure*}%
\centering
\setlength{\tabcolsep}{-3pt}
\begin{tabular}{cccc}
\includegraphics[width=0.55\columnwidth]{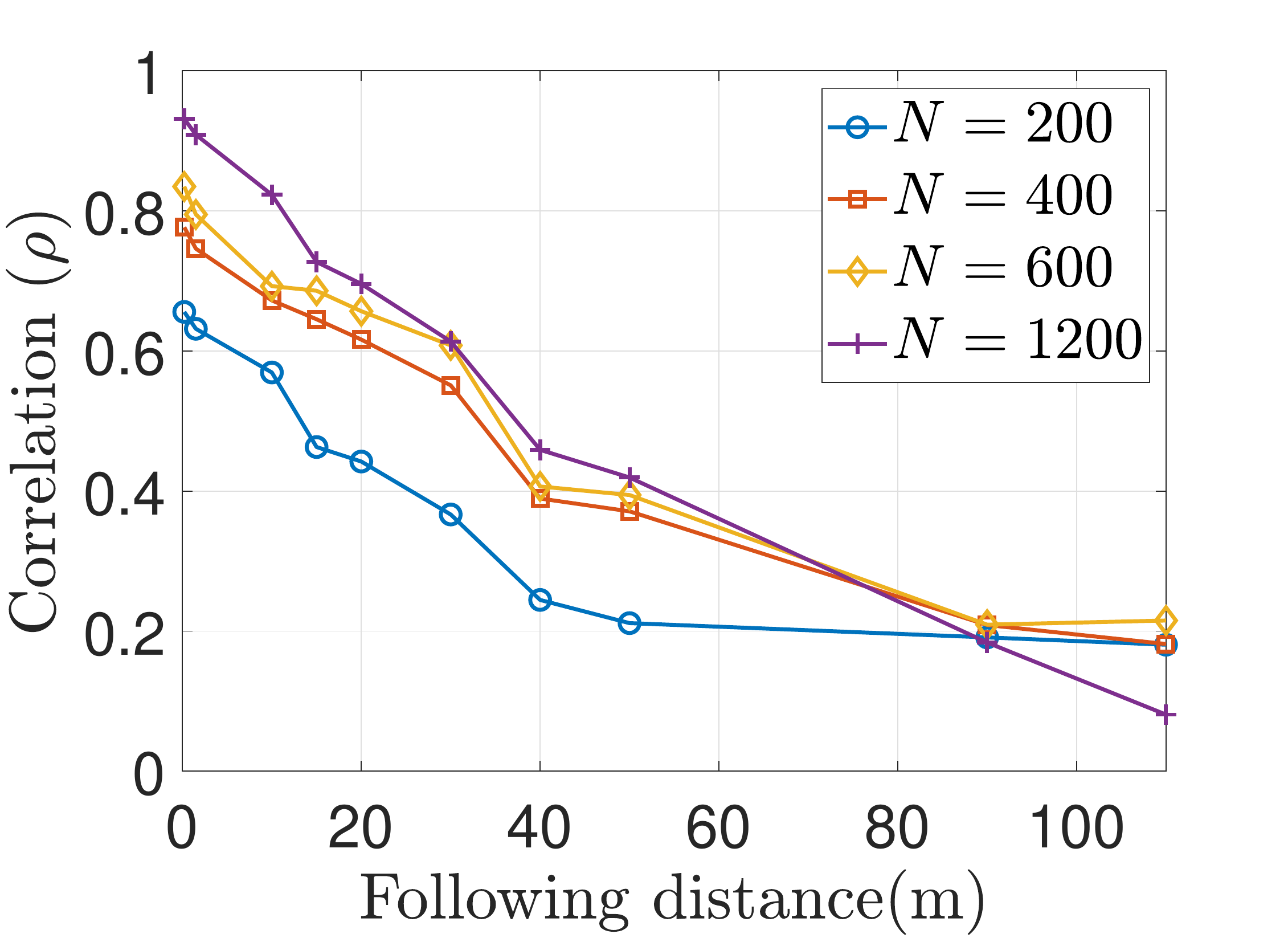}
&    \includegraphics[width=0.55\columnwidth]{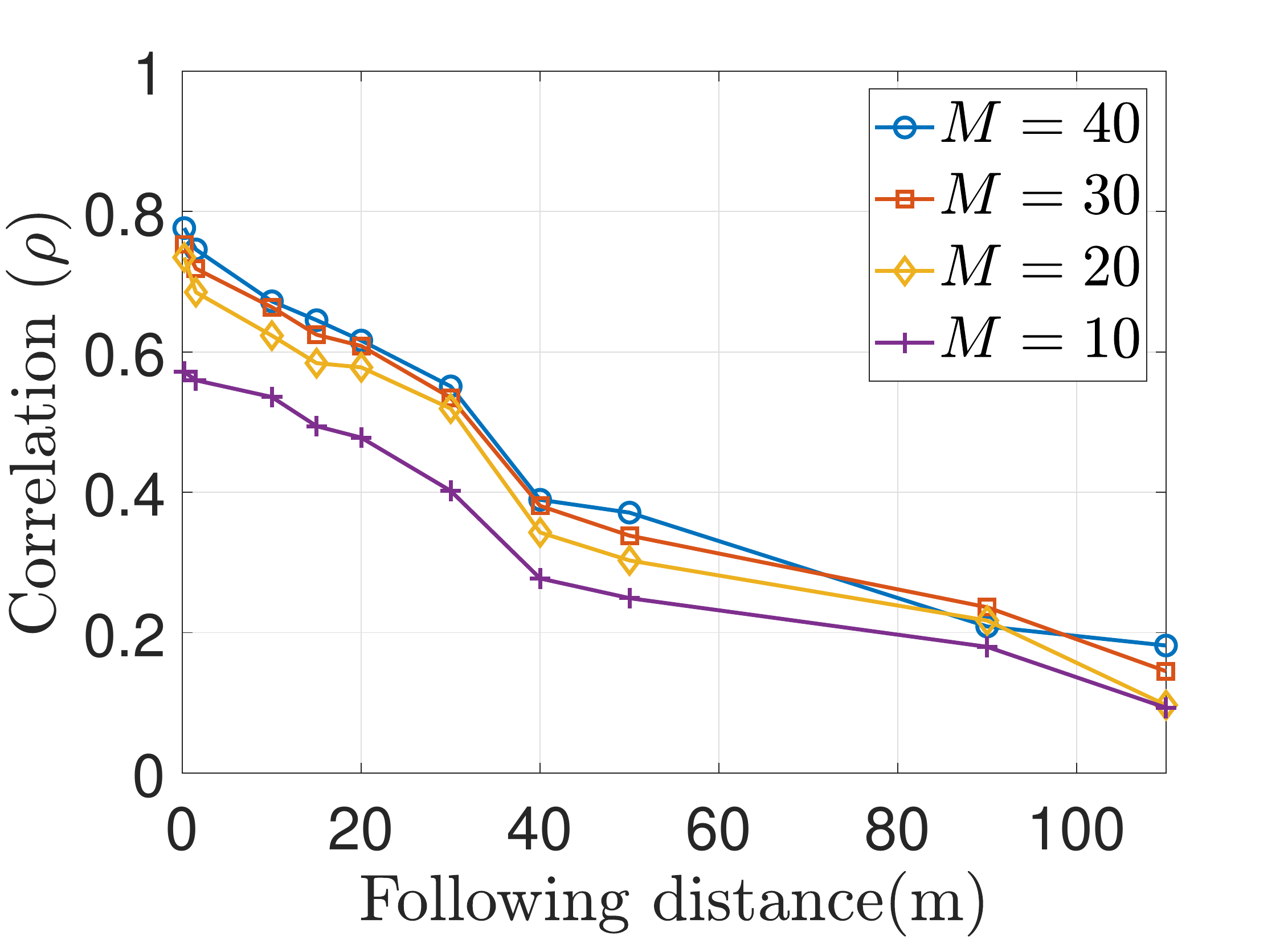}
&    \includegraphics[width=0.55\columnwidth]{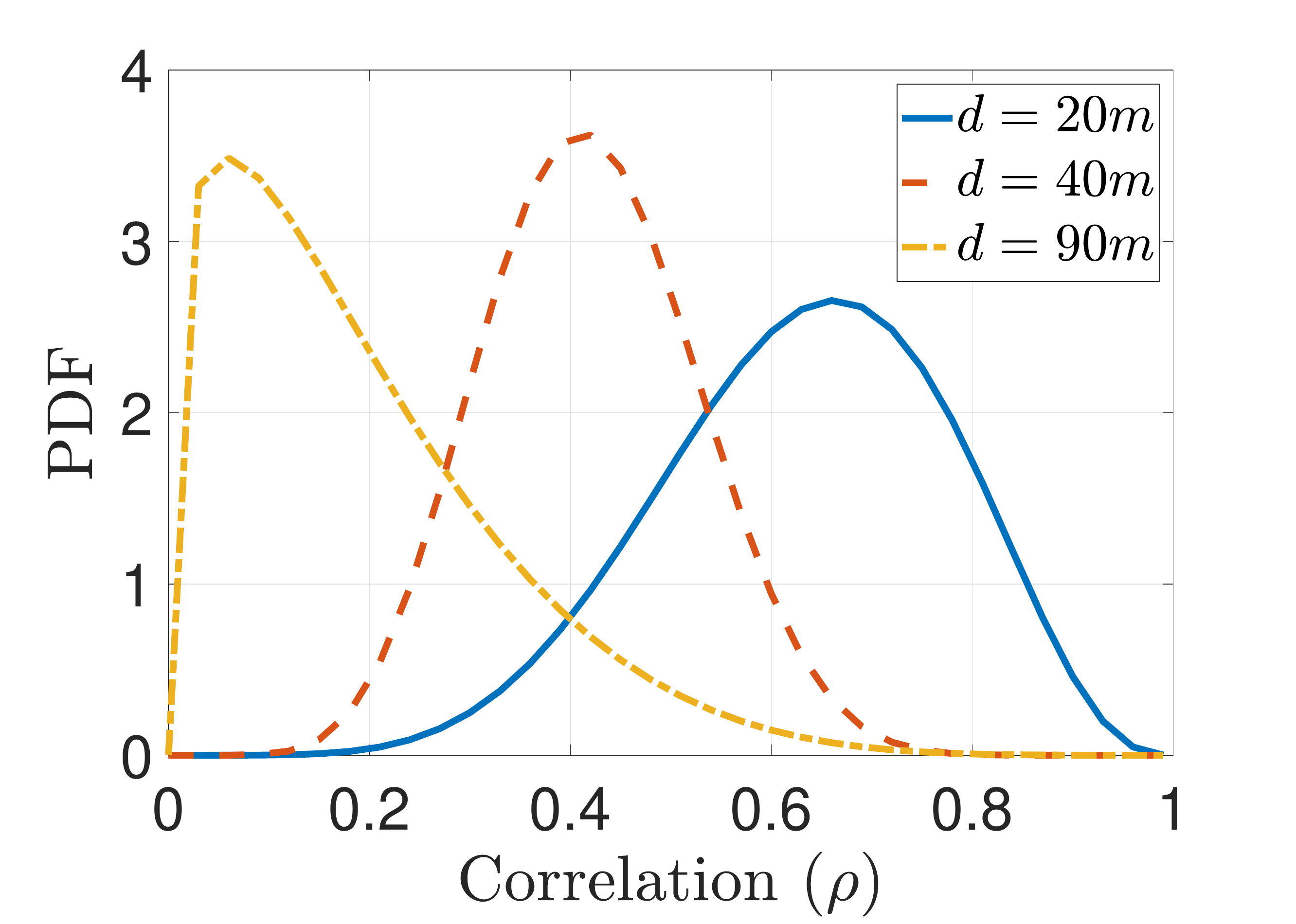}
&    \includegraphics[width=0.55\columnwidth]{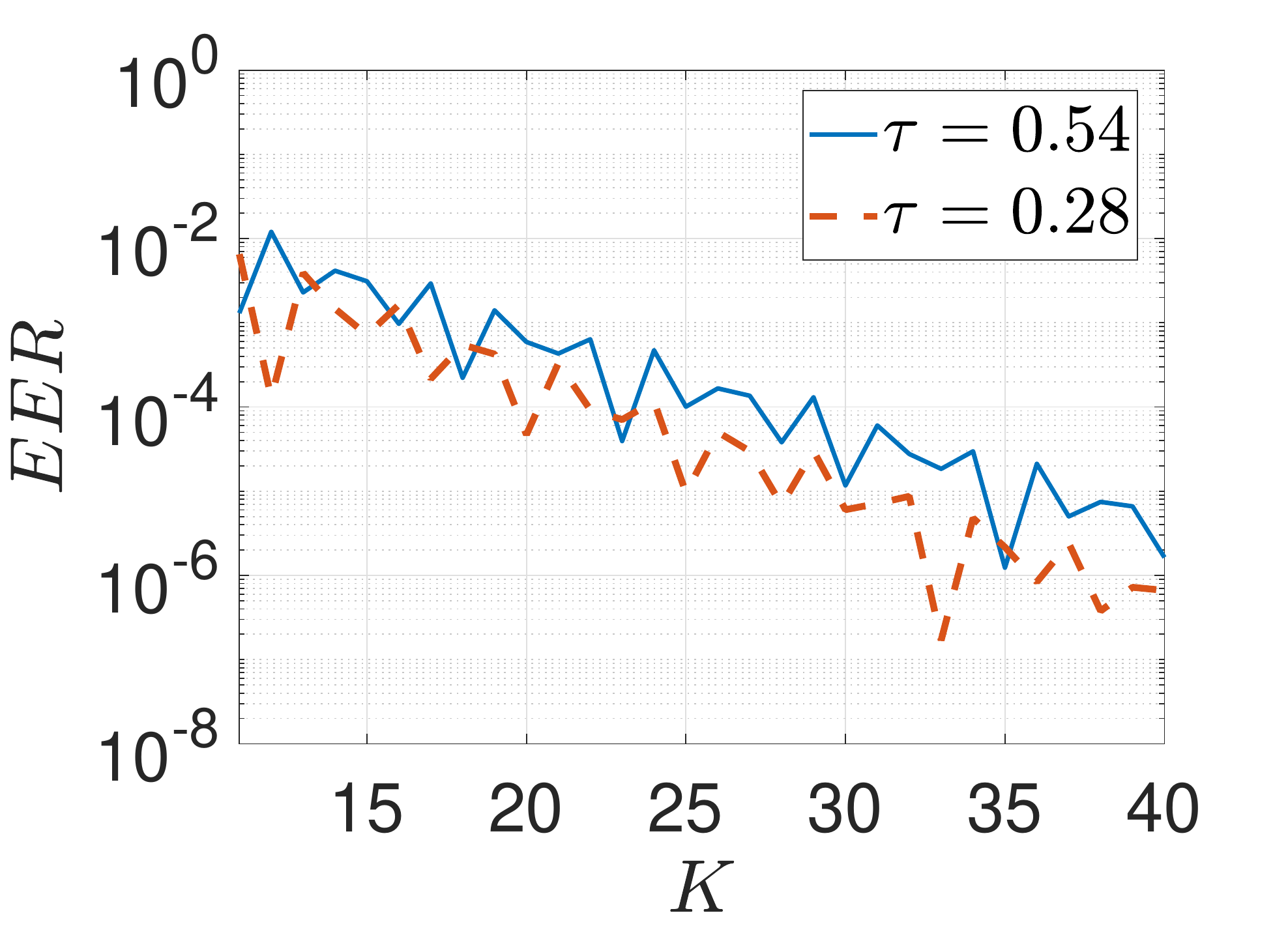}\\
(a) & (b) & (c) & (d)
\end{tabular}
\caption{(a) Average correlation as a function of the following distance for different subset sizes $N$, (b) average correlation as a function of the following distance for different moving average window length, when $N=400$, (c) PDF of the correlation for different following distances, (d) EER as a function of the number of tests $K$ for two values of $\tau.$}
\label{fig:parameters}
\vspace{-0.2in}
\end{figure*}

\subsection{PoF Test Parameter Selection}

The {\em PoF} test is controlled by the selection of the  $N$, $M$, $K$, $\tau$, and  $\alpha$ parameters. For clarity, we summarize the definition of these parameters in Table \ref{table:notation}. In this section, we show how to select the parameter in practice and then evaluate the PoF protocol in three  driving environments. 

To select $N$ and $M$, we performed experiments on a freeway section of length 1.4 miles using Setup 1. The particular section was not accessible to other traffic and was located next to a highway. This presented an ideal situation for safely controlling the following distance. The two vehicles were driven at 30Mph over multiple runs and at different following distances. We collected RSS data using the radio testbed and processed the collected data using various test parameters.

{\bf Selecting $N$.} The subset size $N$ determines the number of samples used per correlation test. It must be long enough to ensure that RSS values exhibit high correlation, but should not prolong the test duration. We experimented with different lengths $N$. For each $N$, we reused the second-half samples in each subset to form the first-half samples for the following subset to reduce the test duration.  Figure~\ref{fig:parameters}(a) shows $\rho$ as a function of the following distance $d$. Generally, when a larger $N$ is selected, the correlation increases (except for ranges over 90m where the two RSS sequences are uncorrelated).  From Fig.~\ref{fig:parameters}(a), we can see that beyond  $N=400$ the gains in correlation are relatively small. Therefore, we fix $N=400$ for all the following evaluations. 

\begin{table}[t]
\caption{Parameters of the PoF Test.}
\centering
\scalebox{0.9}{
\begin{tabular}{cl}
  \hline
  Notation & Definition \\
  \hline
  $N$ & Number of samples in subsets $\Gamma^k_V$ and $\Gamma^k_C$\\
  $M$ & Moving average window size \\
  $K$ & Number of RSS subsets, correlation values, and correlation tests\\
  $\tau$ & Passing threshold for a single correlation test\\
  $\alpha$ & Fraction of correlation tests to pass {\em PoF} verification  \\
  $f_C$ & Passing rate of a single correlation test achieved by $\mathcal{C}$\\
  $F_C$ & Passing rate of $K$ correlation tests achieved by  $\mathcal{C}$\\
  \hline
\end{tabular}
}
\label{table:notation}
\vspace{-15pt}
\end{table}

{\bf Selecting $M$.} The length of the moving average window impacts the elimination of small-scale fading. Intuitively, a larger window leads to a more stable moving average, and a higher correlation $\rho$, but the moving average becomes more predictable.  
In Fig.~\ref{fig:parameters}(b), we show the correlation $\rho$ as a function of the following distance for different $M$. As expected, we observe an increase in correlation with $M$ at short distances, whereas the impact of $M$ is small for large distances because the two RSS sequences are uncorrelated. Moreover, the increase in correlation diminishes after $M=20$. Based on these observations, we set $M=20$.

{\bf Selecting $\tau$, $K$, and  $\alpha$.} Let $f=\Pr(\rho_k(d) \geq \tau)$ denote the probability of passing a single correlation test when an RSS subset $\Gamma_k$ is used to compute $\rho_d(k)$. This probability depends on the selection of $\tau$ and the following distance $d$. For a total of $K$ tests, a PoF is passed if $\sum_{k=1}^K \frac{I(\rho(k)\geq \tau)}{K} \geq \alpha.$ Assuming independent tests due to the use of different subsets, the probability of passing a PoF verification consisting of $K$ correlation tests is 
\begin{equation}
    F=\sum\limits_{x=\lceil \alpha \cdot K \rceil}^{K}
    {K \choose x}(f)^x \cdot (1-f)^{K-x}.
\label{eq:V_c passing rate}
\end{equation}
where ${K \choose x}$ is the binomial coefficient. Let $F_C$ denote the probability of passing the PoF verification for $\Cc$ and $F_M$ to be the passing probability for $\mathcal{M}$. Probability $F_C$ is derived from Eq.~\eqref{eq:V_c passing rate} by substituting the probability $f$ of passing a single correlation test at $d<d_{ref}$, given the selection of $\tau.$ Similarly, $F_M$ is derived from Eq.~\eqref{eq:V_c passing rate} by substituting the probability $f$ of passing a single correlation test at some distance $d>d_{ref}$,
The equal error rate ($EER$) is defined as 
\begin{align}
    EER: \quad 1- F_{C} = F_{M}.
    \label{eq:EER}
\end{align}
We aim at selecting $\tau^*$ that minimizes the $EER$. To understand the interplay between $\tau,$ $K$, $\alpha$ and $F$, we generated the PDF of the correlation $\rho$ for three representative following distances. The respective PDFs are shown in Fig.~\ref{fig:parameters}(c). From the PDF, one can select a desired $\tau$ to satisfy a required passing rate for valid candidates for a given $d_{ref}.$ For instance, we chose $\tau=0.54$ for $d=20m$ and $\tau=0.28$ for $d=40m.$ A $d=90m$ is representative of a following-afar adversary.

Given $\tau$, we performed an exhaustive search over the two remaining free variables $K$ and $\alpha$ to minimize the $EER.$ Here, we limit $K$ to 40 to ensure that a {\em PoF} test adheres to a time limit and also limited $\alpha$ such that $\lceil \alpha K \rceil$ takes {\em integer} values between 1 and $K$. Figure~\ref{fig:parameters}(d) shows the $EER$ as a function of $K$ when the optimal $\alpha$ is selected. As expected, the $EER$ decreases with $K$. Here, a $K$ that satisfied a desired $EER$ requirement can be selected at the expense of a {\em PoF} test duration.  For the freeway experiments, we set $K=20.$

An alternative method for selecting $\tau$ and $\alpha$ under fixed $K$ is to first determine two following distances from the platooning requirements. The first distance is that of the valid candidate, namely $d_{ref}$, whereas the second is that of the afar adversary that we try to prevent against. We then compute the threshold $\tau$ for a single correlation test from the exponential model in Eq.~\eqref{exp model} by setting $d=d_{ref}$. Once $\tau$ is fixed, we select $\alpha$ to maximize the gap between  $F_C$  and $F_M$. Given that the average correlation model may not hold for all driving environments, we opted to use the exhaustive search method to explore the performance of the {\em PoF}. These methods require the use of at least one trusted vehicle besides the verifier in the platoon to calculate passing rate $f$. If there is only one vehicle in the platoon, $\Vc$ has to select $\tau$ from eq. \eqref{exp model}. Except $\tau$, all  other parameters ($N, M, K, \alpha$) can be preset since they are stable in different environments.

\subsection{Evaluation of \textit{PoF} on the Freeway}
\label{sec:freeway}
For the freeway experiments, we employed Setup 1, with the two vehicles driving at approximately at 30Mph. Because the specific freeway section was closed, we were able to control the following distance between the two vehicles. We drove the vehicles at following distances between 10m - 115m.

\begin{figure}[t]
\centering
\includegraphics[width=0.75\linewidth]{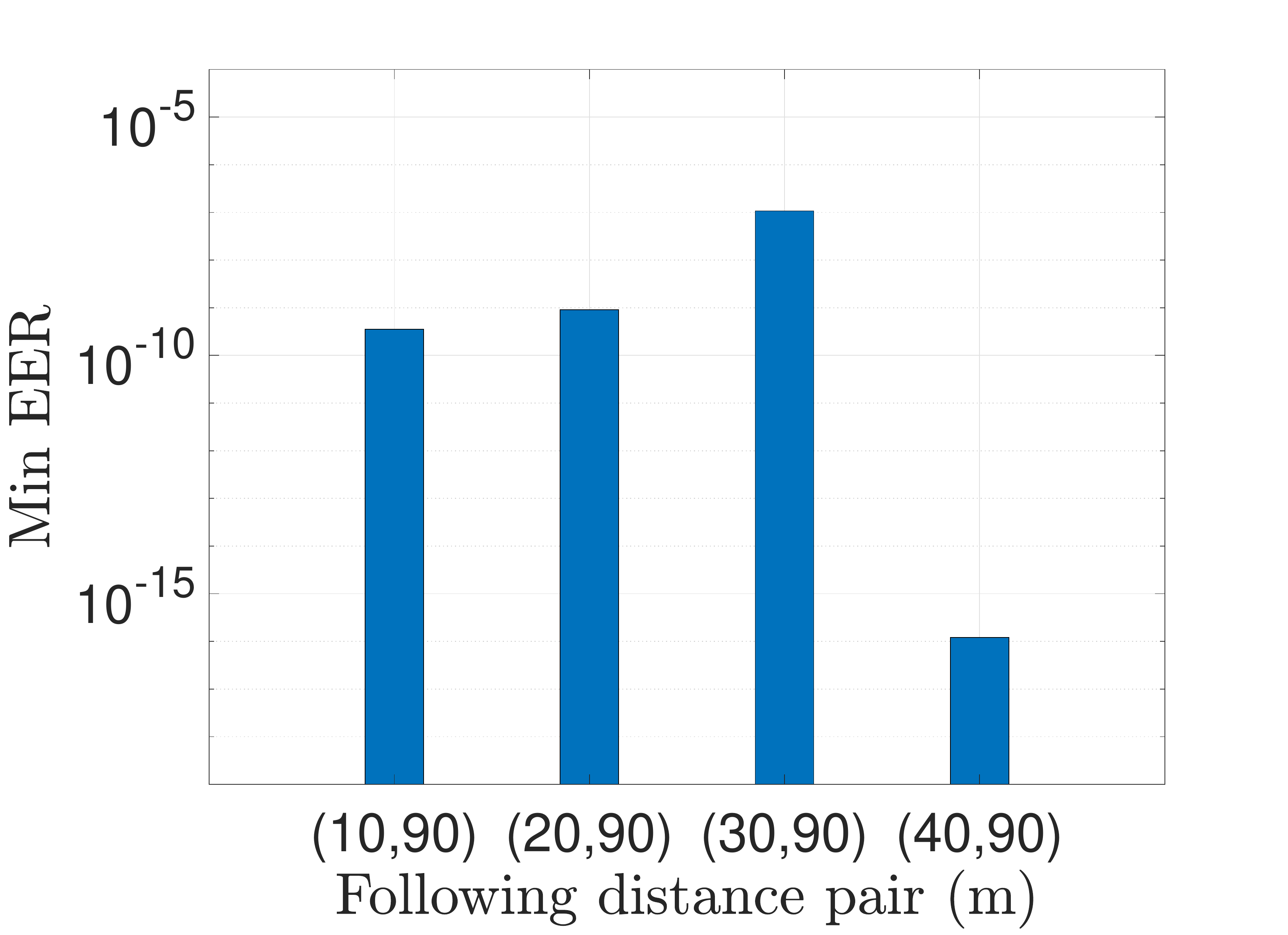}
\caption{Selecting parameters for pairs of following distances when $K=20$ and $0.5<f_{C}<1$, $0<f_{M}<0.5$.
From left to right: (1) $\tau^*=0.38$, $\alpha^*=\frac{11}{20}$, (2) $\tau^*=0.36$, $\alpha^*=\frac{11}{20}$, (3) $\tau^*=0.35$, $\alpha^*=\frac{11}{20}$, and (4) $\tau^*=0.34$, $\alpha^*=\frac{9}{20}$.}
\label{fig:pairs}
\vspace{-0.2in}
\end{figure}

Based on the parameter selection we discussed in the previous section, we set $N=400,$ $M=20$, and $K=20.$ We then performed an exhaustive search to find the optimal threshold values $\tau^{\ast}$ and $\alpha^{\ast}$ that minimize the $EER.$ Figure~\ref{fig:pairs} shows the minimum EER for different values of valid following distance $d_{ref}$ when the adversary follows at 90m. The optimal values of $\tau^{\ast}$ and $\alpha^{\ast}$ that minimize the $EER$ for each following distance are also shown. First, we observe that our method achieves a fairly low $EER$. Moreover, the optimal   $\tau^{\ast}$ and $\alpha^{\ast}$ do not vary significantly with the change of $d_{ref}.$

\begin{figure}[t]
\centering
\includegraphics[width=0.75\linewidth]{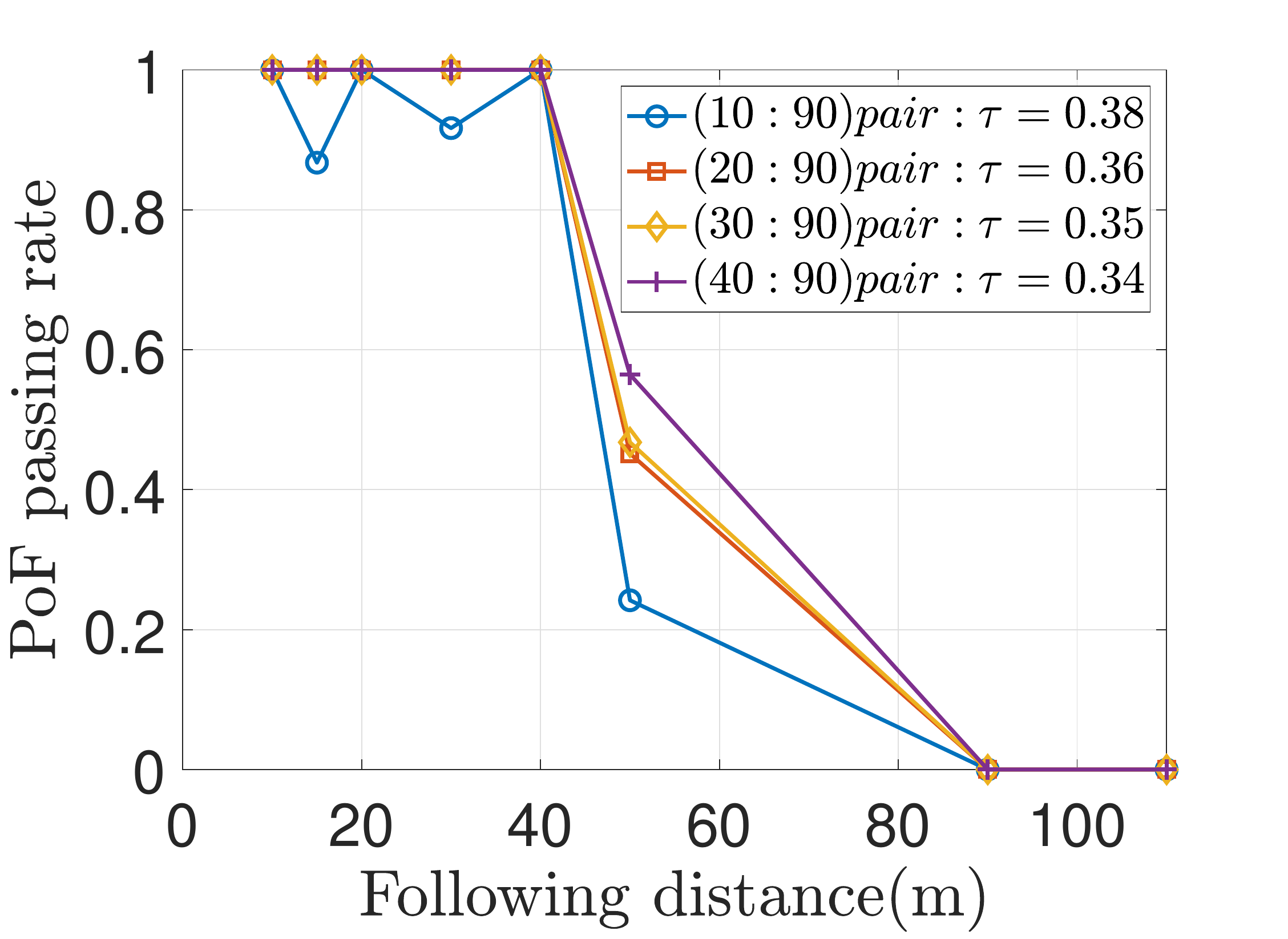}
\caption{PoF passing rate as a function of the following distance for the optimal $\tau^{\ast}$ and $\alpha^{\ast}$ obtained in Fig.~\ref{fig:pairs}. }
\label{fig:theory}
\vspace{-0.2in}
\end{figure}

In Fig.~\ref{fig:theory}, we show the PoF passing rate as a function of the following distance for the optimal  $\tau^{\ast}$ and $\alpha^{\ast}$ values obtained from minimizing the $EER$. When a candidate is within a following distance between 10m-40m the passing rate is close to 1. The passing rate drops to zero for distances larger than 90m. The method leaves a ``guard'' zone between 40-90 where the passing rate is from 0.2 to 0.4. This zone cannot be considered to be secure as an adversary following  in this zone could pass a {\em PoF} test with non-negligible probability. This is because the correlation degrades gracefully with distance and does not exhibit a step-function type of behavior. 

Ideally, for any environment, we estimate the distribution of $\rho$ with different distance, and set $\tau, N, \alpha$. However, in practice, the distance between two vehicles is difficult to control due to the traffic, especially in urban and highway areas, which thwarts us from selecting $\tau$ with CDF of $\rho$. Instead, we are interested in evaluating {\em PoF} for certain given distance bound, from which we select $\tau$ based on the statistically relationship between the passing rate of a single correlation and the threshold $\tau$. After that, $K$ and $\alpha$ can be chosen with  minimum $EER$. For all the following results, we also use 30$\%$ of experimental data as a training sequence for parameter selection, and the remaining 70$\%$ for testing.
\begin{figure*}[t]
\centering
\setlength{\tabcolsep}{-2pt}
\begin{tabular}{cccc}
    \includegraphics[width=0.53\columnwidth]{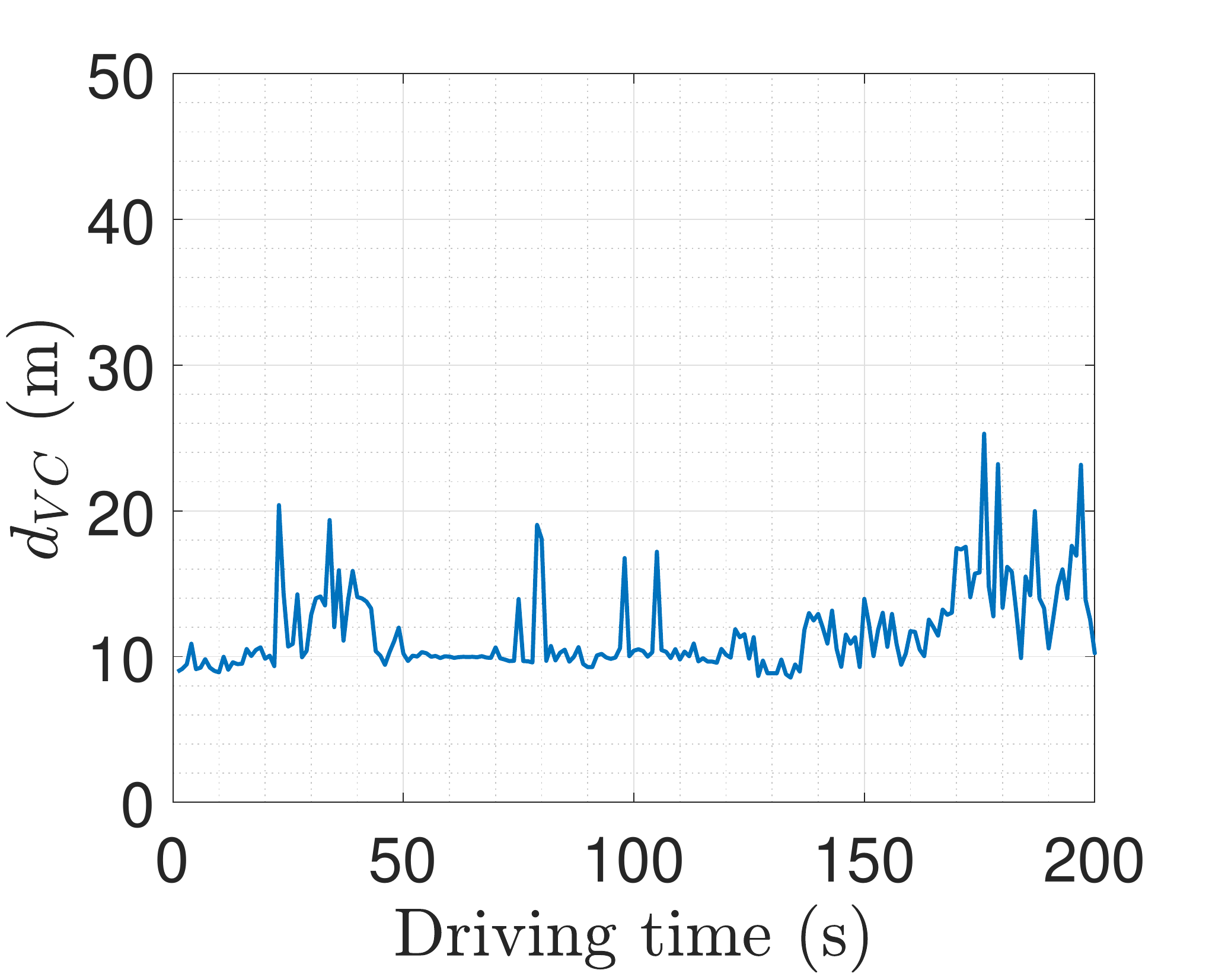}  &
    \includegraphics[width=0.53\columnwidth]{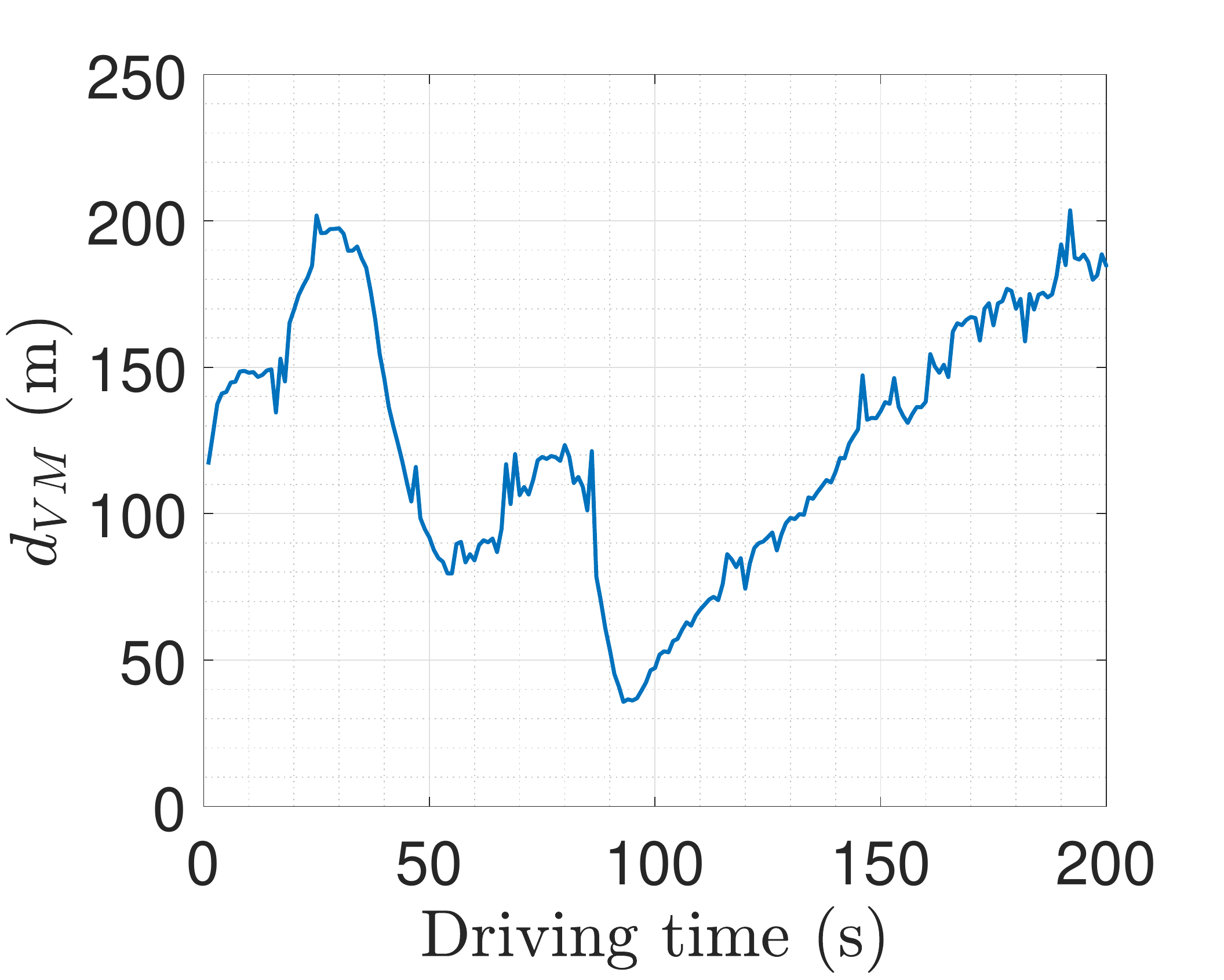} &
    \includegraphics[width=0.53\columnwidth]{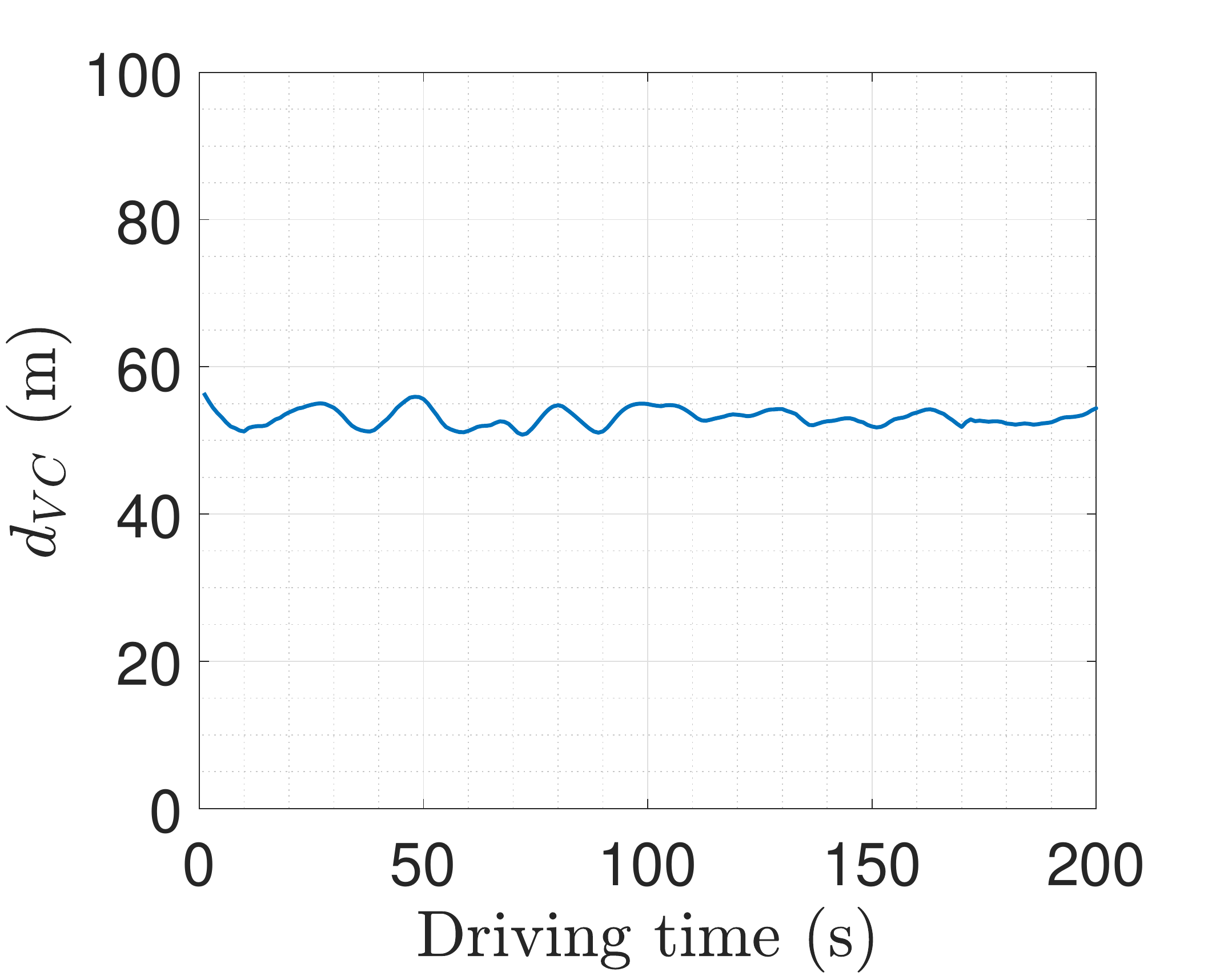}  &
    \includegraphics[width=0.53\columnwidth]{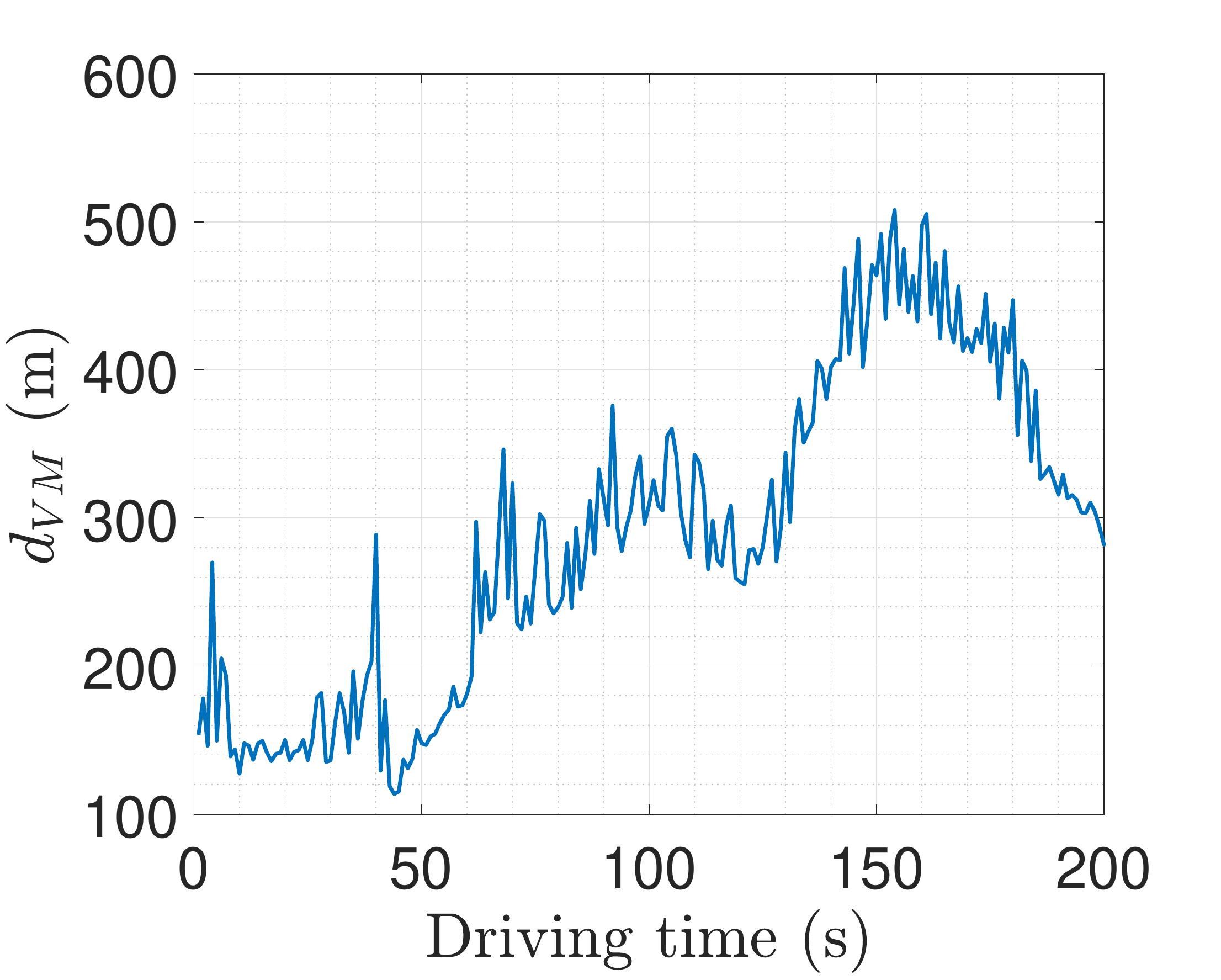}\\
    (a) urban  &(b) urban &(c) highway &(d) highway
    \end{tabular}
\caption{The following distance $d_{VC}$ and $d_{VM}$ for urban and highway environments.}
\label{fig:dist}
\vspace{-0.2in}
\end{figure*}
\begin{figure}[t]
\centering
\setlength{\tabcolsep}{-2pt}
\begin{tabular}{c}
\includegraphics[width=0.8\columnwidth]{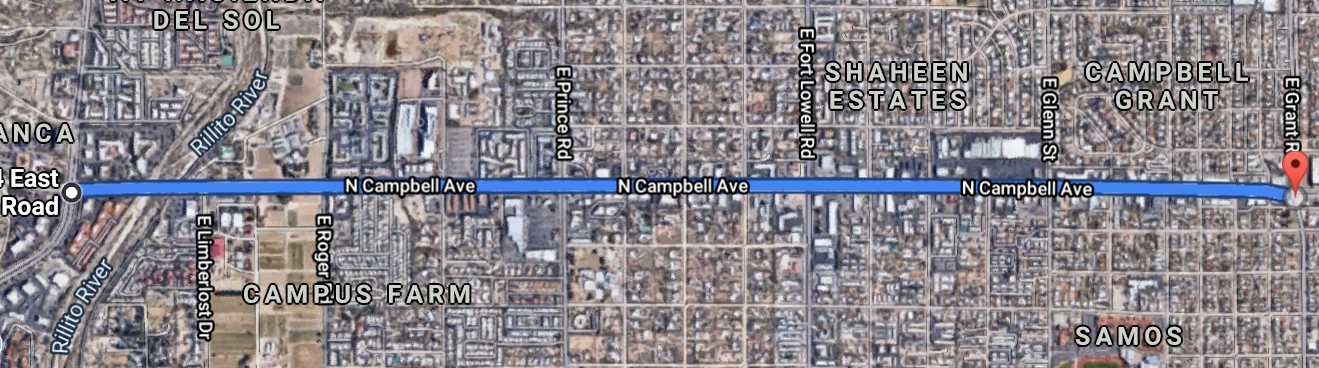}\\
(a) urban\\
\includegraphics[width=0.8\columnwidth]{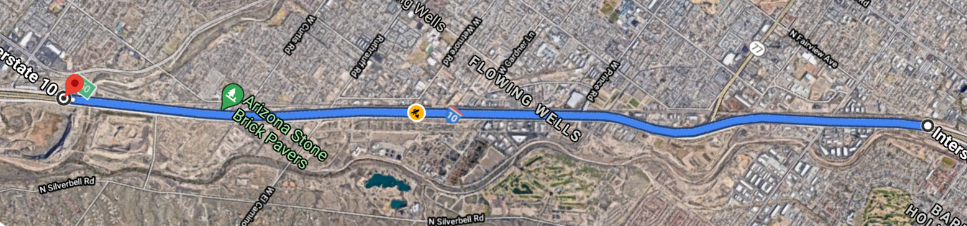}\\
(b) highway 
\end{tabular}
\caption{Driving route for urban and highway environments.}
\label{fig:exp_route}
\vspace{-0.2in}
\end{figure}

\subsection{Evaluation of \textit{PoF} in an Urban Area}
\label{sec:urban}

For the urban area experiments, we used Setup 1 to drive the two vehicles along a 2.5-mile route inside Tucson city from River to Grant road, as shown in Fig. \ref{fig:exp_route}(a), several rounds for one hour. Except  stops at red lights, the vehicles normally run at a speed up to 40Mph and always followed each other. Figure~\ref{fig:dist}(a) shows the following distance $d$ over time. The following distance was fairly stable with an average of about 11 meters. 
Since the mixed traffic did not allow us to precisely control the following distance $d$, we did not obtain the PDF of the correlation for different distances as we did for the freeway experiments. Thus, we   select optimal values of  $\tau$, $K$, and $\alpha$ by minimizing the $EER$.   

\subsubsection{Parameter Selection}
\label{sec:practical issues}
 
\begin{figure*}%
\centering
\setlength{\tabcolsep}{-2pt}
\begin{tabular}{cccc}
    \includegraphics[width=0.52\columnwidth]{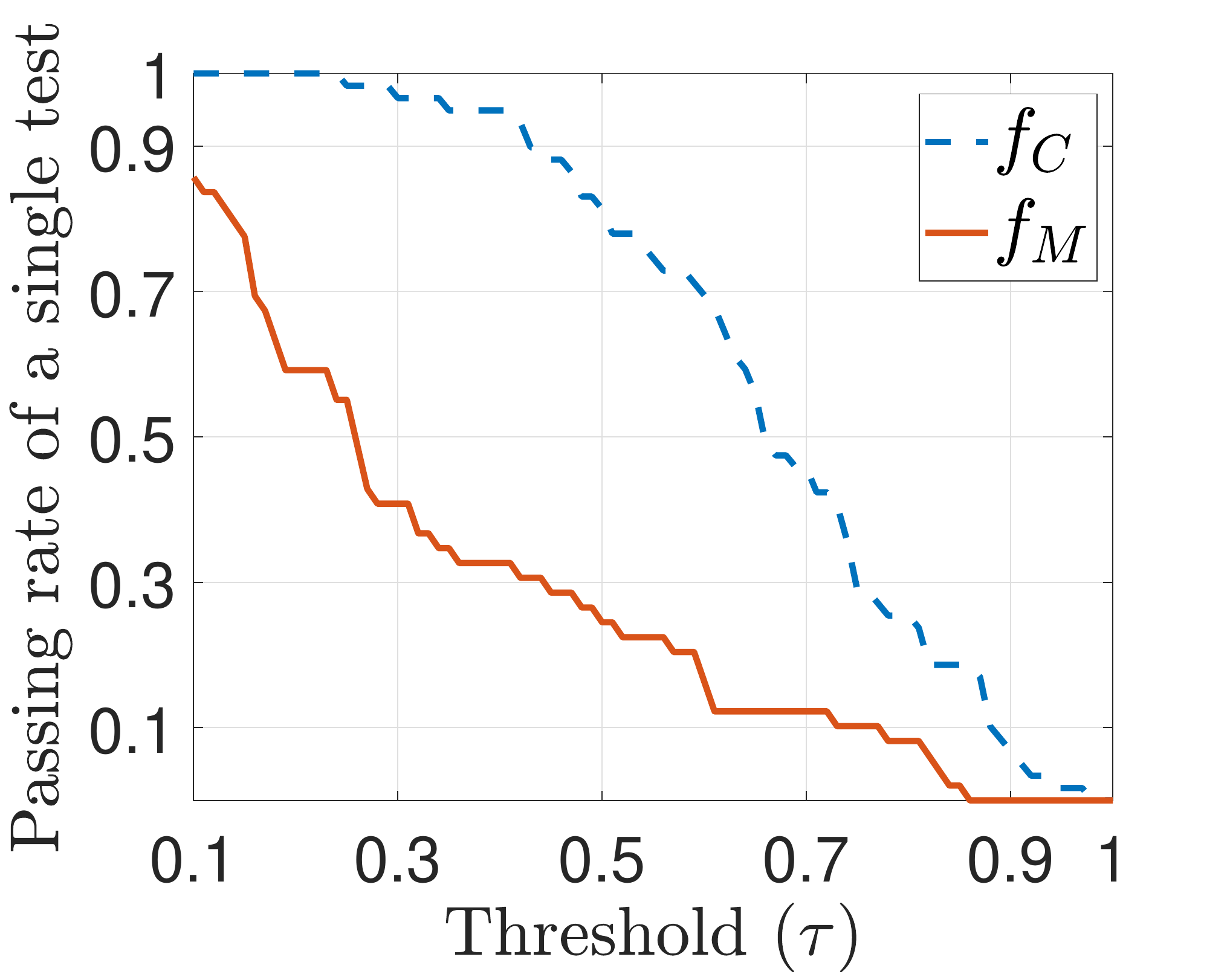} &
    \includegraphics[width=0.52\columnwidth]{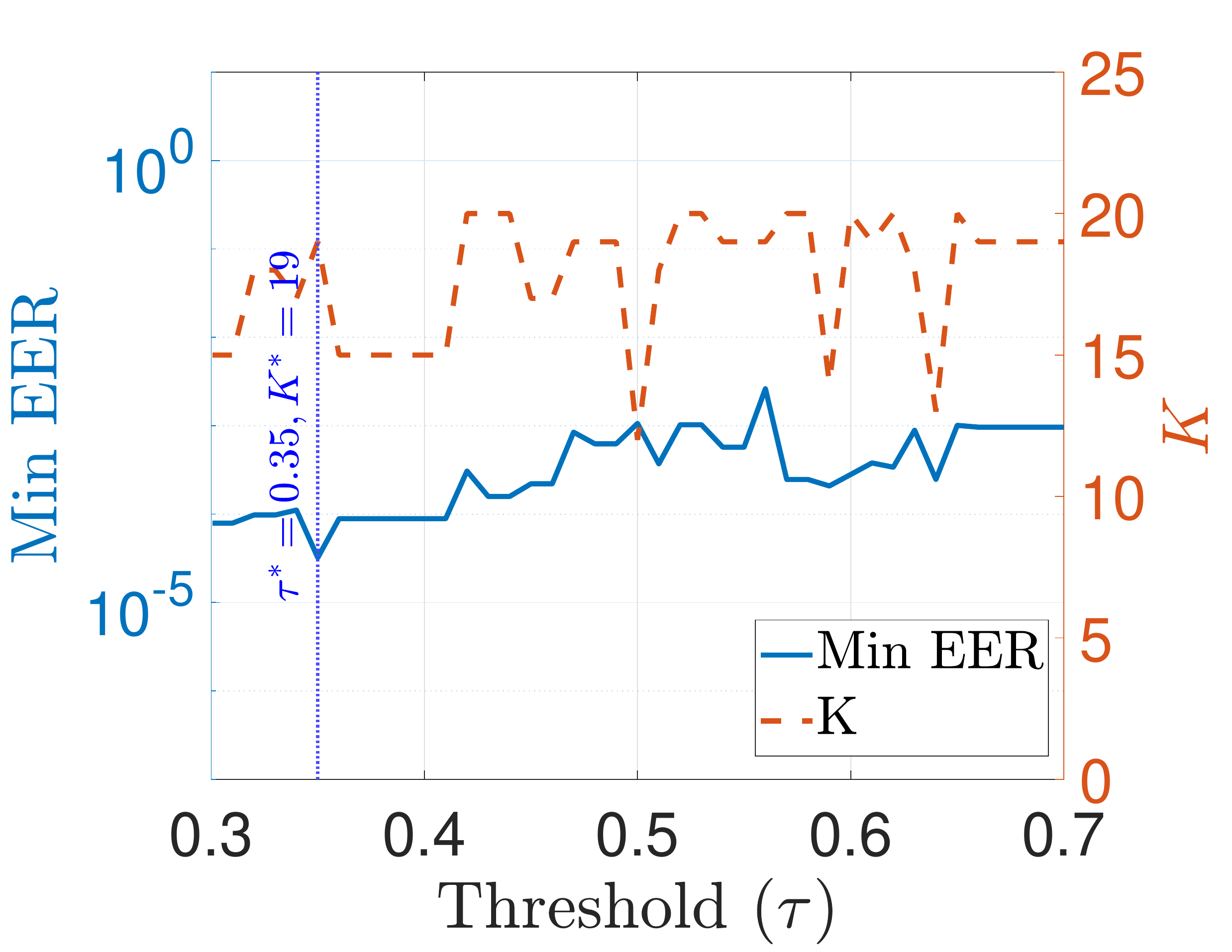} &
    \includegraphics[width=0.52\columnwidth]{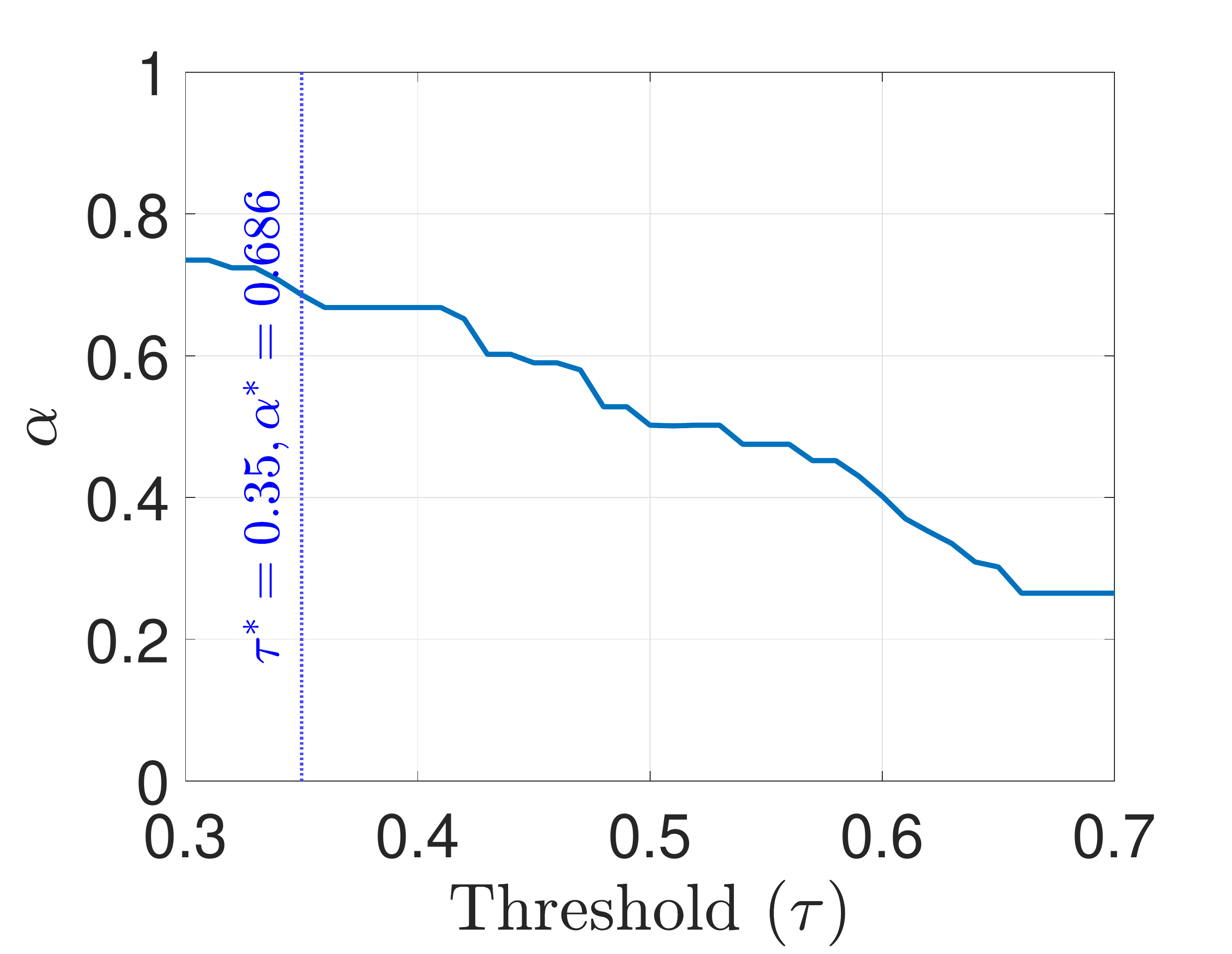} &
    \includegraphics[width=0.52\columnwidth]{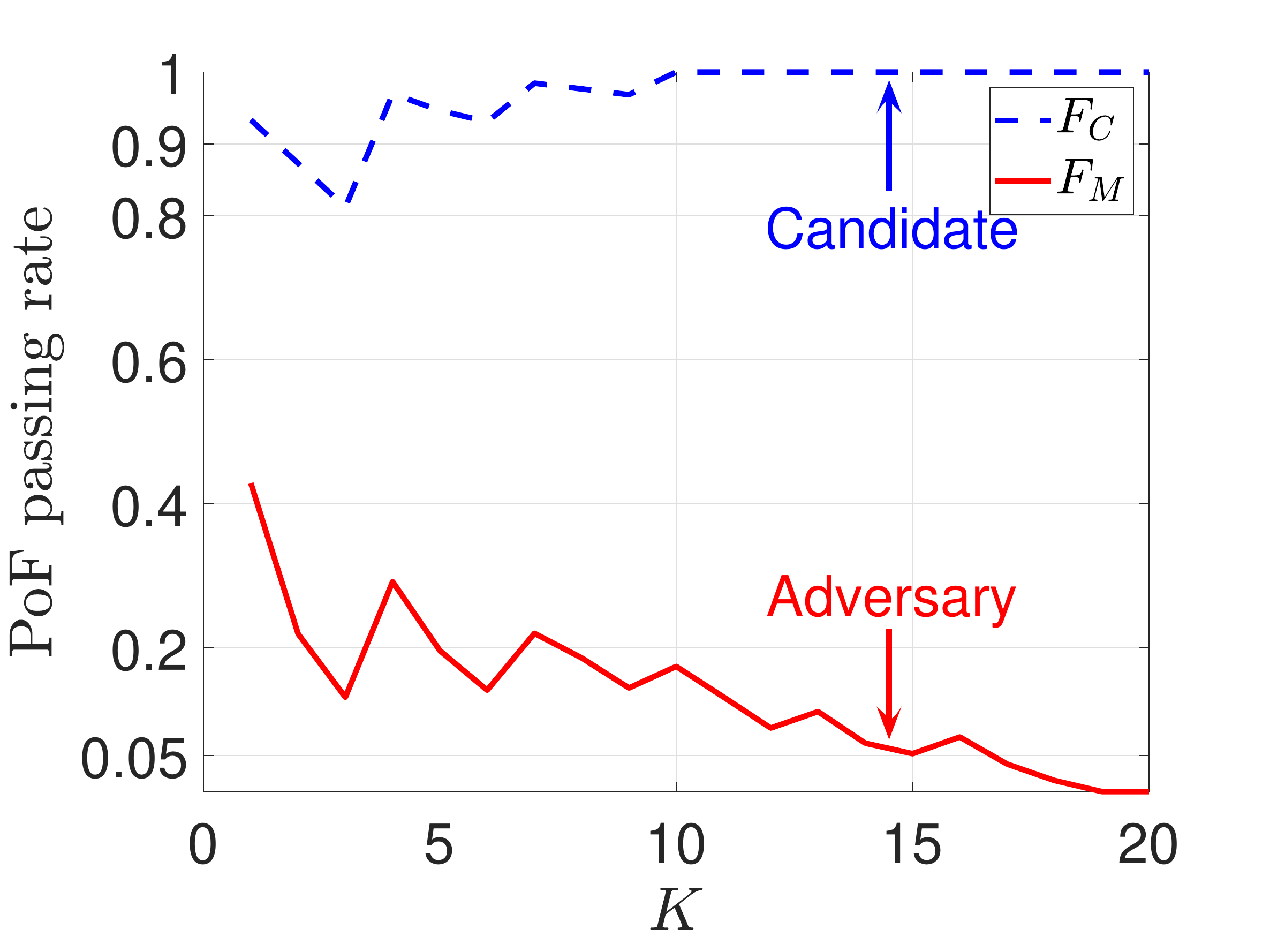}\\
    (a) & (b) & (c) & (d)
    \end{tabular}
\caption{Urban driving, remote adversary: (a) single correlation test passing rates for  $\mathcal{C}$ and $\mathcal{M}$, (b) minimum $EER$  and corresponding optimal $K$, (c) $\alpha$ that minimizes the $EER$. (d) PoF passing rate for a valid candidate and for the adversary.}
\label{fig:urban_2}
\vspace{-0.2in}
\end{figure*}

{\bf Selecting $\tau$.}  According to Sec. IV-B, the single test passing rates for $\mathcal{C}$ and $\mathcal{M}$ should satisfy $f_C > 0.5$ and $f_{M} < 0.5$, respectively.  To select the threshold $\tau$,  we first plot $f_C$ and $f_M$ as functions of the single test correlation threshold $\tau$, based on our experimental data (Fig. \ref{fig:urban_2}(a)). From this we set the range of $\tau$ to be $0.3 <\tau <0.7$, in order to satisfy   $f_C > 0.5$ and $f_{M} < 0.5$.

{\bf Selecting $K$ and $\alpha$.} Recall that we use $K$ tests and $\alpha$ fraction of passing to drive the probability of successful verification for the candidate vehicle $F_C$ and an adversary $F_M$ to one and zero, respectively. 
In Fig. \ref{fig:urban_2}(b), we show the minimum $EER$ for $0.3 <\tau <0.7$ against the corresponding $K$, Values of   $\alpha$ are shown in Fig. \ref{fig:urban_2}(c). We can see that $\tau^{\ast}=0.35$ with the corresponding  $K^{\ast}=19$ and  $\alpha^{\ast}=0.686$ minimize the $EER$.

\subsubsection{Remote adversary}
In the following, we show results for a remote adversary,  who pre-records RSS data on the known platoon route and replays it  to pass the PoF verification. 
We let the Nissan Sentra serve as a remote adversary by prerecording RSS for 70 minutes ahead on the same route of the platoon.  Then the selected parameters above are used in all PoF verification test sets. Each test set will output accept  or reject  after 19 correlation tests. We  run the PoF tests for multiple times (continuous verification) and calculated the PoF test passing rate. In Fig. \ref{fig:urban_2}(d), we show the PoF passing rate versus the number of correlation tests $K$. For each $K$, the PoF passing rate was computed over 10 PoF runs. We observed that the PoF passing rate for $\mathcal{C}$ increases with the number of  correlation tests, while it decreases for $\mathcal{M}$. After 19 correlation tests, $\mathcal{C}$ achieved $100\%$ PoF passing rate, and the verifier rejected the adversary ($F_M = 0$). This shows that in an urban environment,  our {\em PoF} protocol can successfully differentiate a legitimate candidate from a remote adversary.

\subsubsection{Following-afar adversary}
\label{sec:urban type1}
\begin{figure*}%
\centering
\setlength{\tabcolsep}{-0.1pt}
\begin{tabular}{cccc}
    \includegraphics[width=0.52\columnwidth]{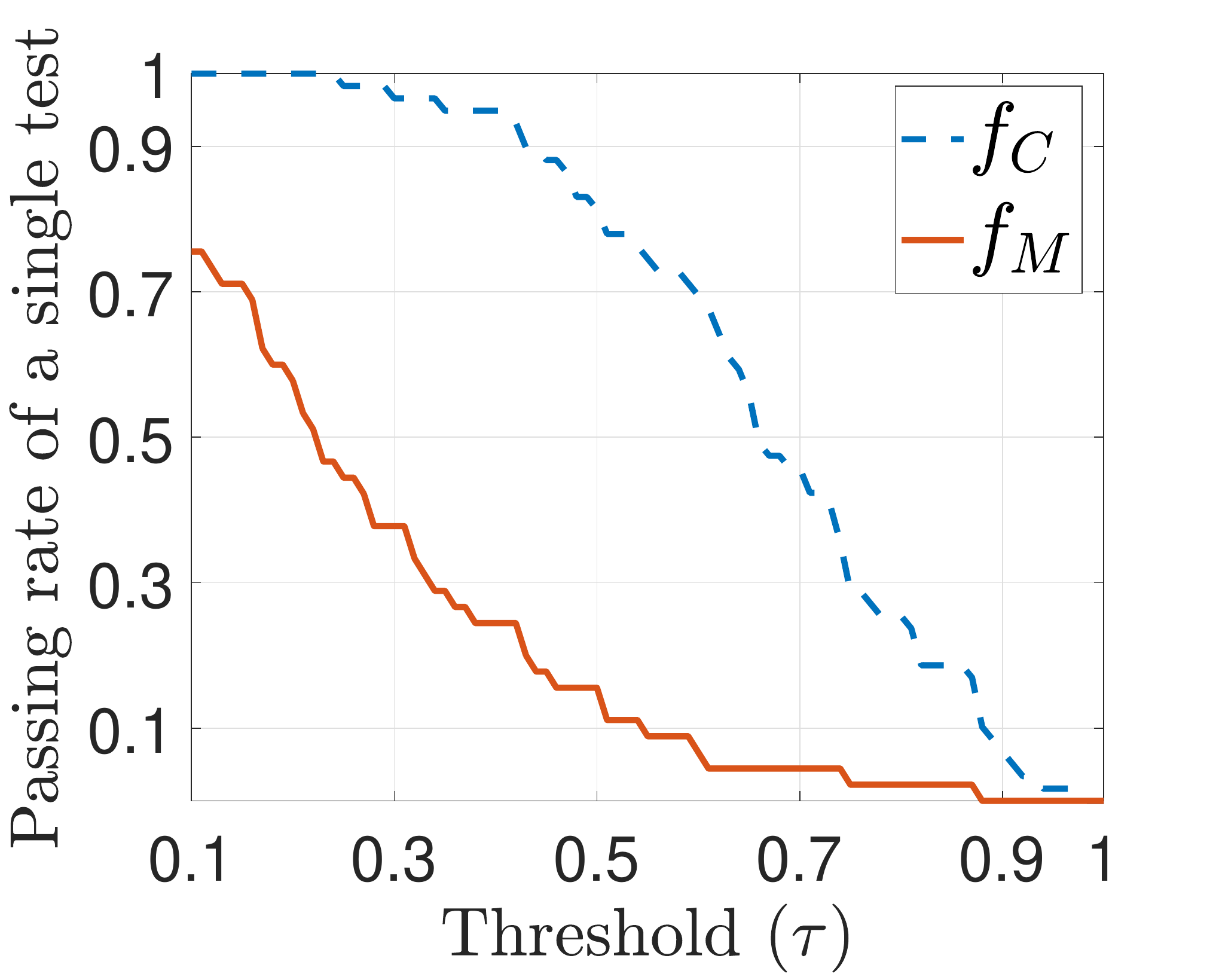}&
    \includegraphics[width=0.52\columnwidth]{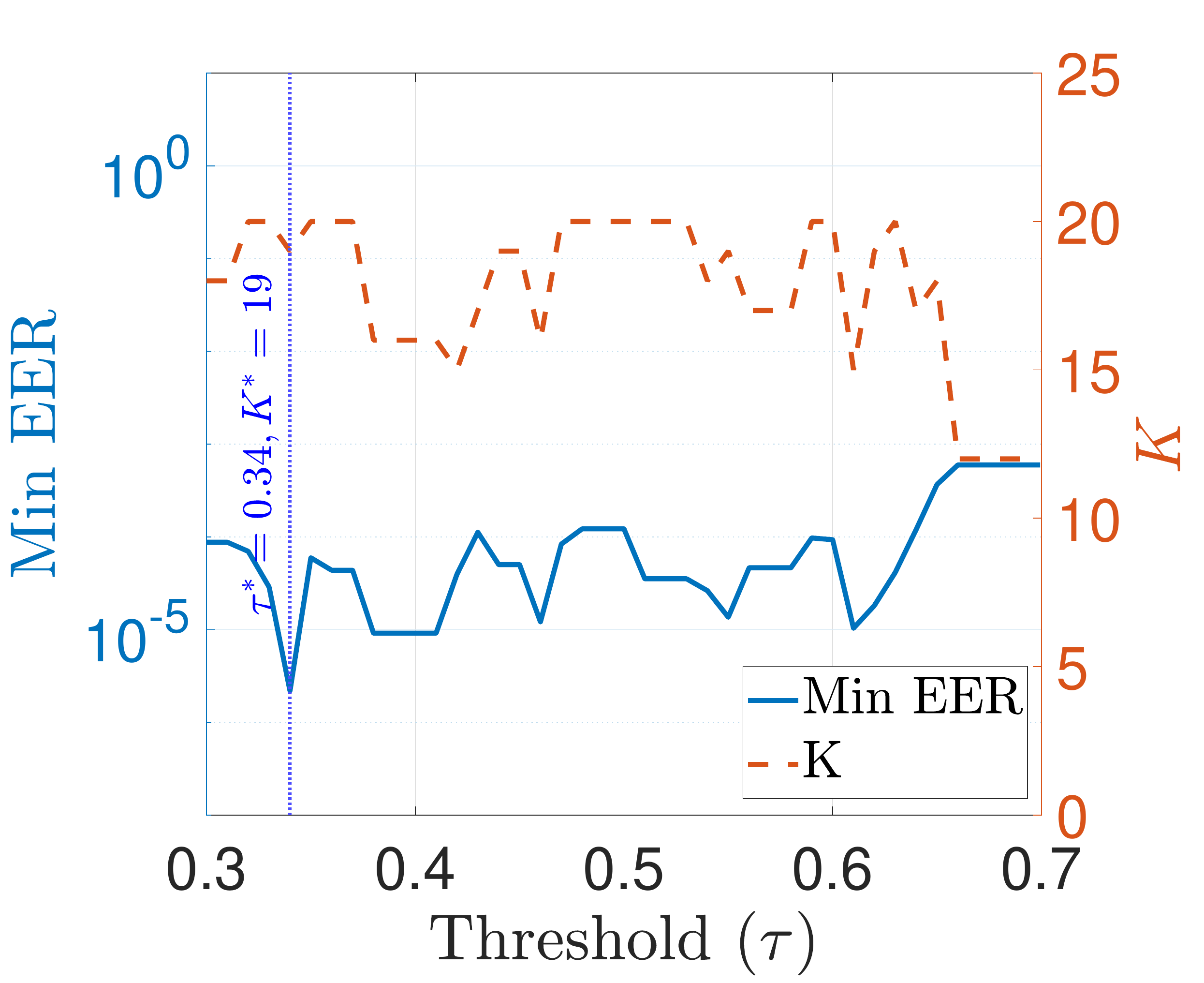}&
    \includegraphics[width=0.52\columnwidth]{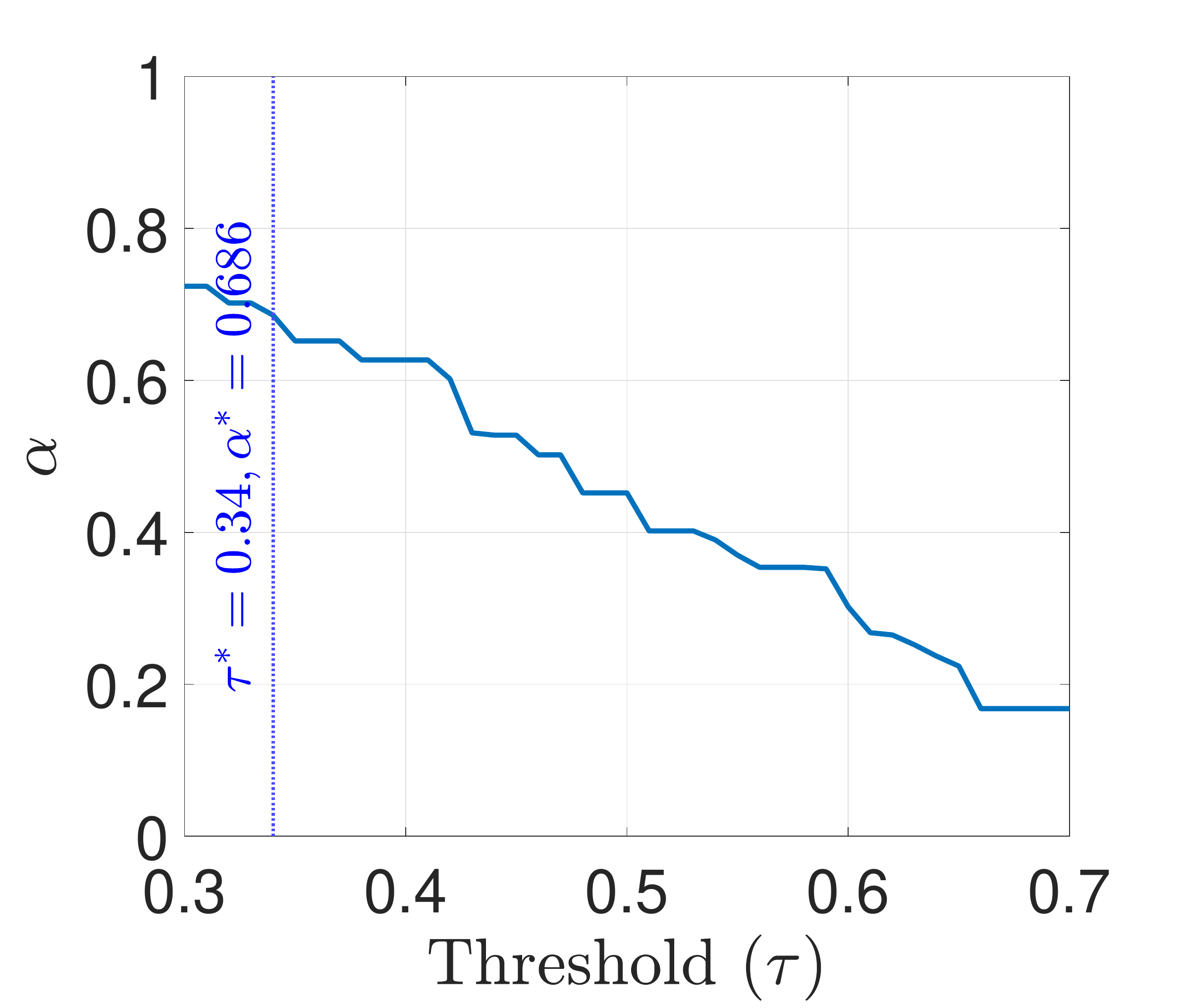}&
    \includegraphics[width=0.52\columnwidth]{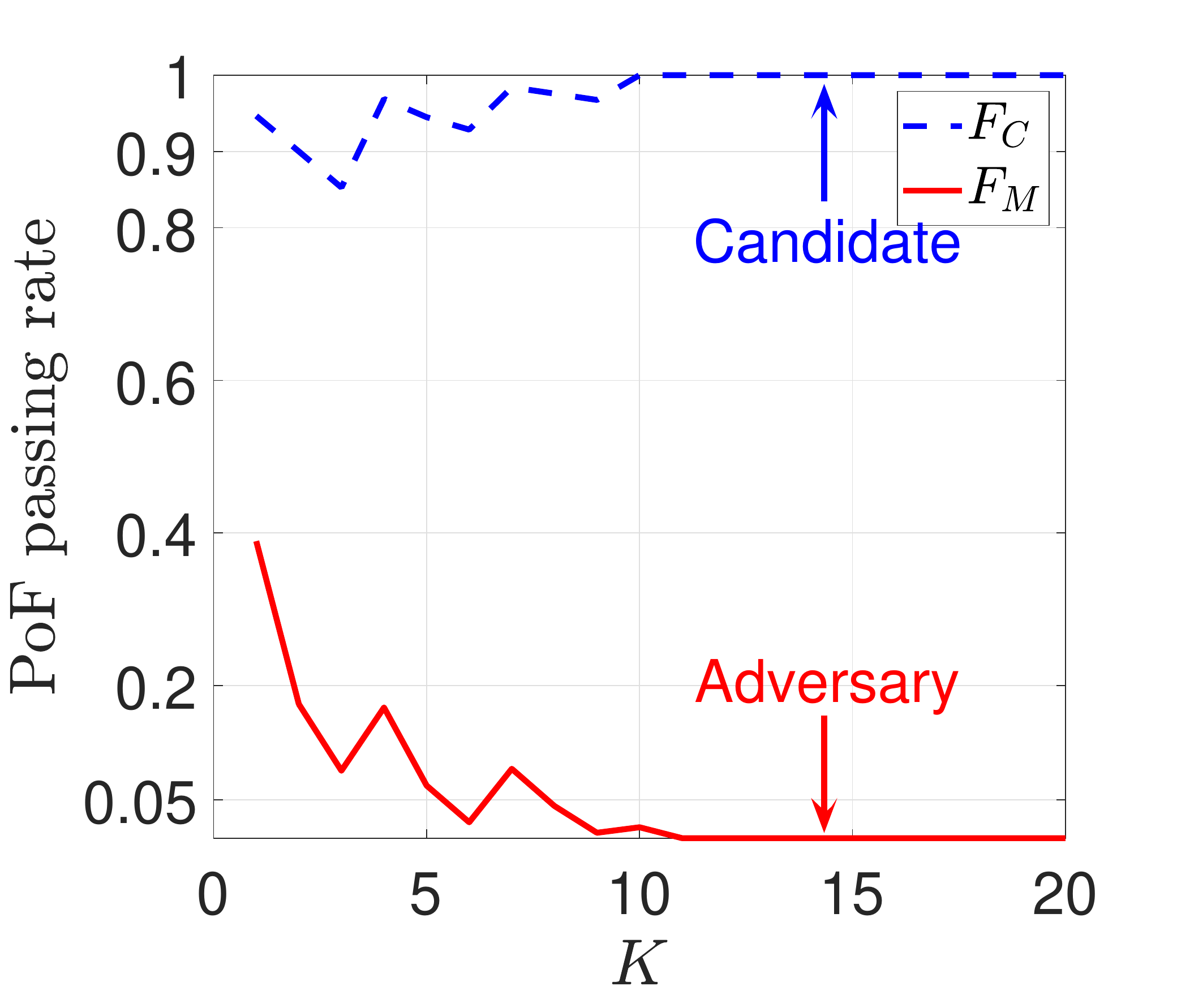}\\
    (a) & (b) & (c) &(d)
\end{tabular}
\caption{Urban driving, following-afar adversary: (a) single correlation test passing rates for  $\mathcal{C}$ and $\mathcal{M}$, (b) minimum $EER$ and corresponding optimal $K$, (c) $\alpha$ that minimizes the $EER$. (d) PoF passing rate for a valid candidate and for the adversary.}
\vspace{-0.2in}
\label{fig:urban_1}
\end{figure*}

We further evaluated  the following-afar adversary model, when $\Mc$ tries to pass verification by transmitting   real-time recorded RSS data. This adversary was realized by driving $\mathcal{M}$ at least 125 meters behind $\mathcal{V}.$ Due to the presence of traffic lights on the city streets, the distance between $\mathcal{M}$ and $\mathcal{V}$ varied during the experiment, as shown in Fig. \ref{fig:dist}(b), with 125 meters being the average.
We selected  the parameters with training sets collected by a following-afar model following the same steps as in Sec. \ref{sec:practical issues}, and show  the details in Fig. \ref{fig:urban_1}(a)-(c). Similarly, we plot  the PoF passing rate for $\mathcal{C}$ and $\mathcal{M}$ over 15 PoF verifications in Fig.~\ref{fig:urban_1}(d).  Again, we can see that the {\em PoF} is secure against the following-afar adversary. 

\subsection{Evaluation of \textit{PoF} on the Highway}
\label{sec:highway}
We ran experiments on a highway environment using Setup 2. Two platooning vehicles were driven on the piece of I-10 highway, shown in Fig. \ref{fig:exp_route}(b), from exit 250 to 257 for 6.4-mile back and forth over a period of 1.5 hours. The distance between $\Vc$ and $\Cc$ was maintained using ACC and remained quite stable (see Fig. \ref{fig:dist}(c)), with a 53.4m average.

\subsubsection{Remote adversary}

\begin{figure*}%
\centering
\setlength{\tabcolsep}{-3pt}
\begin{tabular}{cccc}
    \includegraphics[width=0.51\columnwidth]{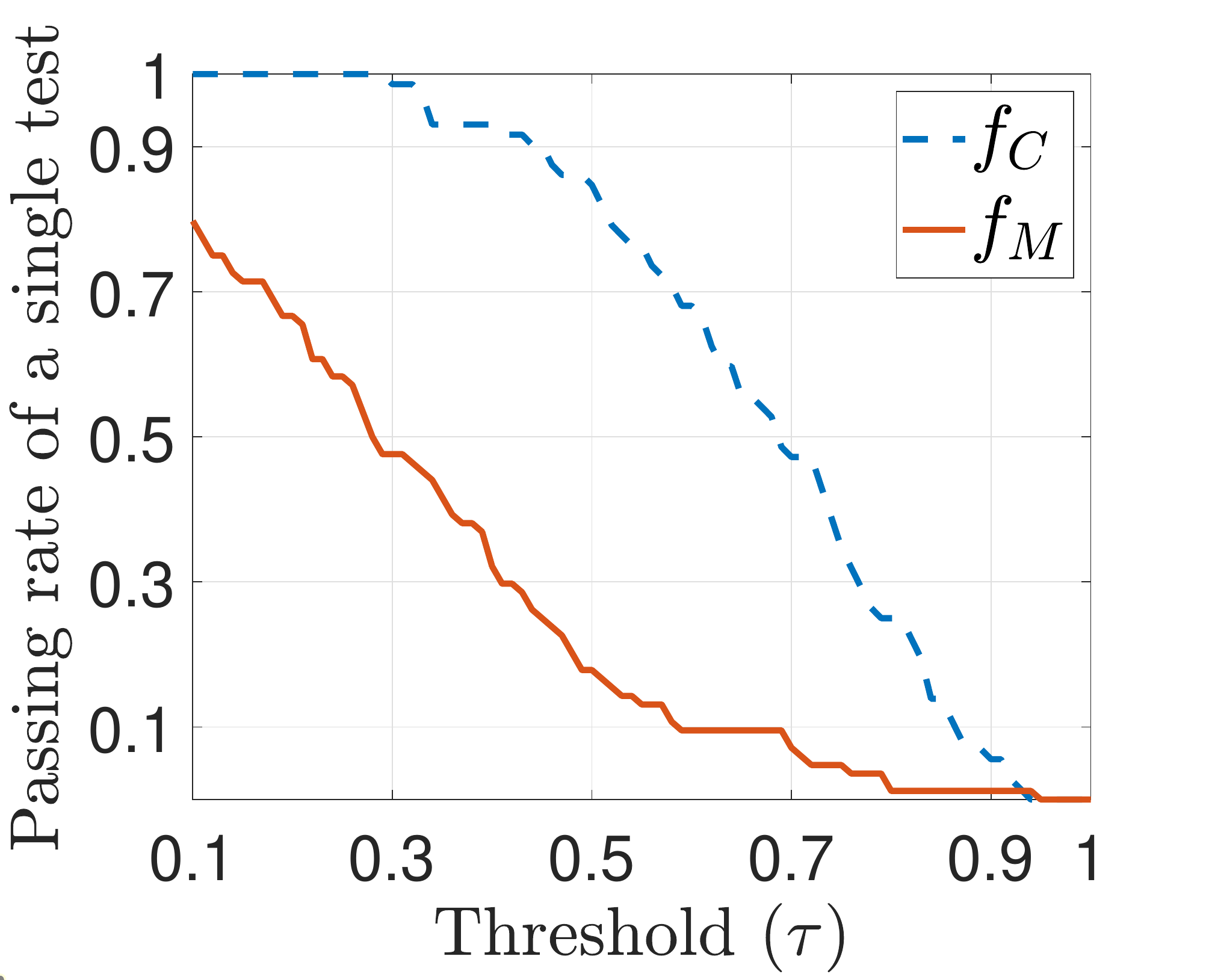}&
    \includegraphics[width=0.52\columnwidth]{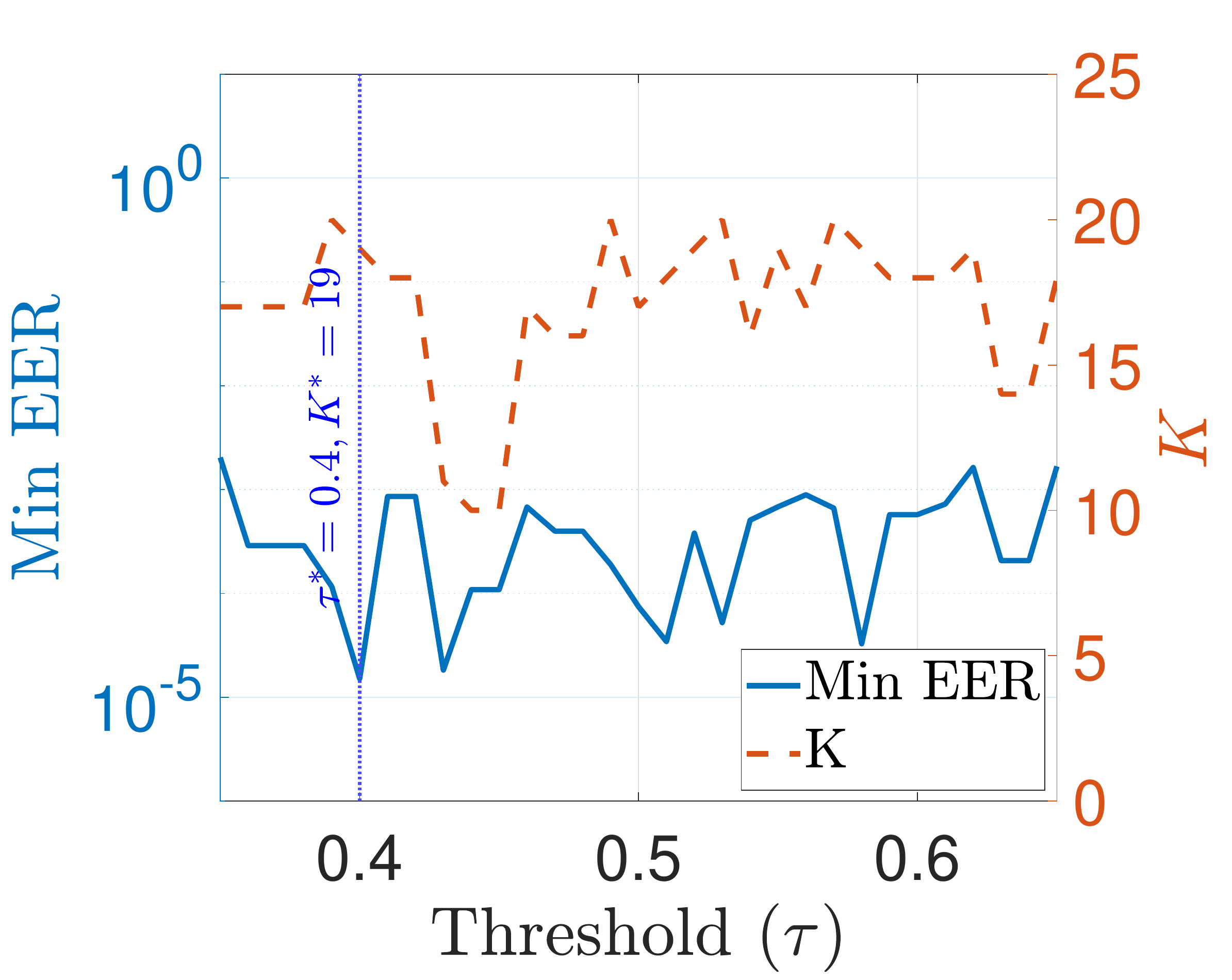}&
    \includegraphics[width=0.52\columnwidth]{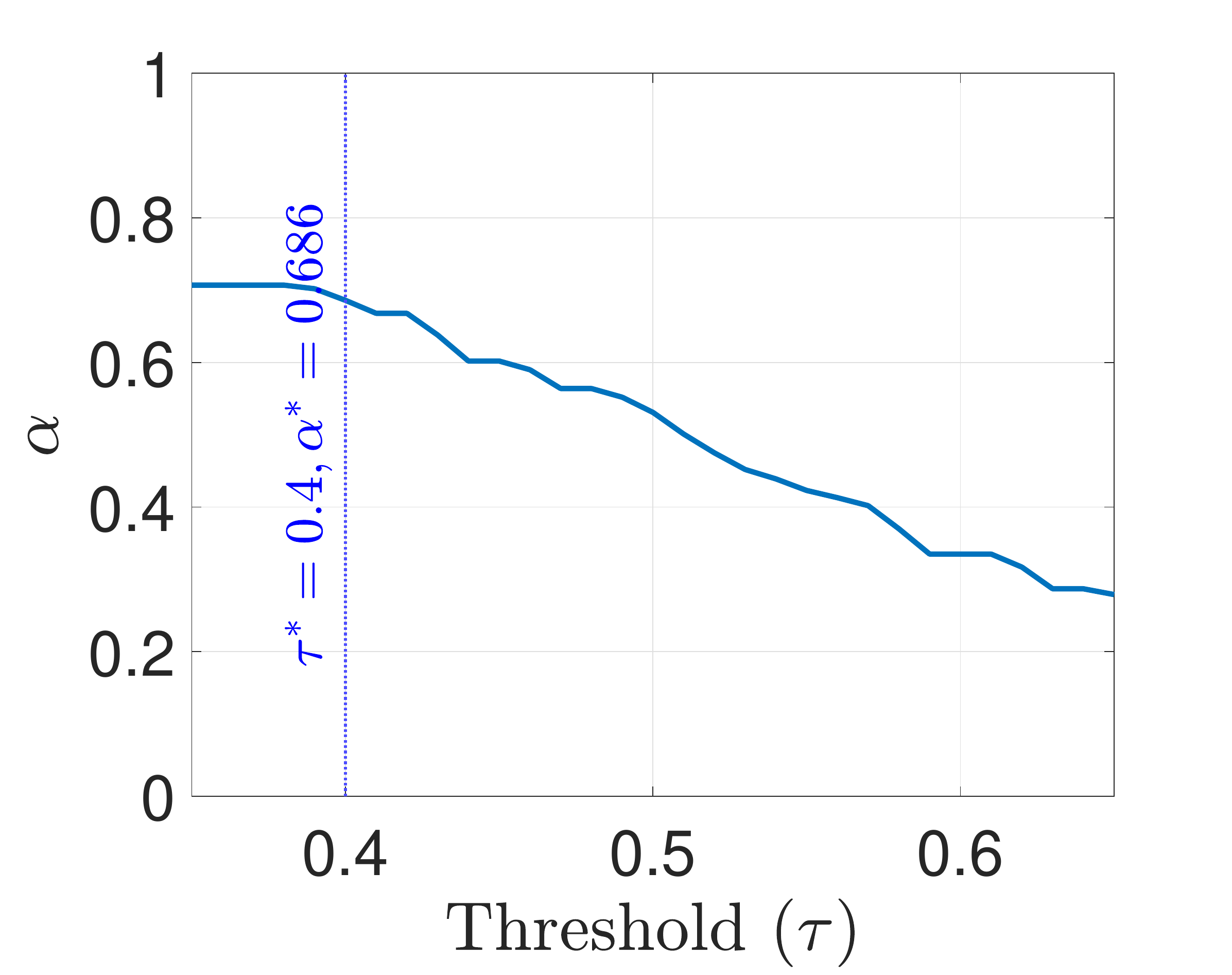}&
    \includegraphics[width=0.52\columnwidth]{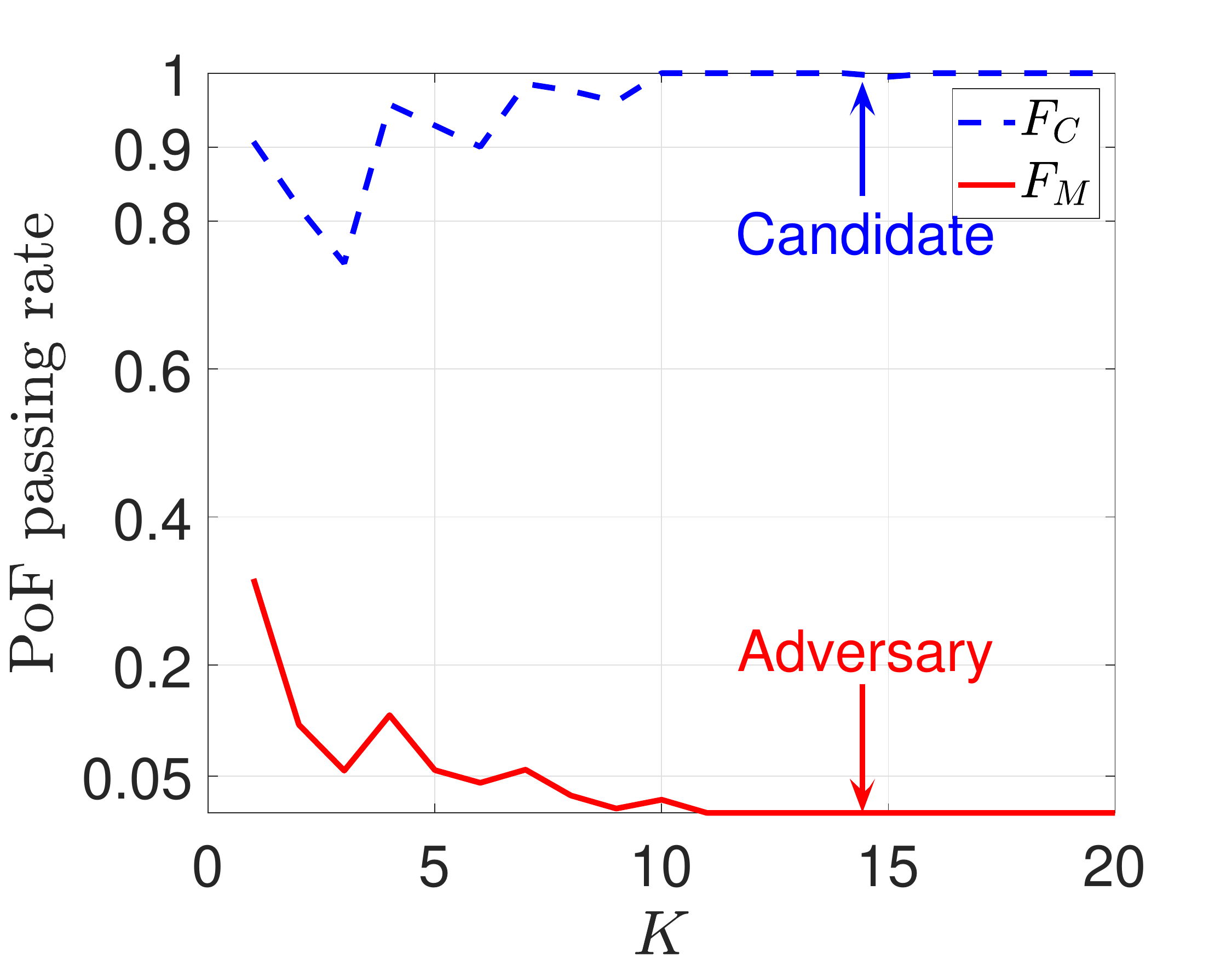}\\
    (a) & (b) & (c) &(d)
\end{tabular}
\caption{Highway driving, remote adversary: (a) single correlation test passing rates for  $\mathcal{C}$ and $\mathcal{M}$, (b) minimum $EER$ and corresponding optimal $K$, (c) $\alpha$ that minimizes the $EER$. (d) PoF passing rate for a valid candidate and for the adversary.}
\label{fig:highway_2}
\vspace{-0.2in}
\end{figure*}

We drove the Nissan Sentra to pre-record the RSS on the same route as the legitimate platoon. Due to the high speed  on the highway, the channel varied much more rapidly  compared to  urban areas. Therefore, for this experiment, we pre-recorded the RSS data 40 minutes ahead of time to mimic a remote adversary. The  results for parameter selection were  plotted in Fig. \ref{fig:highway_2}(a)-(c), where we selected $\tau^{\ast}=0.4$, $K^{\ast}=19$ and $\alpha^{\ast}=0.686$. In Fig. \ref{fig:highway_2}(d), we showed the convergence curve of PoF passing rate for both $\mathcal{C}$ and $\mathcal{M}$  over 13 PoF runs, where the legitimate candidate was verified with certainty for $K\geq10$ and the remote adversary was always detected for $K\geq11$, respectively.

\subsubsection{Following-afar adversary} 
\label{sec:highway type1}
\begin{figure*}%
\centering
\setlength{\tabcolsep}{-0.1pt}
\begin{tabular}{cccc}
    \includegraphics[width=0.51\columnwidth]{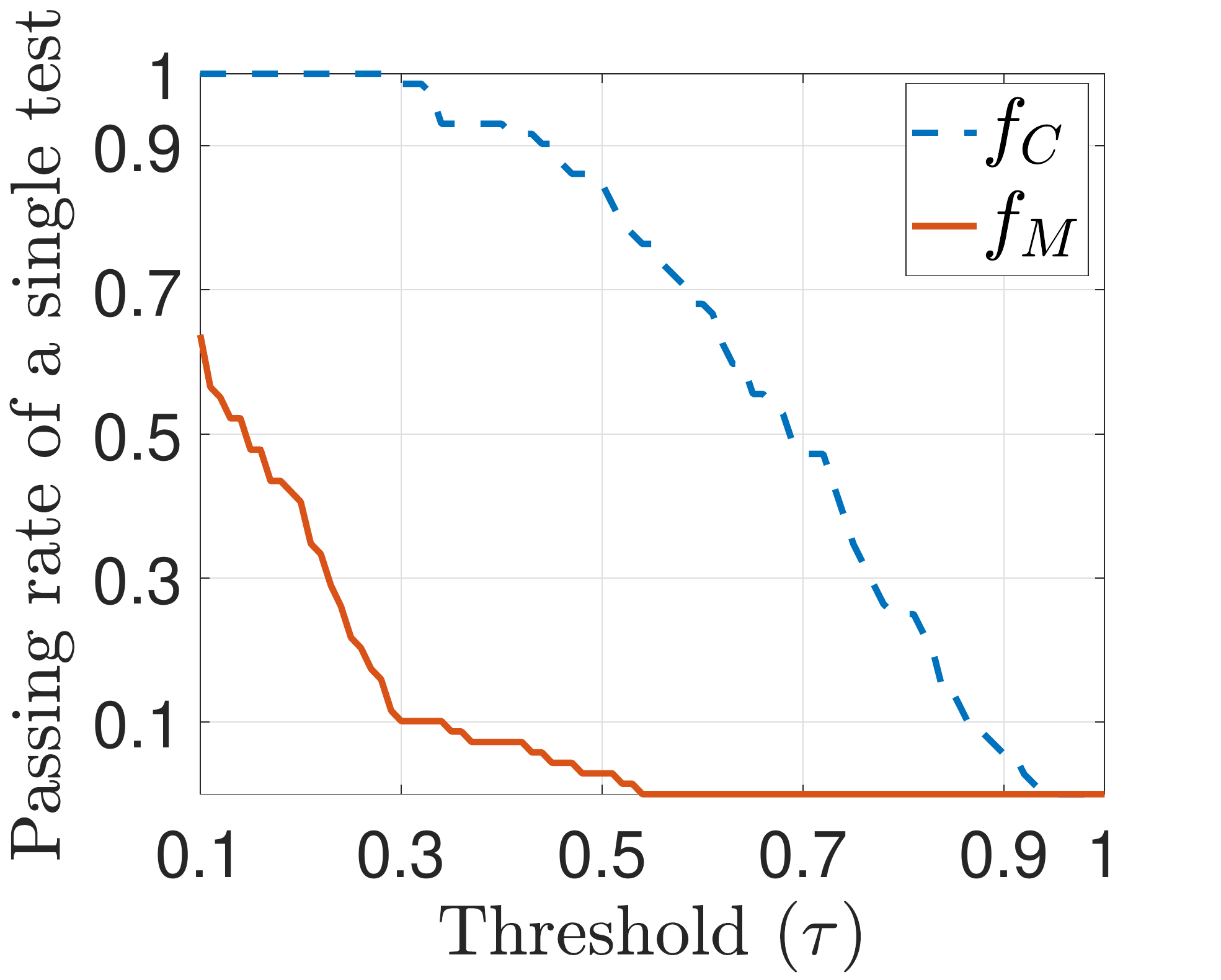}&
    \includegraphics[width=0.52\columnwidth]{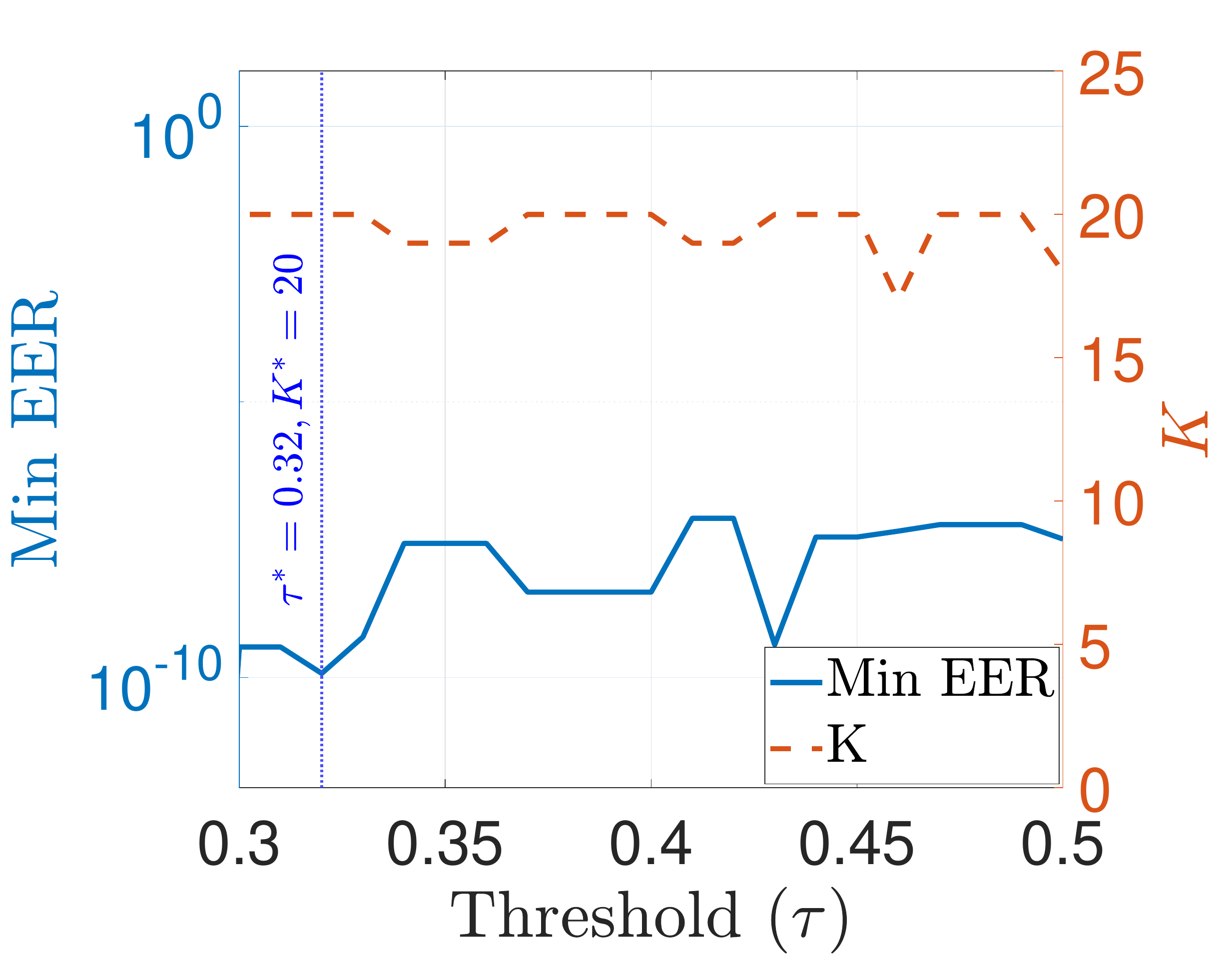}&
    \includegraphics[width=0.52\columnwidth]{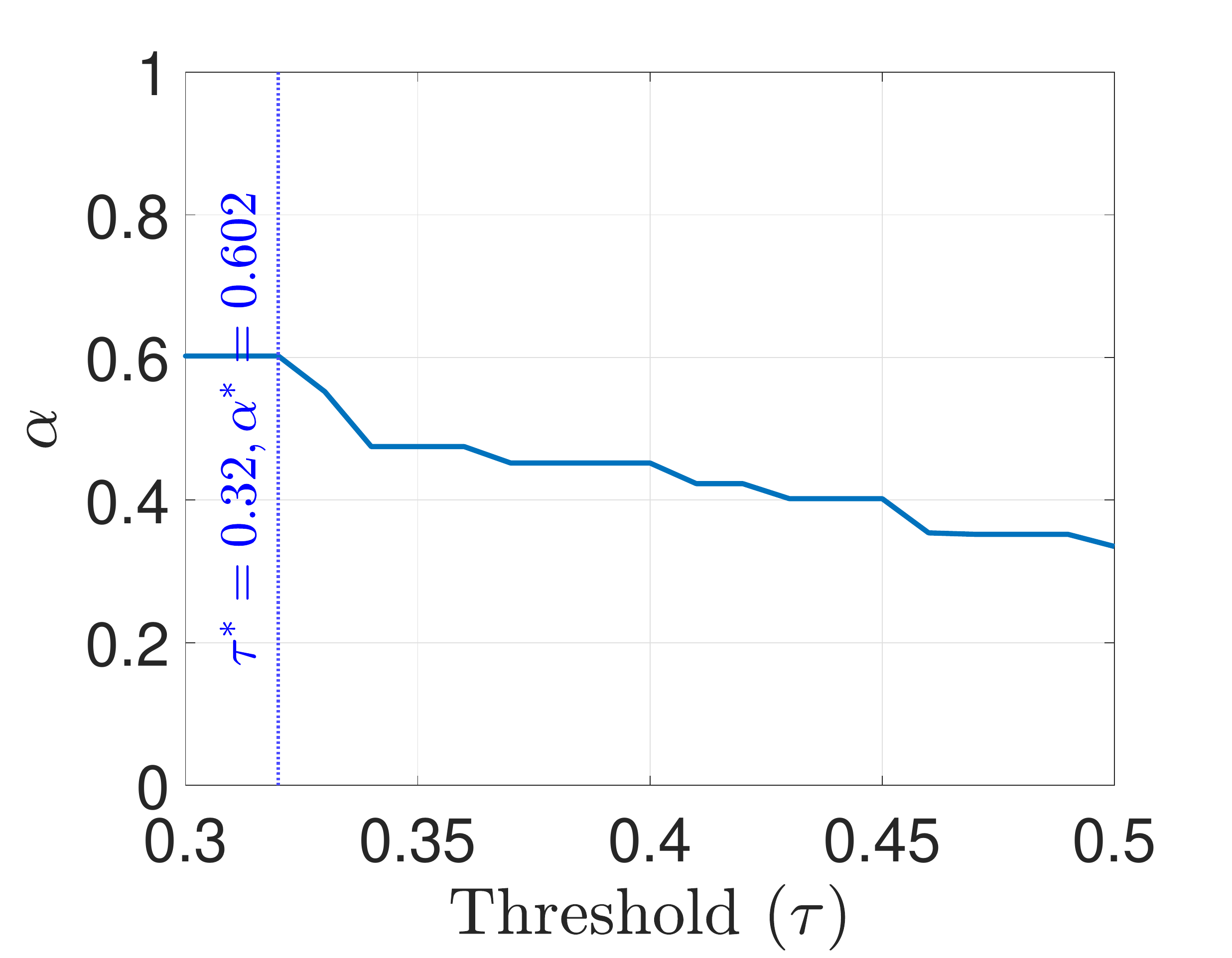}&
    \includegraphics[width=0.52\columnwidth]{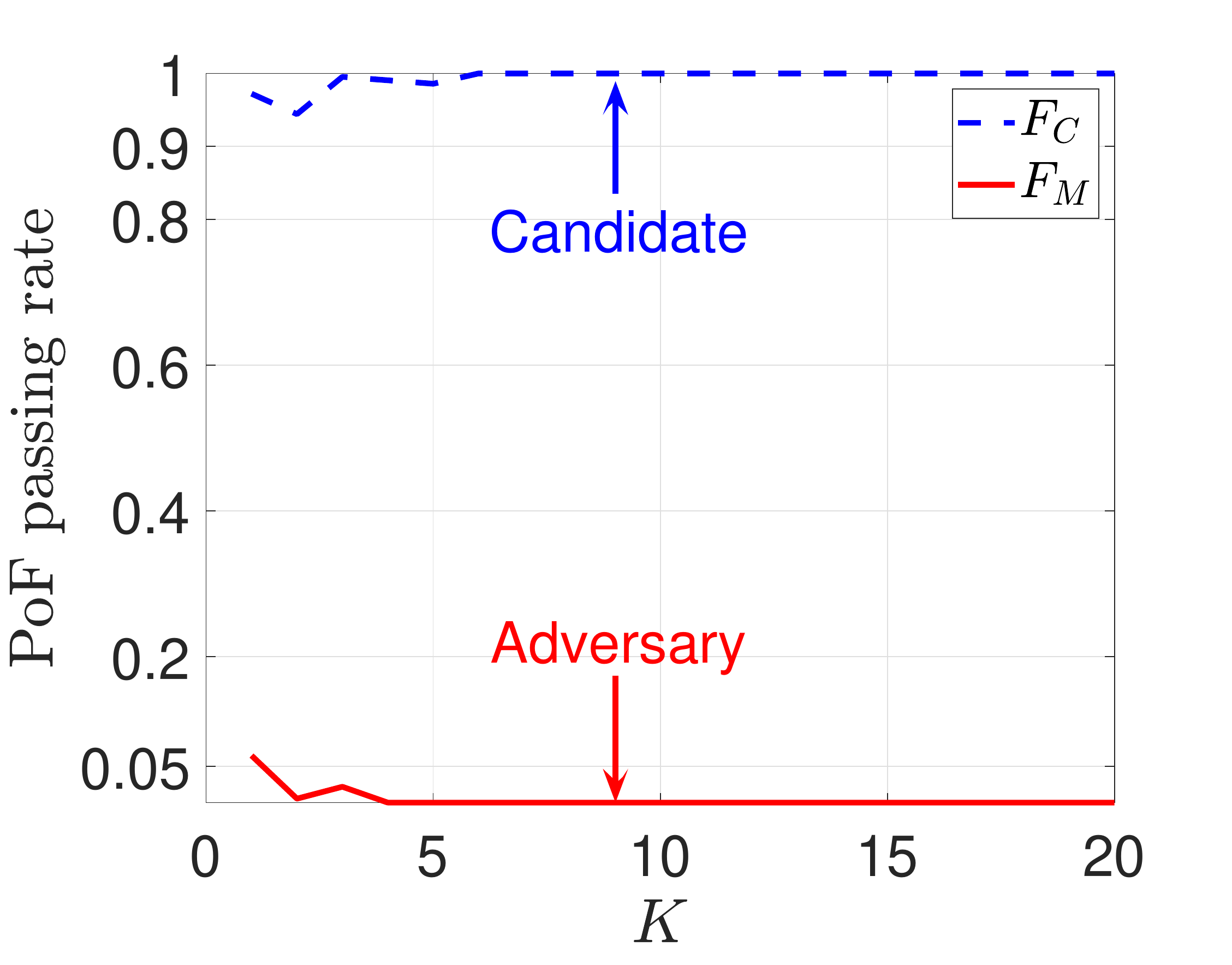}\\
     (a) & (b) & (c) &(d)
\end{tabular}   
\caption{Higway driving, following-afar adversary: (a) single correlation test passing rates for  $\mathcal{C}$ and $\mathcal{M}$, (b) minimum $EER$ and corresponding optimal $K$, (c) $\alpha$ that minimizes the $EER$. (d) PoF passing rate for a valid candidate and for the adversary.}
\label{fig:highway_1}
\vspace{-0.2in}
\end{figure*}

Next, we evaluated the following-afar adversary. The adversary followed the verifier at a distance of at least 100m. The following distance trace is shown in Fig. \ref{fig:dist}(d). The average following distance was 250m and varied significantly due to heavy highway traffic. The results of parameter selection are shown in  Fig. \ref{fig:highway_1}(a)-(c). We selected $\tau^{\ast}=0.32$, $K^{\ast}=20$, and $\alpha^{\ast}=0.602$. The PoF passing rate  averaged over 10 runs is shown in Fig. \ref{fig:highway_1}(d), where we observe that $F_C=1$ and $F_M=0$ when $K\geq6$ and $K\geq4$, respectively. We can see that the PoF yields better performance against the following-afar adversary (lower EER and faster PoF passing rate) than the remote adversary, mainly because the distance between the verifier and the adversary is larger in the highway and vehicles travel at high speeds.

\subsection{Evaluation of \textit{PoF} against Partially-following Adversary}
\label{sec:partially-following}

\begin{figure}[t]
\centering
\setlength{\tabcolsep}{-2pt}
\begin{tabular}{cc}
    \includegraphics[width=0.55\columnwidth]{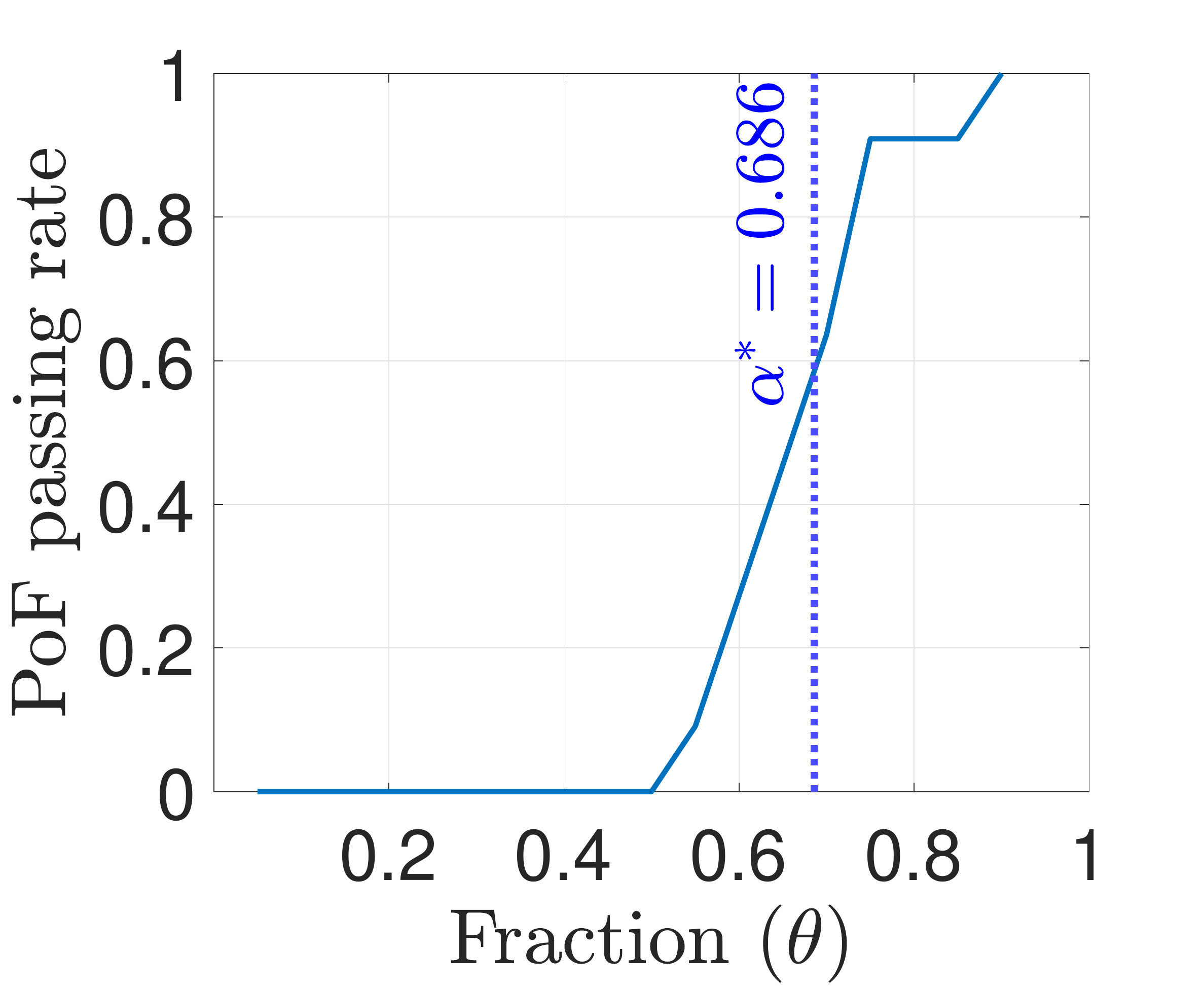} &  \includegraphics[width=0.55\columnwidth]{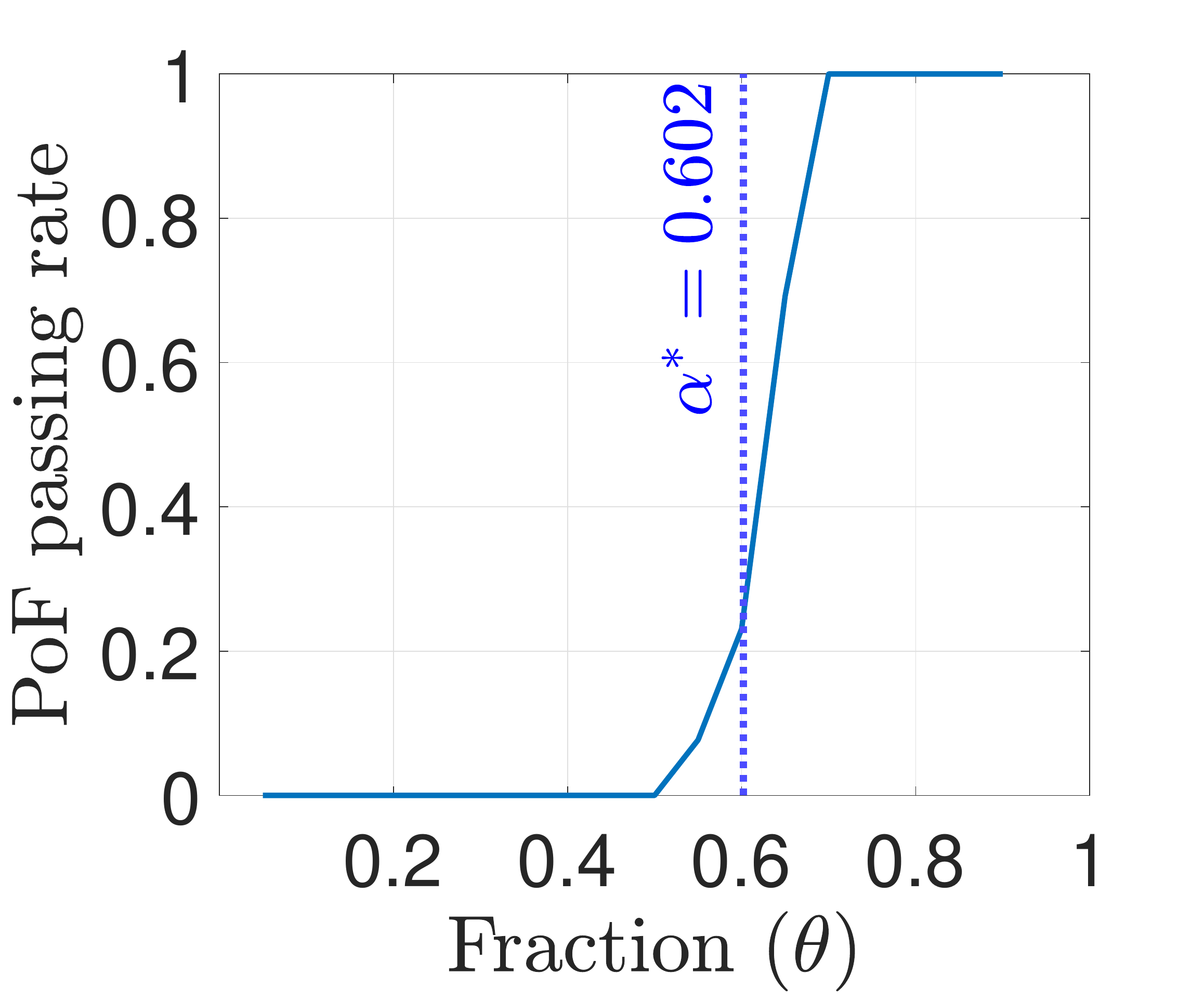}\\
    (a) urban  & (b) highway 
    \end{tabular}
\caption{The PoF passing rate as a function of the fraction of following time $\theta$ for a partially-following adversary.}
\label{fig:partial}
\vspace{-0.2in}
\end{figure}

We further implemented a  partially-following adversary who was within following distance for a fraction $\theta$ of a PoF duration, whereas he remained outside the following distance for the remaining PoF time. In Fig. \ref{fig:partial}, we show the PoF passing rate as a function of $\theta$. The test parameters ($N, K, M, \tau, \alpha$) were set to the values indicated in Sec.~\ref{sec:urban} for the urban environment and Sec.~\ref{sec:highway} for the highway environment,  assuming a following-afar adversary. The passing rates are averaged over 11 PoF runs for the urban and 13 runs for the highway environment, respectively. We observe that to pass the authentication  with non-zero probability, the adversary has to follow the platoon at least 50\% of the PoF duration (in the order of 100 seconds). For the urban environment where the RSS fluctuates more rapidly, the adversary is guaranteed to pass the PoF (passing rate equal to one) if he follows 90\% of the time, whereas this percentage drops to around 70\% on the highway environment. This indicates that a partially-following adversary could be successful, if he dedicates a significant portion of time in truly following the verifier. 
  
\subsection{Evaluation of RSS Randomness}

Another approach for a remote adversary is to predict real-time RSS data from pre-recorded data. We use approximate entropy $AE(\cdot)$ to evaluate the randomness of the moving average values used to compute the correlation. Approximate entropy is preferred to  sample entropy because it is a more accurate randomness measure when the number of samples is limited \cite{pincus1991approximate}. We calculate the approximate entropy of $\overline{\gamma_V}$ following standard steps in \cite{yentes2013appropriate, pincus1991approximate}. For the two parameters $m$ and $R$ required for approximation entropy calculation (i.e.,  the length of compared run of data and the similarity criterion, respectively),  we use typical values $m=2$ and $R=0.2 \times std(\overline{\gamma_V})$ as done in \cite{pincus1991approximate}, where $std(\overline{\gamma_V})$ is the standard deviation of $\overline{\gamma_V}$. The $ApEn(\overline{\gamma_V})$ for the urban and highway environment is 0.4730 and 0.3088, respectively. The $ApEn$ of a perfectly repeatable time series is about 0 \cite{yentes2013appropriate}, and is around 0.6 for  binary expansions of some common irrational numbers \cite{rukhin2000approximate}. Hence, our results indicated that the large-scale fading in dynamic traffic is random and unpredictable enough. Furthermore, it would be difficult for the attacker to pass the verification by predicting the RSS measurement of verifier $\mathcal{V}$.

\subsection{Duration of the PoF Protocol}
The duration of each {\em PoF} protocol is dominated by the RSS collection phase. The number of the RSS samples needed for the test is decided by the parameters we select. Since we reuse $N/2$ RSS samples for two consecutive subsets, it requires $(K+1)\times N/2$ RSS samples to complete $K$ correlation tests, which takes $\frac{(K+1)\times N}{2\times \text{(Sampling Rate)}}$ seconds for data collection. For the urban and highway environments, we fixed $N = 400$, $K= 20$, and our sampling rate was 20Hz. Therefore, about 200 seconds are required for each PoF protocol run. This is a reasonable cost as vehicle platoons are intended to travel for relatively long periods of time.

\section{Future Directions}
\label{sec:discussion}
{\bf Verification of other physical properties.} Our current PoF construct verifies continuous following within the following distance. For strict platooning verification relative vehcile positioning and lane restrictions shall also be verified. To verify the relative position, we can  leverage two  vehicles already accepted by the platoon. Assume vehicle $C$ wishes to prove it is located between $A$ and $B$, then $C$ can send its RSS samples to  both $A$ and $B$. If $C$ is following behind $A$ and $B$, this can also be proved if $C$ is in the range of $B$ but not $A$ or by repeating tests with different $d_{ref}$.

Verifying traveling on the same lane is more challenging. In the RF domain, one can use  features such as the angle-of-arrival (AoA) and Doppler shift. With multi-antenna receivers that may be standard with the advent of 5G, a verifier can determine the candidate's AoA using beamforming. If the candidate is in the same lane, the AoA is either $90^{\circ}$ or $270^{\circ}$. AoA has been previously used  to enhance  WiFi security \cite{xiong2013securearray}, as well as for secure motion verification \cite{sun2020svm}. It is difficult to spoof without deploying artificial reflectors. A high mobility scenario makes it nearly impossible. If only a single-antenna transceiver is available, one can exploit the Doppler shift (DS) which reflects the relative speed. If the candidate follows the verifier closely at the same speed and lane, he should measure similar DS from V2V signals from other vehicles. Any vehicle on another lane would not measure the same DS due to different angles.
 
{\bf PoF from other sensing modalities.} Cameras can also be used to capture the ambient physical environment as the platoon travels on the road. Imagine a verifier and a candidate traveling on the same highway. Using cameras, they can capture, analyze, and cross-correlate images of transient environment features. For instance, the two vehicles could capture and timestamp images of a passing by semi-truck (moving element) with some static feature in the background (building, tree, billboard, etc). This will ensure that the two vehicles are co-traveling within the following distance. 

Another approach is to measure the following distance using  LiDAR. The verifier could randomly perturb the following distance by subtly accelerating and braking. A valid candidate should be able to accurately measure the distance changes and report them to the verifier. If both parties agree on the perturbations, the PoF test is passed.  We will explore these extensions in our future works.

\section{Conclusion}
\label{sec:conclusion}
We proposed a novel security primitive called Proof-of-Following (PoF) that binds the physical property of ``following''  to the candidate vehicle's digital identity. Our PoF protocol allows a candidate vehicle to continuously prove to a verifier vehicle that it follows the verifier within the typical platooning distance. We exploited the large-scale wireless fading from cellular towers as an easy-to-measure solution correlating the motions of vehicles. Our approach has a remarkable advantage in hardware requirements as  only the RF modality is required, which is widely available in outdoor environments. We conducted extensive real-world experiments in the freeway, urban and highway environments to evaluate the performance and security of our PoF protocol. Results showed that PoF is resistant to both pre-recording and following attacks with overwhelming probability and robust performance.

\section*{Acknowledgements}
We thank the anonymous reviewers for their insightful comments. We also like to thank Dr. Jonathan Sprinkle and Dr. Matt Bunting for their help with the highway experiments. This work was supported by ARO grant W911NF-19-1-0050.

\bibliographystyle{IEEEtranS}
\bibliography{protocol}

\begin{thebibliography}{10}
\providecommand{\url}[1]{#1}
\csname url@samestyle\endcsname
\providecommand{\newblock}{\relax}
\providecommand{\bibinfo}[2]{#2}
\providecommand{\BIBentrySTDinterwordspacing}{\spaceskip=0pt\relax}
\providecommand{\BIBentryALTinterwordstretchfactor}{4}
\providecommand{\BIBentryALTinterwordspacing}{\spaceskip=\fontdimen2\font plus
\BIBentryALTinterwordstretchfactor\fontdimen3\font minus
  \fontdimen4\font\relax}
\providecommand{\BIBforeignlanguage}[2]{{%
\expandafter\ifx\csname l@#1\endcsname\relax
\typeout{** WARNING: IEEEtranS.bst: No hyphenation pattern has been}%
\typeout{** loaded for the language `#1'. Using the pattern for}%
\typeout{** the default language instead.}%
\else
\language=\csname l@#1\endcsname
\fi
#2}}
\providecommand{\BIBdecl}{\relax}
\BIBdecl

\bibitem{NI}
\BIBentryALTinterwordspacing
 [Online]. Available: \url{https://www.ettus.com}
\BIBentrySTDinterwordspacing

\bibitem{3GPP}
\BIBentryALTinterwordspacing
3GPP. Testing {C-V2X} access layer for its devices. [Online]. Available:
  \url{https://www.3gpp.org/news-events/2120-c-v2x}
\BIBentrySTDinterwordspacing

\bibitem{secureV2X}
\emph{3rd Generation Partnership Project;Technical Specification Group Services
  and System Aspects;Security aspect for LTE support of Vehicle-to-Everything
  (V2X) services Rel-16, V16.0.0}, 3GPP Std. TS 33.185, Jul. 2020.

\bibitem{algans2002experimental}
A.~Algans, K.~I. Pedersen, and P.~E. Mogensen, ``Experimental analysis of the
  joint statistical properties of azimuth spread, delay spread, and shadow
  fading,'' \emph{IEEE Journal on Selected Areas in Communications}, vol.~20,
  no.~3, pp. 523--531, 2002.

\bibitem{Avoine2018survey}
G.~Avoine, M.~A. Bing\"{o}l, I.~Boureanu, S.~\v{c}apkun, G.~Hancke,
  S.~Karda\c{s}, C.~H. Kim, C.~Lauradoux, B.~Martin, J.~Munilla, A.~Peinado,
  K.~B. Rasmussen, D.~Singel\'{e}e, A.~Tchamkerten, R.~Trujillo-Rasua, and
  S.~Vaudenay, ``Security of distance-bounding: A survey,'' \emph{ACM Comput.
  Surv.}, vol.~51, no.~5, Sep. 2018.

\bibitem{bian2019reducing}
Y.~Bian, Y.~Zheng, W.~Ren, S.~E. Li, J.~Wang, and K.~Li, ``Reducing time
  headway for platooning of connected vehicles via {V2V} communication,''
  \emph{Transportation Research Part C: Emerging Technologies}, vol. 102, pp.
  87--105, 2019.

\bibitem{bissmeyer2012central}
N.~Bi{\ss}meyer, J.~Njeukam, J.~Petit, and K.~M. Bayarou, ``Central misbehavior
  evaluation for vanets based on mobility data plausibility,'' in \emph{Proc.
  of VANET}, 2012, pp. 73--82.

\bibitem{brands1993distance}
S.~Brands and D.~Chaum, ``Distance-bounding protocols,'' in \emph{Workshop on
  the Theory and Application of of Cryptographic Techniques}.\hskip 1em plus
  0.5em minus 0.4em\relax Springer, 1993, pp. 344--359.

\bibitem{conti2020context}
M.~Conti and C.~Lal, ``Context-based co-presence detection techniques: A
  survey,'' \emph{Computers \& Security}, vol.~88, p. 101652, 2020.

\bibitem{guan2015measurements}
K.~Guan, B.~Ai, Z.~Zhong, C.~F. L{\'o}pez, L.~Zhang, C.~Briso-Rodr{\'\i}guez,
  A.~Hrovat, B.~Zhang, R.~He, and T.~Tang, ``Measurements and analysis of
  large-scale fading characteristics in curved subway tunnels at 920 {MH}z,
  2400 {MH}z, and 5705 {MH}z,'' \emph{IEEE Transactions on Intelligent
  Transportation Systems}, vol.~16, no.~5, pp. 2393--2405, 2015.

\bibitem{guanetti2018control}
J.~Guanetti, Y.~Kim, and F.~Borrelli, ``Control of connected and automated
  vehicles: State of the art and future challenges,'' \emph{Annual Reviews in
  Control}, vol.~45, pp. 18--40, 2018.

\bibitem{gudmundson1991correlation}
M.~Gudmundson, ``Correlation model for shadow fading in mobile radio systems,''
  \emph{Electronics Letters}, vol.~27, no.~23, pp. 2145--2146, 1991.

\bibitem{halevi1996practical}
S.~Halevi and S.~Micali, ``Practical and provably-secure commitment schemes
  from collision-free hashing,'' in \emph{Annual International Cryptology
  Conference}, 1996, pp. 201--215.

\bibitem{han2018you}
J.~Han, A.~J. Chung, M.~K. Sinha, M.~Harishankar, S.~Pan, H.~Y. Noh, P.~Zhang,
  and P.~Tague, ``Do you feel what {I} hear? enabling autonomous iot device
  pairing using different sensor types,'' in \emph{2018 IEEE S\&P}, pp.
  836--852.

\bibitem{han2017convoy}
J.~Han, M.~Harishankar, X.~Wang, A.~J. Chung, and P.~Tague, ``Convoy: Physical
  context verification for vehicle platoon admission,'' in \emph{Proc. of
  HotMobile}, 2017, pp. 73--78.

\bibitem{hayashi2013casa}
E.~Hayashi, S.~Das, S.~Amini, J.~Hong, and I.~Oakley, ``Casa: context-aware
  scalable authentication,'' in \emph{Proc. of SOUPS}, 2013, pp. 1--10.

\bibitem{he2014shadow}
R.~He, Z.~Zhong, B.~Ai, and C.~Oestges, ``Shadow fading correlation in
  high-speed railway environments,'' \emph{IEEE Transactions on Vehicular
  Technology}, vol.~64, no.~7, pp. 2762--2772, 2014.

\bibitem{IEEE:WAVE}
\emph{IEEE Standard for Wireless Access in Vehicular Environments
  (WAVE)--Certificate Management Interfaces for End Entities}, IEEE Std. IEEE
  1609.2.1, 2020.

\bibitem{jia2015survey}
D.~Jia, K.~Lu, J.~Wang, X.~Zhang, and X.~Shen, ``A survey on platoon-based
  vehicular cyber-physical systems,'' \emph{IEEE communications surveys \&
  tutorials}, vol.~18, no.~1, pp. 263--284, 2015.

\bibitem{juuti2017stash}
M.~Juuti, C.~Vaas, I.~Sluganovic, H.~Liljestrand, N.~Asokan, and I.~Martinovic,
  ``Stash: Securing transparent authentication schemes using prover-side
  proximity verification,'' in \emph{IEEE SECON}, 2017, pp. 1--9.

\bibitem{Kamel2020Simulation}
J.~Kamel, M.~R. Ansari, J.~Petit, A.~Kaiser, I.~B. Jemaa, and P.~Urien,
  ``Simulation framework for misbehavior detection in vehicular networks,''
  \emph{IEEE Transactions on Vehicular Technology}, vol.~69, no.~6, pp.
  6631--6643, 2020.

\bibitem{ko2021approach}
B.~Ko and S.~H. Son, ``An approach to detecting malicious information attacks
  for platoon safety,'' \emph{IEEE Access}, vol.~9, pp. 101\,289--101\,299,
  2021.

\bibitem{li2020t2pair}
X.~Li, Q.~Zeng, L.~Luo, and T.~Luo, ``T2pair: Secure and usable pairing for
  heterogeneous iot devices,'' in \emph{Proc. of CCS}, 2020, pp. 309--323.

\bibitem{luu2020spacing}
D.~L. Luu, C.~Lupu, L.~S. Ismail, and H.~Alshareefi, ``Spacing control of
  cooperative adaptive cruise control vehicle platoon,'' in \emph{Proc. of IEEE
  AQTR}, 2020, pp. 1--6.

\bibitem{lyamin2016study}
N.~Lyamin, Q.~Deng, and A.~Vinel, ``Study of the platooning fuel efficiency
  under {ETSI ITS-G5} communications,'' in \emph{Proc. of IEEE 19th ITSC},
  2016, pp. 551--556.

\bibitem{mathur2011proximate}
S.~Mathur, R.~Miller, A.~Varshavsky, W.~Trappe, and N.~Mandayam, ``Proximate:
  proximity-based secure pairing using ambient wireless signals,'' in
  \emph{Proc. of {MobiSys}}, 2011, pp. 211--224.

\bibitem{miettinen2014context}
M.~Miettinen, N.~Asokan, T.~D. Nguyen, A.-R. Sadeghi, and M.~Sobhani,
  ``Context-based zero-interaction pairing and key evolution for advanced
  personal devices,'' in \emph{Proc. of CCS}, 2014, pp. 880--891.

\bibitem{Nguyen2020Enhancing}
V.-L. Nguyen, P.-C. Lin, and R.-H. Hwang, ``Enhancing misbehavior detection in
  5{G} vehicle-to-vehicle communications,'' \emph{IEEE Transactions on
  Vehicular Technology}, vol.~69, no.~9, pp. 9417--9430, 2020.

\bibitem{pincus1991approximate}
S.~M. Pincus, ``Approximate entropy as a measure of system complexity,''
  \emph{Proc. of the National Academy of Sciences}, vol.~88, no.~6, pp.
  2297--2301, 1991.

\bibitem{qualcommV2X}
\BIBentryALTinterwordspacing
I.~Qualcomm~Technologies. Connecting vehicles to everything with {C-V2X}.
  [Online]. Available: \url{https://www.qualcomm.com/research/5g/cellular-v2x}
\BIBentrySTDinterwordspacing

\bibitem{recommendation1997guidelines}
I.-R. Recommendation, ``Guidelines for evaluation of radio transmission
  technologies for {IMT}-2000,'' \emph{Rec. ITU-R M. 1225}, 1997.

\bibitem{rukhin2000approximate}
A.~L. Rukhin \emph{et~al.}, ``Approximate entropy for testing randomness,''
  \emph{Journal of Applied Probability}, vol.~37, no.~1, pp. 88--100, 2000.

\bibitem{schurmann2011secure}
D.~Sch{\"u}rmann and S.~Sigg, ``Secure communication based on ambient audio,''
  \emph{IEEE TMC}, vol.~12, no.~2, pp. 358--370, 2011.

\bibitem{senarath2007multi}
G.~Senarath, ``Multi-hop relay system evaluation methodology (channel model and
  performance metric),'' \emph{http://ieee802.
  org/16/relay/docs/80216j-06\_013r3. pdf}, 2007.

\bibitem{shi2013bana}
L.~Shi, M.~Li, S.~Yu, and J.~Yuan, ``Bana: Body area network authentication
  exploiting channel characteristics,'' \emph{IEEE Journal on Selected Areas in
  Communications}, vol.~31, no.~9, pp. 1803--1816, 2013.

\bibitem{so2019physical}
S.~So, J.~Petit, and D.~Starobinski, ``Physical layer plausibility checks for
  misbehavior detection in {V2X} networks,'' in \emph{Proc. of WiSec}, 2019,
  pp. 84--93.

\bibitem{song2021organization}
M.~Song, F.~Chen, and X.~Ma, ``Organization of autonomous truck platoon
  considering energy saving and pavement fatigue,'' \emph{Transportation
  Research Part D: Transport and Environment}, vol.~90, p. 102667, 2021.

\bibitem{sun2017data}
M.~Sun, M.~Li, and R.~Gerdes, ``A data trust framework for vanets enabling
  false data detection and secure vehicle tracking,'' in \emph{Proc. of
  CNS}.\hskip 1em plus 0.5em minus 0.4em\relax IEEE, 2017, pp. 1--9.

\bibitem{sun2020svm}
M.~Sun, Y.~Man, M.~Li, and R.~Gerdes, ``{SVM}: secure vehicle motion
  verification with a single wireless receiver,'' in \emph{Proc. of WiSec},
  2020, pp. 65--76.

\bibitem{szyszkowicz2010feasibility}
S.~S. Szyszkowicz, H.~Yanikomeroglu, and J.~S. Thompson, ``On the feasibility
  of wireless shadowing correlation models,'' \emph{IEEE Transactions on
  Vehicular Technology}, vol.~59, no.~9, pp. 4222--4236, 2010.

\bibitem{tippenhauer2015uwb}
N.~O. Tippenhauer, H.~Luecken, M.~Kuhn, and S.~Capkun, ``{UWB}
  rapid-bit-exchange system for distance bounding,'' in \emph{Proc. of WiSec},
  2015, pp. 1--12.

\bibitem{turri2016cooperative}
V.~Turri, B.~Besselink, and K.~H. Johansson, ``Cooperative look-ahead control
  for fuel-efficient and safe heavy-duty vehicle platooning,'' \emph{IEEE
  Transactions on Control Systems Technology}, vol.~25, no.~1, pp. 12--28,
  2016.

\bibitem{ADSs}
\BIBentryALTinterwordspacing
\emph{Automated Vehicles Comprehensive Plan}, U.S. Department of
  Transportation, Washington, DC, USA, 2021. [Online]. Available:
  \url{https://www.transportation.gov/sites/dot.gov/files/2021-01/USDOT_AVCP.pdf}
\BIBentrySTDinterwordspacing

\bibitem{vaas2018get}
C.~Vaas, M.~Juuti, N.~Asokan, and I.~Martinovic, ``Get in line: Ongoing
  co-presence verification of a vehicle formation based on driving
  trajectories,'' in \emph{Proc. of the IEEE EuroS\&P}, 2018, pp. 199--213.

\bibitem{platoon2020}
G.~Wood, ``Truck platooning expected to make inroads in 2020,'' \url{
  https://www.oemoffhighway.com/electronics/smart-systems/automated-systems/article/21114230/truck-platooning-expected-to-make-inroads-in-2020}.

\bibitem{wu2018survey}
Y.~Wu, A.~Khisti, C.~Xiao, G.~Caire, K.-K. Wong, and X.~Gao, ``A survey of
  physical layer security techniques for 5{G} wireless networks and challenges
  ahead,'' \emph{IEEE Journal on Selected Areas in Communications}, vol.~36,
  no.~4, pp. 679--695, 2018.

\bibitem{xiong2013securearray}
J.~Xiong and K.~Jamieson, ``Secure{A}rray: Improving {W}i{F}i security with
  fine-grained physical-layer information,'' in \emph{Proc. of MobiCom}, 2013,
  pp. 441--452.

\bibitem{yentes2013appropriate}
J.~M. Yentes, N.~Hunt, K.~K. Schmid, J.~P. Kaipust, D.~McGrath, and
  N.~Stergiou, ``The appropriate use of approximate entropy and sample entropy
  with short data sets,'' \emph{Annals of Biomedical Engineering}, vol.~41,
  no.~2, pp. 349--365, 2013.

\bibitem{zhang2020distributed}
D.~Zhang, Y.-P. Shen, S.-Q. Zhou, X.-W. Dong, and L.~Yu, ``Distributed secure
  platoon control of connected vehicles subject to {DoS} attack: Theory and
  application,'' \emph{IEEE Transactions on Systems, Man, and Cybernetics:
  Systems}, 2020.

\end{thebibliography}

\appendix
\normalsize

\subsection{Resistance to MitM Attacks}
\label{Appendix-pof}

\begin{figure}[t]
\centering
\setlength{\tabcolsep}{-2pt}
\begin{tabular}{c}
    \includegraphics[width=0.95\columnwidth]{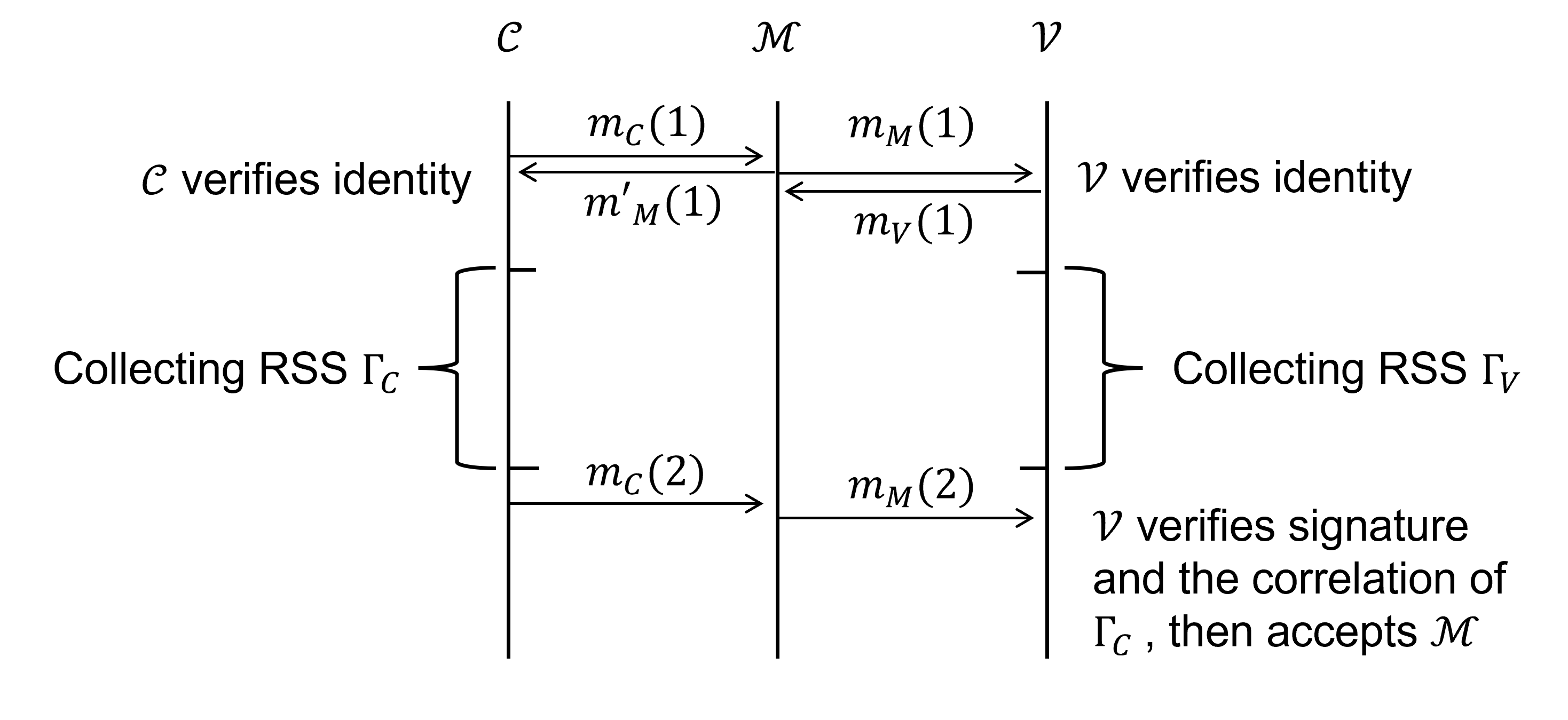}
    \end{tabular}
\caption{A successful MiTM attack on the PoF protocol when $\mathcal{C}$ does not know the verifier's identity a priori.}
\label{Success_MitM}
\vspace{-0.2in}
\end{figure}

When the verifier's identity is not known a priori to the candidate, a MiTM attack is possible if the adversary $\mathcal{M}$ spoofs a verifier. Let a candidate $\mathcal{C}$ follow a legitimate verifier $\mathcal{V}$ at a distance $d_{VC}<d_{ref}$. Let  a following-afar adversary $\mathcal{M}$ attempt to launch a MiTM attack, as shown in Fig.~\ref{fig:MitM_topology}. For the PoF protocol in Fig.~\ref{algorithm}, $\mathcal{M}$ can succeed in a  MiTM attack via the steps shown in Fig.~\ref{Success_MitM}. 
The candidate $\mathcal{C}$ initiates a platoon join request by sending message
\[
m_C(1) \leftarrow \text{REQ}, ID_C, pk_C, cert_C.
\]
Note that the join request is in plaintext as it is not directed to a specific verifier (alternatively, the candidate may respond to a probe from nearby verifiers, similar to the reception of SSIDs from nearby Wi-Fi networks, but the end result is the same in terms of knowing the identity of $\mathcal{V}$.) The adversary $\mathcal{M}$ corrupts $m_C(1)$ (e.g., via jamming) to prevent $\mathcal{V}$ from receiving it and poses as a verifier by responding with 
\[
\resizebox{.95\hsize}{!}{$m^{\prime}_M(1)\leftarrow E_{pk_C}[sig_{sk_M}(REPLY,ID_M),
REPLY,ID_M,cert_M,pk_M].$}
\]
 At the same time, $\Mc$ opens a parallel session with $\mathcal{V}$ by sending a request
\[
m_M(1) \leftarrow REQ,ID_M,pk_M,cert_M
\]
to join the platoon of the legitimate verifier $\mathcal{V}$. After $\mathcal{V}$ replies to the request from $\mathcal{M}$, both $\mathcal{V}$ and $\mathcal{C}$ initiate the collection of the RSS samples at the same time. The candidate sends $m_C(2)$ to $\mathcal{M},$ which contains $\Gamma_C$. The adversary obtains $\Gamma_C$ by decrypting it with $sk_M$ and then forwards 
\[
m_M(2) \leftarrow E_{pk_V}[sig_{sk_C}(\Gamma_C, ID_C), \Gamma_C, ID_C]
\]
to $\mathcal{V}$. As $\mathcal{C}$ follows $\mathcal{V}$, the RSS samples provided by $\mathcal{M}$ are highly correlated with those collected by $\mathcal{V}$ and $\mathcal{V}$ accepts. This results in $\mathcal{M}$ being admitted to the platoon, despite the fact that $\mathcal{M}$ {\em does not follow $\mathcal{V}$ within distance $d_{ref}.$}

\begin{figure*}[t]
        \centering
        \scalebox{0.74}{
        \begin{tabular}{ccc}
            Candidate $\mathcal{C}$& &  Verifier $\mathcal{V}$  \\
            \hline
            \multicolumn{1}{l}{\bf Given:} & &\\
             $ID_{C},pk_{C},sk_{C},pk_{CA},cert_C$ & &$ID_{V},pk_{V}, sk_{V},pk_{CA},cert_V$ \\
              & & Parameters ($N$, $M$, $\tau$, $K$, $\alpha$, $\Delta t$) \\

            \multicolumn{1}{l}{{\bf Discovery:}}& &\\
            &$\xleftarrow{ Broadcast\ m_{V}(1)}$& $m_V(1) \leftarrow ID_V, pk_V, cert_V$ \\
            
            \multicolumn{1}{l}{{\bf Initialization:}}& &\\
            
            Verify: $ver_{pk_{CA}}(cert_V,ID_V,pk_V)\overset{?}{=} \text{true}$\\
            
             $m_C(1) \leftarrow E_{pk_V}[sig_{sk_C}(\text{REQ}, ID_C), \text{REQ}, ID_C, pk_C, cert_C]$ &  &  \\
            
            &$\xrightarrow{~m_{C}(1)~}$& Decrypt: $D_{sk_V}[m_C(1)] \leftarrow sig_{sk_C}(\text{REQ}, ID_C)$, $\text{REQ}$, $ID_C, pk_C, cert_C$\\

            & & Verify: $ver_{pk_{CA}}(cert_C,ID_C,pk_C)\overset{?}{=} \text{true}$\\
            
            & & Verify: $ver_{pk_C}[sig_{sk_C}(\text{REQ}, ID_C)],\text{REQ}, ID_C] \overset{?}{=} \text{true}$  \\
            
            & & $m_V(2) \leftarrow E_{pk_C}[sig_{sk_V}(\text{REPLY}, ID_V), \text{REPLY}, ID_V]$\\
            
            Decrypt: $D_{sk_C}[m_V(2)] \leftarrow sig_{sk_V}(\text{REPLY}, ID_V), \text{REPLY}, ID_V$   &  $\xleftarrow{~m_{V}(2)~}$&\\

            Verify: $ ver_{pk_V}[sig_{sk_V}(\text{REPLY}, ID_V),\text{REQ}, ID_V] \overset{?}{=} \text{true}$ & &\\
            
            \multicolumn{1}{l}{{\bf Collection:}}& &\\
            SYNC & & \\
            Collect $\Gamma_C$ & & Collect $\Gamma_V$\\
            
            \multicolumn{1}{l}{{\bf Commit:}}& &\\ $c \leftarrow commit(\Gamma_C, ID_C, r)$ & &\\
            
            $m_C(2) \leftarrow E_{pk_V}[sig_{sk_C}(c), c]$ & &\\
            &$\xrightarrow{~m_{C}(2)~}$& Decrypt: $D_{sk_V}[m_C(2)]=sig_{sk_C}(c), c$\\
            & & Verify: $ver_{pk_C}[sig_{sk_C}(c), c]\overset{?}{=} \text{true}$\\
            
            \multicolumn{1}{l}{{\bf Open:}}& &\\
        
            Delay $\Delta t$ & &\\
             
             $m_C(3) \leftarrow E_{pk_V}[sig_{sk_C}(\Gamma_C, ID_C, r), \Gamma_C, ID_C, r]$ & &  \\
            
            &$\xrightarrow{~m_{C}(3)~}$ & Decrypt: $D_{sk_V}[m_C(3)] = sig_{sk_C}(\Gamma_C, ID_C,r)$, $\Gamma_C$, $ID_C,r $\\
            
            & & Verify: $ver_{pk_C}[Sig_{sk_C}(\Gamma_C, ID_C,r)],\Gamma_C, ID_C,r] \overset{?}{=} \text{true}$ \\
            
            & & Verify: $c\overset{?}{=}commit(\Gamma_C, ID_C,r)$\\

            & & Verify: $t_{commit}-t_V(n)\overset{?}{<}\epsilon$\\ 
          
          \multicolumn{1}{l}{{\bf Verification:}}& &\\
          & & Align $\Gamma_C$, $\Gamma_V$,\\ 
           
          & & Form $\Gamma_C^k$, $\Gamma_V^k$,\\
           
          & & Compute $\{\rho(1), \rho(2), \cdots, \rho(K)\}$,\\
           
          & & Verify: $\sum_{k=1}^{K} \frac{I(\rho(k) \geq \tau)}{K} \geq \alpha \overset{?}{=} \text{true}$ \\

          \multicolumn{1}{l}{{\bf Continuous following verification:}}& &\\ 
          & & Repeat collection and verification\\ 

            \hline
        \end{tabular}}
         \caption{The PoF protocol with a commitment phase that assumes the identity of the verifier is unknown to the candidate.}
     \label{fig:commit_PoF}
    \vspace{-0.2in}
\end{figure*}

\noindent {\bf A MiTM-resistant PoF protocol.} 
To defeat this type of MiTM attack, we amend our protocol to include a commitment scheme with a delayed opening phase that renders the RSS samples obtained by the adversary stale. The commitment scheme satisfied both the hiding and binding properties and can be implemented with any of the known methods such as using one-way functions (e.g., hashe functions) \cite{halevi1996practical}. The updated version of our protocol is shown in Fig.~\ref{fig:commit_PoF}. The changes compared with  the original protocol are as follows. Since the candidate is not targeting a specific verifier ($pk_V$ is not preloaded to $\mathcal{C}$), we have included a discovery phase where the candidate responds to a probe by a verifier. Initially, a verifier would broadcast its credentials to allow discovery by candidates by sending 
\[
m_V(1) \leftarrow ID_V, pk_V, cert_V.
\]
$\mathcal{C}$ will first verify the public key of the verifier and respond with the same join request message as in the original protocol
\[
m_C(1) \leftarrow \text{REQ}, ID_C, pk_C, cert_C.
\]
Upon the verification of the candidate's public key, the verifier will respond with the reply message indicating the start time for the RSS sample collection. After the collection of the RSS sample set $\Gamma_C,$ the candidate commits to $\Gamma_C$ by setting
\[
c \leftarrow commit(\Gamma_C, ID_C, r)
\]
where $commit$ is a commitment function satisfying the hiding and binding properties. The hiding property prevents the adversary from collecting the RSS values from $\Cc$ until $\Cc$ opens the commitment. The binding property prevents the adversary from committing early to the verifier and then changing his committed value once it receives the RSS samples from $\Cc.$  The commitment sent by $\Cc$ commits to the RSS set $\Gamma_C,$ the $ID_C,$ and is randomized by the nonce $r$. Specifically, $\Cc$ sends 
\[
m_C(2) \leftarrow E_{pk_V}[sig_{sk_C}(c), c]
\]
to $\mathcal{V}$ who decrypts the commitment value $c$ and verifies $\mathcal{C}$'s signature, but cannot obtain $\Gamma_C$ at this stage. $\mathcal{C}$ waits for a period $\Delta t$ before it opens the commitment. The delay $\Delta t$ is set to the time that it takes for RSS samples in $\Gamma_C$ to decorrelate with the samples that can be collected after the commitment has been opened. Upon passing of time $\Delta t,$ $\Cc$ opens the commitment by sending 
\[
m_C(3) \leftarrow E_{pk_V}[sig_{sk_C}(\Gamma_C, ID_C, r), \Gamma_C, ID_C, r]
\]
to $\mathcal{V}$. $\mathcal{V}$ verifies $\mathcal{C}$'s signature and also verifies the commitment by checking if $c=commit(\Gamma_C, ID_C, r).$ Moreover, $\mathcal{V}$ ensures that the opening of the commitment occurred in a timely manner. Let $t_{commit}$ denote the time that $\mathcal{V}$ receives the commitment and $t_V(n)$ be the timestamp of the last RSS sample in $\Gamma_V$. $\Vc$ checks if the commitment occurred right after the collection of the RSS samples has terminated. 
\[
t_{commit} - t_V(n) < \epsilon,
\]
where $\epsilon$ is some small value that accounts for the synchronization error between $\mathcal{C}$ and $\mathcal{V}$ and the transmission delay of the commitment (the propagation delay is relatively negligible). Here, $\epsilon << \Delta t$. As we will see in our MiTM security analysis, this prevents $\mathcal{M}$ from committing late in his parallel session with $\mathcal{V}$ so it can receive the opening message from $\mathcal{C}$ first and then send his own commitment to $\mathcal{V}$. 
The remaining of the protocol proceeds in the same way as  the original PoF with the verifier correlating the RSS values $\Gamma_C$ that are received when the commitment is opened with the RSS values that $\mathcal{V}$ has collected over the same period of time. \\

\begin{figure}[t]
\centering
\setlength{\tabcolsep}{-2pt}
\begin{tabular}{c}
\includegraphics[width=0.95\columnwidth]{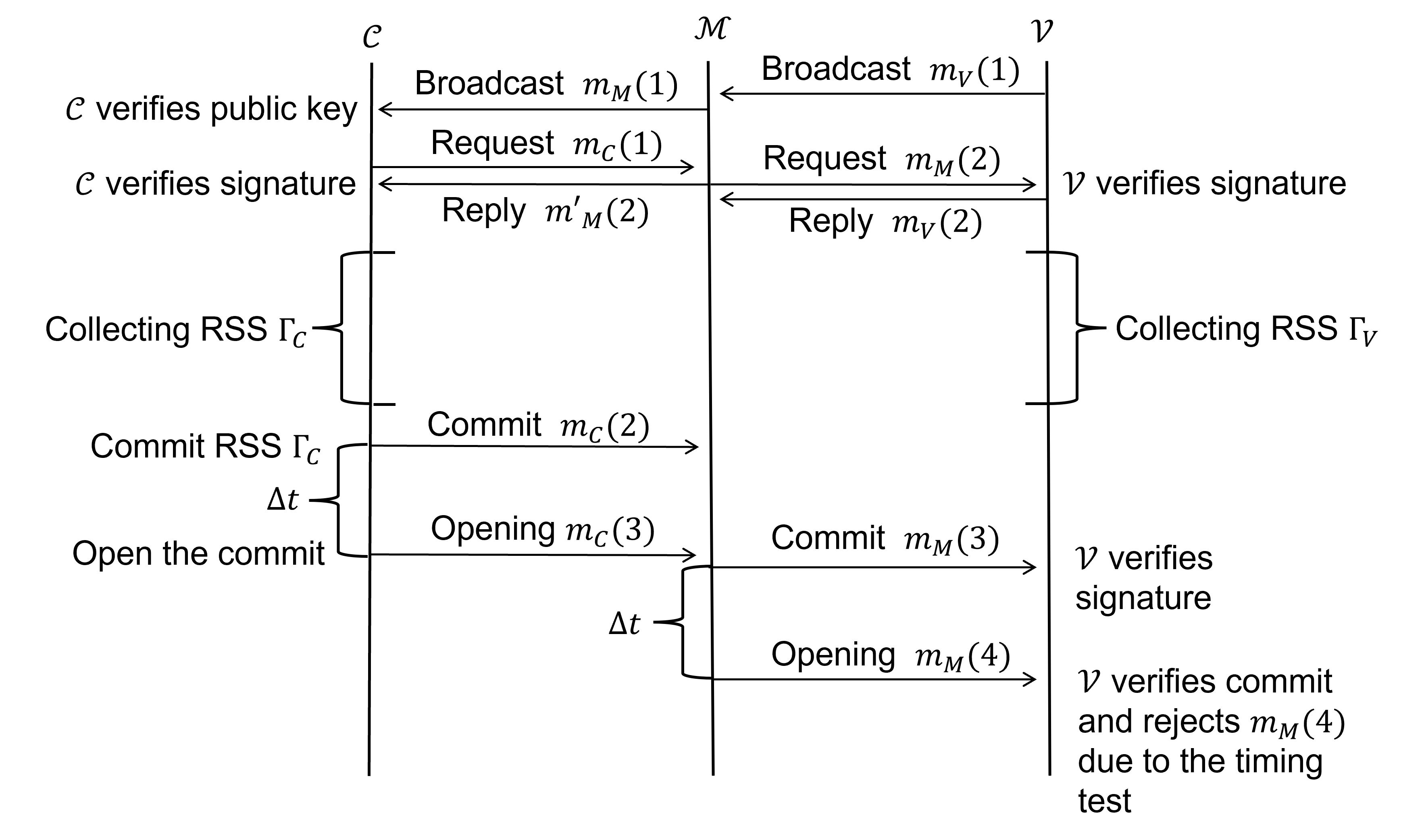}\\
(a) \\
\includegraphics[width=0.95\columnwidth]{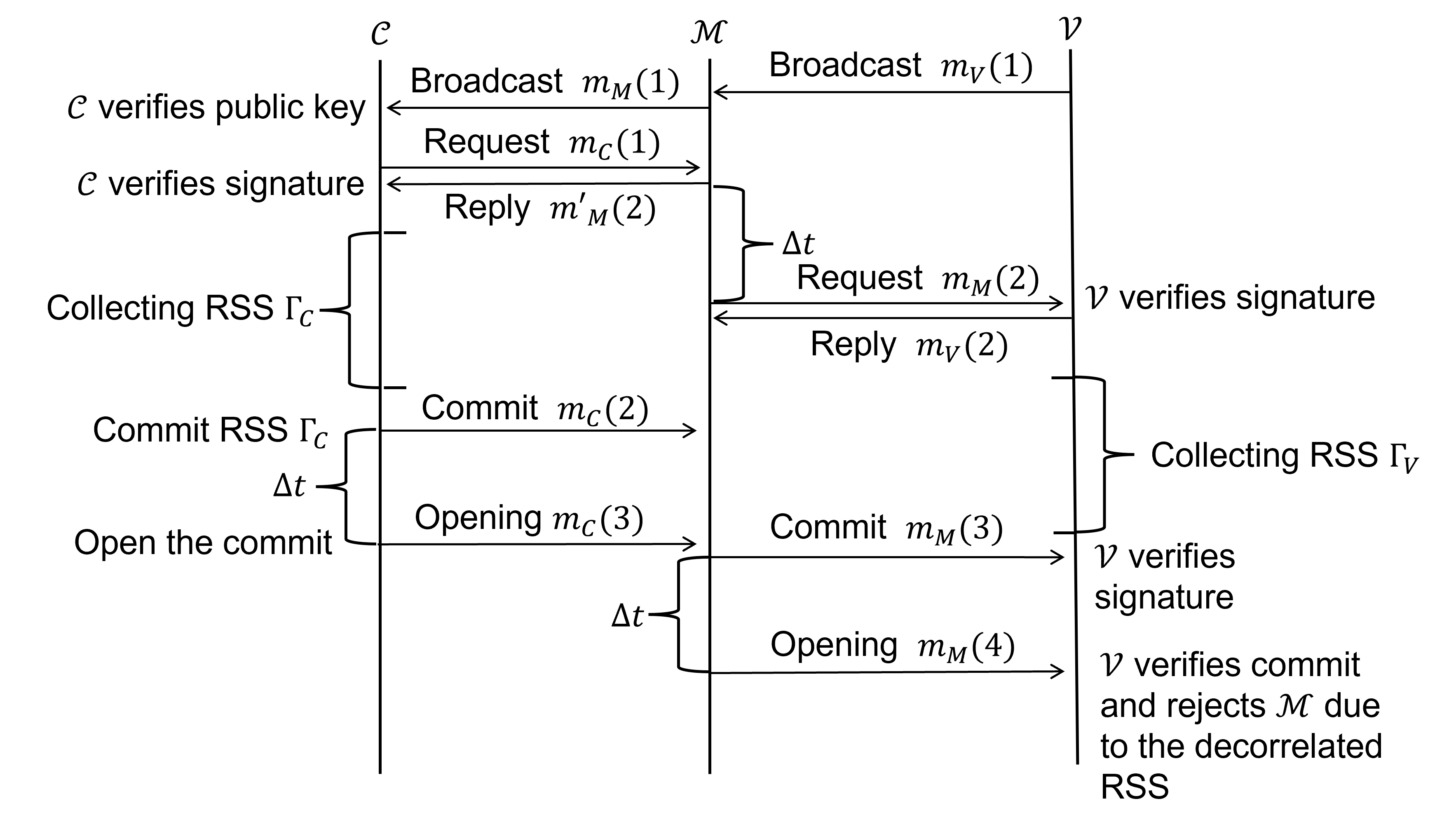}\\
(b)
\end{tabular}
\caption{MiTM attacks on the PoF protocol when $ID_V$ is unknown to $\mathcal{C}$ beforehand. In (a), $\mathcal{M}$ opens a parallel session with $\mathcal{C}$ and $\mathcal{V}$ at the same time, but fails the timing test because it commits to $\mathcal{V}$ after $\mathcal{C}$ opens his commitment. In (b), $\mathcal{M}$ delays the parallel session with $\mathcal{V}$ by $\Delta t$ to satisfy the timing test. However, the samples collected from $\mathcal{C}$ and $\mathcal{V}$ are decorrelated due to the delay $\Delta t.$}
\label{fig:MitM3}
\vspace{-0.2in}
\end{figure}

\noindent {\bf Resistance to MiTM attacks.} 
We now demonstrate that the commitment-based PoF protocol is resistant to MiTM attacks. We consider two possible MiTM attack timelines. In the timeline shown in Fig.~ \ref{fig:MitM3}(a), $\mathcal{M}$ opens two parallel sessions as follows. Upon initialization of the PoF protocol, $\mathcal{M}$ jams $m_V(1)$ reactively to prevent the communication of $\mathcal{C}$ with a legitimate verifier. $\mathcal{M}$ immediately spoofs a verifier by sending 
\[
m_M(1) \leftarrow ID_M, pk_M, cert_M
\]
to $\mathcal{C}.$ The $\Cc$ verifies the public key of $\mathcal{M}$ and responds with
\[
m_C(1) \leftarrow E_{pk_M}[sig_{sk_C}(\text{REQ}, ID_C), \text{REQ}, ID_C, pk_C, cert_C].
\]
At the same time, the adversary opens a parallel session with $\mathcal{V}$ by sending a request
\[
\resizebox{.95\hsize}{!}{$m_M(2) \leftarrow E_{pk_V}[sig_{sk_M}(\text{REQ}, ID_M),\text{REQ}, ID_M, pk_M, cert_M].$}
\]
The verifier checks the signature in the request from $\mathcal{M}$ and assumes that $\mathcal{M}$ wants to join the platoon. $\Vc$ replies with reply 
\[
\resizebox{.95\hsize}{!}{$m_V(2) \leftarrow E_{pk_M}[sig_{sk_V}(REPLY,ID_V),
\text{REPLY},ID_V,pk_V,cert_V].$}
\]
Upon receiving the reply, $\mathcal{M}$ replies to $\mathcal{C}$ with 
\[
\resizebox{.95\hsize}{!}{$m^{\prime}_M(2) \leftarrow E_{pk_C}[sig_{sk_M}(REPLY,ID_M),REPLY,ID_M,pk_M,cert_M].$}
\]
Due to the short succession of messages $m_V(2)$ and $m'_M(2)$, the candidate and the verifier are syncronized and collect RSS samples over the same period. Note that although $\mathcal{M}$ can also collect RSS samples, these will be uncorrelated with the RSS samples collected by $\mathcal{V}$ because $\mathcal{M}$ is far away from $\mathcal{V}.$ Upon the completion of the RSS sample collection, $\mathcal{C}$ will send the commit message  $m_C(2)$ to $\mathcal{M}$.
  \[\resizebox{.95\hsize}{!}{$
  m_C(2) \leftarrow E_{pk_M}[sig_{sk_C}(c),c], ~\text{where}~ c\leftarrow commit(\Gamma_C, ID_C, r)$}
  \]
  Upon receiving the commitment $c$, the adversary cannot obtain the RSS values $\Gamma_C$ due to the hiding property. Moreover, if $\mathcal{M}$ tries to pass along $c=commit(\Gamma_C, ID_C, r)$ to $\mathcal{V}$, because of the binding property, in the opening phase, it is infeasible for $\mathcal{M}$ to make $commit(\Gamma_C, ID_M, r’)=c$, using the RSS revealed by $\mathcal{C}$, its own ID, and another nonce $r'$. Similarly, if $\mathcal{M}$ forges a commitment with  its own RSS, say $c'=commit(\Gamma_M, ID_M, r')$, it cannot change its RSS to the one sent by $\mathcal{C}$ such that $commit(\Gamma_C, ID_M, r'')=c'$ in the opening phase. Thus, the adversary can only wait for $\Delta t$ until $\mathcal{C}$ opens the commitment and then send
   \[
   \resizebox{.95\hsize}{!}{$m_M(3) \leftarrow E_{pk_V}[sig_{sk_C}(c'),c'], ~\text{where}~ c^{\prime}\leftarrow commit(\Gamma_C, ID_M, r^{\prime}).$}
  \]
  to the verifier. However, the adversary will fail the timing test that requires $t_{commit}-t_V(n)<\epsilon$. That is, the commitment $c'$ must arrive at $\mathcal{V}$ before time $\epsilon$ from the collection of the last RSS sample. Since  $\mathcal{C}$ opens the commitment after time $\Delta t$, the adversary does not have a valid $\Gamma_C$ to generate $c'$ in time.
  
  An alternative strategy for the adversary is shown in Fig.~\ref{fig:MitM3}(b). In this strategy, $\mathcal{M}$ opens two parallel sessions with $\mathcal{C}$ and $\mathcal{V}$ posing as a verifier and as a candidate, respectively. However, the opening of the second session is delayed by $\Delta t.$ This allows $\Mc$ to receive the opening of the commitment from $\mathcal{C}$ in time to pass the timing test at $\mathcal{V}.$ However, the RSS samples collected by $\mathcal{V}$ are now delayed by $\Delta t$ compared to the samples collected by $\mathcal{C}$. In addition, for a mobile verifier  $\mathcal{V}$ that  moves at a speed of $v$, the distance between $\mathcal{C}$ and $\mathcal{V}$ has been increased by $\Delta t \times v$ when $\mathcal{V}$ starts to collect RSS. If $\Delta t$ is selected to be large enough, $\Gamma_C$ and $\Gamma_V$ will be decorrelated and therefore $\mathcal{M}$ will fail the verification step.
  
\medskip
\begin{figure}[t]
\centering
\setlength{\tabcolsep}{-2pt}
\begin{tabular}{cc}
    \includegraphics[width=0.5\columnwidth]{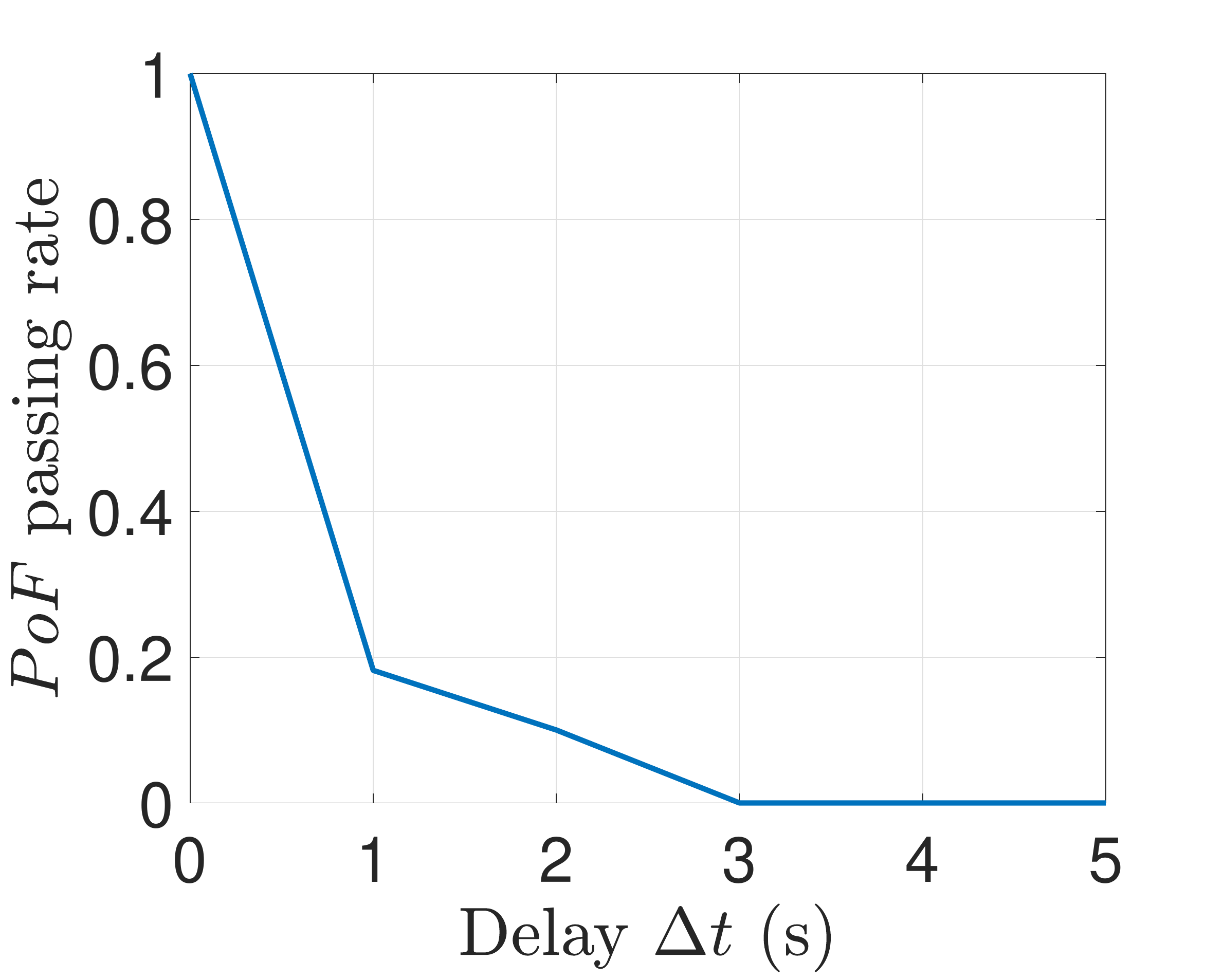} &  \includegraphics[width=0.5\columnwidth]{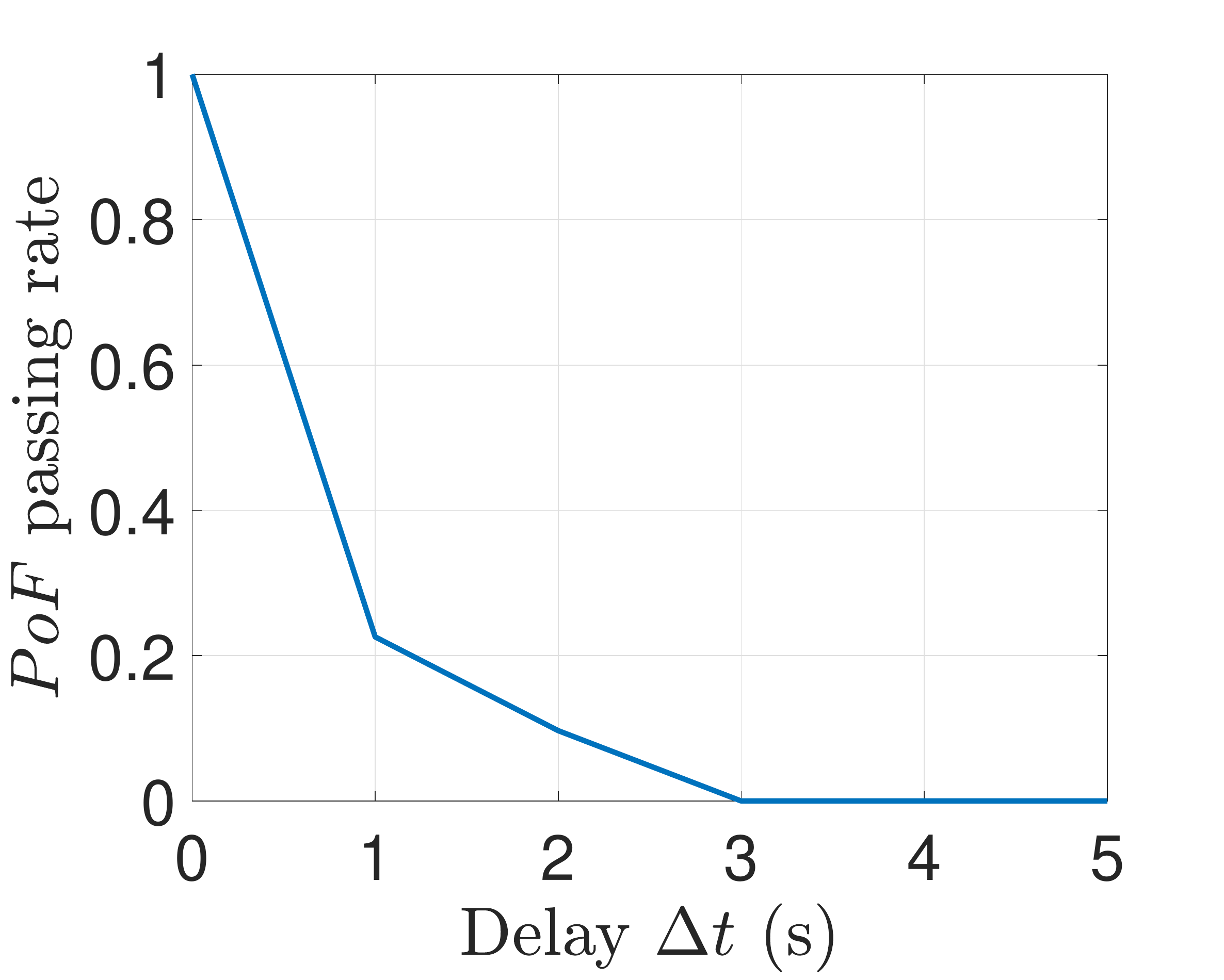}\\
    (a) urban & (b) highway
    \end{tabular}
\caption{PoF passing rate as a function of $\Delta t$ for two environments.}
\label{fig:deltat}
\vspace{-0.2in}
\end{figure}
\noindent {\bf Setting the opening delay $\Delta t$.} To select the opening delay $\Delta t$, we evaluated the PoF protocol passing rate as a function of the delay $\Delta t$ in the collection of RSS samples for the urban and highway environments described in the paper. The driving routes are the same as those shown in Fig. \ref{fig:exp_route}. Figure~\ref{fig:deltat} show the PoF passing rate as a function of $\Delta t.$ The passing rate is averaged over 22 PoF runs in the urban environment and 33 runs on the highway. All the parameters ($N, K, M, \tau, \alpha$) used are the same to Sec. \ref{sec:urban} and Sec. \ref{sec:highway} in different environment against following-afar adversary.

The results show that when $\Gamma_V$ and $\Gamma_C$ are misaligned by at least $\Delta t = 3$ seconds, the passing rate becomes zero. By including a commitment scheme on the RSS values with a delayed opening of at least three seconds, a MiTM attack can be defeated even if the identity of the $\Vc$ is not known to $\mathcal{C}$.

\end{document}